\documentclass[apj]{emulateapj}
\usepackage{graphicx}
\usepackage{epstopdf}
\usepackage{amsmath,hyperref}

\shorttitle{The Radial Velocity Experiment (RAVE): Fourth data release}
\shortauthors{G.~Kordopatis et al.}

\begin{document}

\newcommand{\teff}{T$_{\rm eff}$}
\newcommand{\logg}{$\log~g$}
\newcommand{\feh}{$\rm [Fe/H]$}
\newcommand{\meta}{${\rm [M/H]}$}
\newcommand{\vrad}{$V_{\rm HRV}$}
\newcommand{\afe}{${\rm [\alpha/Fe]}$}
\newcommand{\kms}{km~s$^{-1}$}
\newcommand{\totalspectranumber}{482~430}
\newcommand{\totalstarnumber}{425~561}
\newcommand{\totalatmnumber}{450~055}
\newcommand{\totalchemnumber}{313~338}
\newcommand\temp{$T_{\fontsize{6}{6}\selectfont \mbox{eff}}$}\normalfont
\newcommand\met{[M/H]}
\newcommand\metc{[m/H]$_{\fontsize{6}{6}\selectfont \mbox{chem}}$}\normalfont

\defcitealias{Kordopatis11a}{K11}
\defcitealias{Boeche11}{B11}
\defcitealias{Zwitter10}{Z10}

\title{The Radial Velocity Experiment (RAVE): Fourth data release}

\author{G.~Kordopatis$^{1}$,~  
	G.~Gilmore$^{1}$,~ 
   	M.~Steinmetz$^{2}$,~ 
	C.~Boeche$^{3}$,~ 
	G.M.~Seabroke$^{4}$,~ 
	A.~Siebert$^{5}$,~ 
	T.~Zwitter$^{6,7}$,~  
	J.~Binney$^{8}$,\\
	P.~de~Laverny$^{9}$,~ 
	A.~Recio$-$Blanco$^{9}$,~
	M.E.K.~Williams$^{2}$,~ 
	T.~Piffl$^{2}$,~ 
	H.~Enke$^{2}$,~ 
	S.~Roeser$^{3}$,~ 
	A.~Bijaoui$^{9}$,~ 
	R.F.G.~Wyse$^{10}$,~ \\
	K.~Freeman$^{11}$,~ 
		U.~Munari$^{12}$,~ 
		I.~Carillo${^2}$,~
	B.~Anguiano$^{13,14}$,~
D.~Burton$^{11,15}$,~
R.~Campbell$^{16}$,~
C.J.P.~Cass$^{15}$,~\\
K.~Fiegert$^{15}$,~
M.~Hartley$^{15}$,~
Q.A.~Parker$^{13,14,17}$,~
W.~Reid$^{14,17}$,~
	A.~Ritter$^{18}$,~
K.S.~Russell$^{15}$,~
 M.~Stupart$^{15}$,~
F.G.~Watson$^{15}$,~\\
  O.~Bienaym\'e$^{5}$,~
	J.~Bland$-$Hawthorn$^{19}$,~
		 O.~Gerhard$^{20}$,~
	B.K.~Gibson$^{21}$,~
		E.~K.\ Grebel$^{3}$,~
	A.~Helmi$^{22}$,~\\
	     	J.F.~Navarro$^{23}$,~
	C.~Conrad$^{2}$,~
	B.~Famaey$^{5}$,~
	C.~Faure$^{5}$,~
	A.~Just$^{3}$,~
	J.~Kos$^{6}$,~
	G.~Matijevi{\v c}$^{6}$,~
	 P.J.~McMillan$^{8}$,~\\
	I.~Minchev$^{2}$,~
	R.~Scholz$^{2}$,~
	S.~Sharma$^{19}$,~
	A.~Siviero$^{2}$,~
	 E.~Wylie de Boer$^{11}$,~
	M.~\v{Z}erjal$^{6}$}

\email{gkordo@ast.cam.ac.uk}
	\affil{${^1}$Institute of Astronomy, University of Cambridge, Madingley Road, Cambridge, CB3 0HA, UK\\
$^{2}$Leibniz-Institut f\"ur Astrophysik Potsdam, An der Sternwarte 16, D-14482 Potsdam, Germany\\
$^{3}${Astronomisches Rechen-Institut, Zentrum f\"ur Astronomie der Universit\"at Heidelberg, M\"onchhofstr.\ 12-14, D-69120 Heidelberg, Germany}\\
$^{4}${Mullard Space Science Laboratory, University College London, Holmbury St. Mary, Dorking, Surrey, RH5 6NT, UK}\\
$^{5}${Observatoire Astronomique de Strasbourg, Universit\'e de Strasbourg, CNRS, UMR 7550, 11 rue de l'Universit\'e, 67000 Strasbourg, France}\\
$^{6}${Faculty of Mathematics and Physics, University of Ljubljana, Jadranska 19, 1000 Ljubljana, Slovenia}\\
$^{7}${Center of Excellence SPACE-SI, Askerceva cesta 12, 1000 Ljubljana, Slovenia}\\
$^{8}${Rudolf Peierls Centre for Theoretical Physics, Keble Road, Oxford, OX1 3NP, UK}\\
$^{9}${Laboratoire Lagrange, UMR~7293, Universit\'e de Nice Sophia Antipolis, CNRS, Observatoire de la C\^ote d'Azur, BP4229, 06304, Nice, France}\\
$^{10}${Johns Hopkins University, 3400~N Charles Street, Baltimore, MD~21218, USA}\\
$^{11}${Research School of Astronomy and Astrophysics, Australian National University, Cotter Rd., Weston, ACT 2611, Australia}\\
$^{12}${INAF National Institute of Astrophysics, Astronomical Institute of Padova, 36012 Asiago (VI), Italy}\\
$^{13}${Australian Astronomical Observatory, PO Box 915, North Ryde NSW 1670}\\
$^{14}${Department of Physics and Astronomy, Macquarie University, Sydney, NSW, 2109} \\
$^{15}${Anglo-Australian Observatory, PO Box 296, Epping, NSW 1710, Australia}\\
$^{16}${Western Kentucky University, Bowling Green, Kentucky, USA} \\
$^{17}${Macquarie Research Centre for Astronomy, Astrophysics and Astrophotonics, Sydney, NSW, 2109, Australia}\\ 
$^{18}${National Central University, 300 Zhongda Rd., Zhongli City, Taoyuan County, 325 Taiwan (R.O.C.)}\\
$^{19}${Sydney Institute for Astronomy, School of Physics A28, University of Sydney, NSW 2006, Australia}\\
$^{20}${Max-Planck-Institut fuer Ex. Physik,  Giessenbachstrasse,  D-85748 Garching b. Muenchen,	Germany}\\
$^{21}${ Jeremiah Horrocks Institute, University of Central Lancashire, Preston, PR1 2HE, UK}\\
$^{22}${Kapteyn Astronomical Institute, University of Groningen, PO Box 800, NL-9700 AV Groningen, the Netherlands}\\
$^{23}${ Department of Physics and Astronomy, University of Victoria, Victoria, BC, Canada}}


\begin{abstract}
We present the stellar atmospheric parameters (effective temperature, surface gravity, overall metallicity), radial velocities, individual abundances and distances determined for \totalstarnumber~stars, which constitute the fourth public  data release of the RAdial Velocity Experiment (RAVE). The stellar atmospheric parameters are computed  using a new pipeline, based on the algorithms of MATISSE and DEGAS. The spectral degeneracies and the 2MASS photometric information are now better taken into consideration, improving the parameter determination compared to the previous RAVE data releases.   The individual  abundances for six elements (magnesium, aluminium, silicon, titanium, iron and nickel) are also given, based on a special-purpose pipeline which is also improved compared to that available for the RAVE DR3 and Chemical DR1 data releases. Together with photometric information and proper motions, these data can be retrieved from the RAVE collaboration website and the Vizier database.
\end{abstract}

\keywords{catalogs --- stars:~fundamental parameters  --- stars:~abundances --- techniques:~spectroscopic --- surveys}
\section{Introduction}
The assembly history of the Milky Way can be obtained by analysing the positions, the kinematics, the ages and the chemical compositions of  large statistical samples of Galactic stars \citep{Freeman02}. 
In addition to the identification and characterisation of hierarchical signatures \citep[e.g.][]{Helmi99, Abadi03, Sales09}, the precise measurement of the age-metallicity relation in the solar neighbourhood, for a very large sample of stars, allows us to establish, among much else,  the strength and the importance of radial migration in the Galaxy, perhaps a key ingredient for Galactic evolution \citep{Sellwood02, Vanderkruit11}.
  
Ideally, one would need stellar spectra and precise distances to achieve such a goal. Nevertheless, even in the case where parallaxes are not available, it is still possible to estimate statistically valuable ages and distances of the stars by measuring from their spectra  their atmospheric parameters (effective temperature, \teff, surface gravity, \logg, overall metallicity, \meta \footnote{The stellar overall metallicity is defined as ${\rm [M/H]}=\log \left ( \frac{N(M)}{N(H)} \right )_\star -  \log \left ( \frac{N(M)}{N(H)} \right )_\odot$ where N represents the number density and M all the elements heavier than \ion{He}{0}. }) and projecting them afterwards on theoretical stellar isochrones \citep{Breddels10, Zwitter10, Burnett11, Kordopatis11b, Binney13}.

In the last decade, the advent of multi-fibre spectrographs, combined with large telescopes, has allowed the astronomical community to obtain such  very large amounts of spectra in order to explore the evolution of our Galaxy. Until the release of the first substantial catalogue of Gaia (estimated to be available in early 2017), the already current main large spectroscopic surveys of the Milky Way are the {\it Sloan Extension for Galactic Understanding and Exploration} (SEGUE), the {\it RAdial Velocity Experiment} (RAVE), the {\it APO Galactic Evolution Experiment} (APOGEE), the {\it LAMOST Experiment for Galactic Experiment and understanding} (LEGUE), the {\it GALactic Archaeology with HERMES} (GALAH)  and the {\it Gaia-ESO Survey} (GES).
 
RAVE\footnote{\url{http://www.rave-survey.org}} began observations in 2003, and since then has released three {  data releases (noted DR hereafter)}: DR1 in 2006, DR2 in 2008 and DR3 in 2011 \citep{Steinmetz06, Zwitter08, Siebert11}.  
Furthermore, three catalogues with spectrophotometrically derived distances were published \citep{Breddels10, Zwitter10, Burnett11} and one catalogue with abundances for the individual elements Magnesium (Mg), Aluminium (Al), Silicon (Si), Calcium (Ca), Titanium (Ti), Iron (Fe), and Nickel (Ni)  \citep{Boeche11}.
RAVE is a magnitude-limited survey of stars randomly selected in the southern celestial hemisphere. The original design was to only observe stars in the interval $9<I<12$ but the actual selection function includes stars both brighter and fainter (see Sect.~\ref{Sect:Input_catalogue}). The spectra are obtained from the 6dF facility on the 1.2m Anglo-Australian Observatory's Schmidt telescope in Siding Spring, Australia, where three field plates with 150 robotically positioned fibres are used in turn.
The effective resolution of RAVE is $R=\lambda/\Delta \lambda \sim7500$ and the wavelength range coverage is around the infrared  ionised Calcium triplet (IR \ion{Ca}{2}, $\lambda \lambda 8410-8795$\AA), one of the widely used wavelength ranges for Galactic archaeology.
Up to now, previous RAVE catalogues have released 83~072 radial velocities for 77~461 stars, and 41~672 sets of atmospheric parameters for 39~833 stars.  These produced many valuable studies.

 \cite{Seabroke08} used the symmetry of the velocity distributions to rule out the presence of the Sagittarius stream or the Virgo overdensity in the solar neighbourhood. 
Recently, RAVE data allowed \cite{Williams11} to discover the Aquarius stream and  \cite{Antoja12} to identify additional moving groups in the Galactic disc, consistent with dynamical models of the effects of the bar and the spiral arms.
\cite{Pasetto12a, Pasetto12b} used RAVE data in order to constrain the solar motion relative to the local standard of rest, the rotational lag of the thick disk component, and the components of the thin disk velocity ellipsoids in the solar neighbourhood.
 Additionally, \cite{Siebert11a,Siebert12} measured the mean galactocentric radial velocity of stars in the extended solar neighbourhood and constrained the parameters of Milky Way spiral structure.
 \cite{Williams13} identified differences in the velocity distribution between the north and the south of the Galactic plane with indications of a rarefaction-compression pattern, suggestive of wave-like behaviour.
Furthermore, \cite{Ruchti11,Wilson11} and \cite{Fulbright10} studied the chemo-dynamical information of the thick disc and metal-poor stars of the Galaxy, whereas \cite{Matijevic10,Matijevic11} used RAVE to study single-lined and double-lined binary stars.

Here we present the new DR4 catalogue, which includes \totalspectranumber~spectra. In order to obtain the atmospheric parameters, an updated version of the \citet{Kordopatis11a} pipeline is used, which  combines the DEGAS decision-tree method \citep[{\it DEcision tree alGorithm for AStrophysics}, ][]{Bijaoui12} as well as the MATISSE projection algorithm \citep[{\it MATrix Inversion for Spectral SynthEsis},][]{Matisse} and which takes into account, for the first time, the {\it Two Micron All Sky Survey} (2MASS)  photometric information. This allows us to treat more efficiently the spectral degeneracies than the 
maximum a posteriori 
method used in the previous data releases, reducing  parameter combinations which were found in astrophysically non-justified parts of the (\teff--\logg) space. 

Furthermore, this DR4 catalogue takes advantage of a multitude of new calibration data sets that have been collected recently in order to obtain reliable metallicities. The calibration libraries consist of RAVE and RAVE-like spectra for which there are parameter determinations derived from high-resolution spectroscopy, and stars selected from open and globular clusters of well-known metallicities. 
In addition to the stellar atmospheric parameters published for the newly observed targets, these calibration spectra and the new pipeline have also been applied in order to re-estimate and re-calibrate the parameters  from the previous data-releases.  Then, given the new effective temperatures, surface gravities and metallicities, individual abundances are also computed for the entire data set, using an updated version of the \cite{Boeche11} chemical pipeline. Finally, new distances are also inferred, using the methods presented in \citet{Zwitter10} and \citet{Binney13}.

The paper is structured as follows: 
First, in Sect.~\ref{Sect:Input_catalogue} we present the new input catalogue of RAVE, and in Sect.~\ref{sect:pipeline}, we present the pipeline that is used in order to obtain the atmospheric parameters, by recalling the basic equations  and the updates of the MATISSE and DEGAS algorithms that are in the core of this new parameterisation pipeline. Then,  in Sect.~\ref{sect:Validation_pipeline}, we show which calibration data sets are used and discuss how the calibration relation is obtained. Section~\ref{sect:Chemi_pipe} presents the updates on the chemical pipeline that is used in order to measure the individual abundances. Based on the presented pipelines and the calibration relation that is established, we present in Sect.~\ref{sect:dr4} the DR4 atmospheric parameters catalogue, as well as a comparison with the previous DR3 parameters, in particular for the metallicities. 
Sections~\ref{sect:proper_motions}, \ref{sect:Vrad}, \ref{sect:distances} and \ref{sect:APASS} present the DR4 catalogue for the proper motions, the radial velocities and the stellar distances, as well as a description of  the {\it AAVSO Photometric All-Sky Survey} (APASS) photometry which is recommended to be used as it becomes available. Finally, Sect.~\ref{sect:conclusion} sums up. 

\section{New RAVE input catalogues}
\label{Sect:Input_catalogue}

The RAVE wavelength window of 8410--8795\AA~  implies that an $I$-band selection of the targets is the most appropriate.  However, when RAVE started observing in April 2003, there was no southern sky $I$-band photometric survey spanning the RAVE magnitude range.  Therefore, the original input catalogue was constructed by deriving $I$ magnitudes from Tycho-2 photometry and filling in Tycho-2's incompleteness at the faint end of the RAVE magnitude range with SuperCOSMOS photographic $I$ magnitudes (see DR1 paper for more details).  The {\it DEep Near Infrared Survey of the Southern Sky} (DENIS) DR2 was available in May 2005 \citep{deniscat2003}, which included Gunn-$I$ photometry at 0.82~nm, but did not have sufficient coverage (55\% of the southern sky) to be used as the basis for a new RAVE input catalogue.  RAVE DR1, DR2 and DR3 were solely observed from the original input catalogue.  DR3 was the last release to be solely observed from the original input catalogue, thereby concluding the pilot survey. 

DENIS DR3 was available in September 2005 \citep{denis2005} and did have sufficient coverage (80\% of the southern sky) to be used as the basis for a new RAVE input catalogue.  DENIS entries within the RAVE magnitude range ($9<I<12$ mag) were selected, including saturated entries (DENIS saturates at $I < 9.8$~mag -- see later discussion).  Entries with $I$ magnitude error $= 1$ indicate a non-estimated error so these entries were rejected.  Then the remaining entries were cross-matched with 2MASS using a 1~arcsec box search.  This search region was chosen because both DENIS and 2MASS were calibrated on the USNO-A2.0 catalogue, which has an astrometric accuracy of 0.5~arcsec.  The nearest DENIS DR3 catalogue entry to a 2MASS star provided the $I$ magnitude for that star (2MASS does not include $I$-band photometry).  DENIS catalogue entries not within a 1~arcsec search box of a 2MASS star were rejected.   

At the edges of the DENIS CCD detector, both the astrometry and photometry become less accurate. For each different scan of DENIS strip overlap regions, the catalogue reports every detection of a source. If a star in this region is imaged more than once, each detection is included in the catalogue because, although the multiple detections have almost identical magnitudes and positions, they cannot be associated with the same source.  Cross-matching DENIS with 2MASS not only removes these multiple detections from the new input catalogue but also provides more accurate astrometry, leading to fibre placement better matching stellar positions on the sky, which results in higher signal-to-noise spectra.  DENIS includes spurious sources due to charge bleeding and diffraction spikes, which do not have matches in 2MASS and so are removed from the new DENIS input catalogue, increasing its efficiency.  

\begin{figure}
\begin{center}
\includegraphics[width=0.5\textwidth]{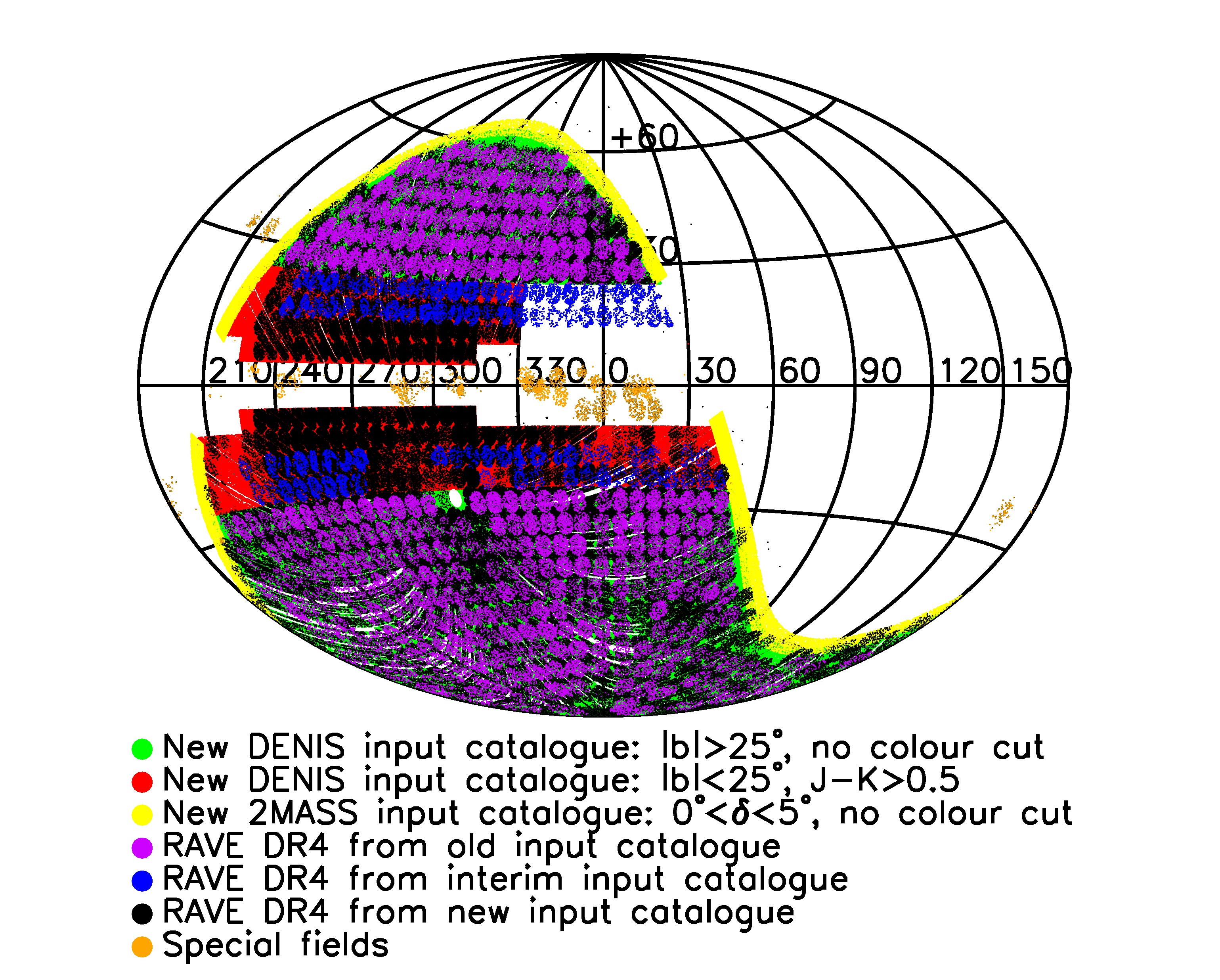}
\end{center}
\caption{Aitoff projection of Galactic co-ordinates of the new input catalogues, colour-coded into no colour cut and colour cut samples.  Overlaid are the RAVE DR4 stars, colour-coded according to their source input catalogue. Their pattern is due to the 6dF field of view (5.9$^{\circ}$).  The original input catalogue was observed with field centre co-ordinates separated by 5.7$^{\circ}$.  The new input catalogue was observed with field centre co-ordinates separated by 5.0$^{\circ}$.}
\label{f:input_aitoff}
\end{figure}

Comparison of Figure~\ref{f:input_aitoff} with \citet{Siebert11} Figure 17 shows that, in addition to covering the sky area of the original RAVE input catalogue, the new DENIS input catalogue has the major new feature of extending to lower Galactic latitudes ($b$).  The aim of the extension towards the Galactic anti-rotation direction ($225^{\circ} < l < 315^{\circ}$, $5^{\circ} < |b| < 25^{\circ}$) is to observe the outer Galactic disc.  Distances probed at the faint magnitude limit ($I$ = 12 mag) are expected to be  up to $\sim$5~kpc for K-type giants (assuming $M_{I} = +1$~mag and no extinction).  Therefore, at low Galactic latitudes in this direction, giants just reach the nominal `edge' of the stellar disc. Symmetric sampling about the Galactic plane will constrain the disc warp and flare.  Distances probed at $I = 12$~mag are only up to $\sim$400~pc for G-type dwarfs (assuming $M_{I} = +4$ mag).  Therefore, the most efficient way to observe the outer disc is to apply a colour cut ($J - K > 0.5$~mag) in this region to avoid observing G-type dwarfs and preferentially observe K-type giants.  The Besan\c{c}on Galactic model \citep{Robin03} predicts that in the RAVE magnitude range at $J - K > 0.5$~mag 70\% of stars are KM-type giants and 30\% are KM-type dwarfs.  Although DENIS includes $J$ and $K$ photometry, the colour cut is performed using 2MASS $J$ and $K$.  This is because of the aforementioned edge affect on the DENIS CCDs and due to higher levels of sky image noise in DENIS than 2MASS.  The noise comes from the thermal infra-red background radiation emitted by the instrument itself.  2MASS optics avoid this by including a cold stop, which DENIS optics do not have.

After observing the new DENIS input catalogue from 2006 to 2010, more targets closer to the Galactic plane were required to maintain RAVE's observing efficiency due to sky regions not always being observable from the UK Schmidt Telescope (UKST) at different times of the year.  The aim was still to target red giants by selecting $J - K > 0.5$~mag, thereby rejecting young foreground stars, which have weak Paschen lines that yield less accurate radial velocities.  This selection works with reddening of $E(B-V) < 0.35$ mag.  Therefore to preserve a homogeneous selection function with this colour cut, the new DENIS input catalogue was extended closer to the Galactic plane in regions where $E(B-V) < 0.35$ mag.  The northern Galactic hemisphere of the Galactic bulge has $E(B-V) > 0.35$ mag and so $b < 25^{\circ}$ at $l > 330^{\circ}$ and $l < 30^{\circ}$ is not included in the new extended DENIS input catalogue.  $10^{\circ} < b < 25^{\circ}$ at $l < 225^{\circ}$ and at $315^{\circ} < l < 330^{\circ}$ has $E(B-V) < 0.35$ mag and so is included.  In the southern hemisphere of the bulge, $-10^{\circ} < b < -25^{\circ}$ at $l < 225^{\circ}$, $l > 315^{\circ}$ and $l < 30^{\circ}$ all have $E(B-V) < 0.35$ mag and so are included.

DR4 includes observations of the interim input catalogue, which extended the original input catalogue from $|b| < 25^{\circ}$ to $|b| < 15^{\circ}$ at all $l$.  These observations were taken before the new DENIS input catalogue was available, which is why they sample the northern bulge (blue dots in Figure~\ref{f:input_aitoff}) outside of the new extended DENIS input catalogue colour cut footprint (red dots in Figure~\ref{f:input_aitoff}).  However, there are many observations of the interim input catalogue (blue dots in Figure~\ref{f:input_aitoff}) within the colour cut footprint.  It is important to note that these do not include the colour cut and so include all colours.  These fields can be identified using the Galactic co-ordinates in Figure~\ref{f:input_aitoff} and obsdate $\le$ 20060312.  The special fields outside of the colour cut footprint (orange dots in Figure~\ref{f:input_aitoff}) are specific science and calibration fields and do not include the colour cut.  

More bright targets north of the celestial equator ($\delta \le 5^{\circ}$) were required, again, to maintain RAVE's observing efficiency in bright time.  DENIS's sky coverage is $\delta \le 2^{\circ}$.  Therefore, to extend the input catalogue to $\delta = 5^{\circ}$ required DENIS $I$ to be constructed from 2MASS $J$ and $K$ using:

\begin{equation}
I_{\textrm{DENIS}} = J_{\textrm{2MASS}} + 0.054 + 1.18(J_{\textrm{2MASS}} - K_{\textrm{2MASS}})
\end{equation}

where $\Delta_{\textrm{DENIS}} = 0.15$ mag.  This was done for 2MASS sources with $0^{\circ} \le \delta \le 5^{\circ}$ and $0^{h} < \alpha < 6^{h}$, $7^{h}30 < \alpha < 17^{h}$ and $19^{h}30 < \alpha < 24^{h}$ (yellow dots in Figure~\ref{f:input_aitoff}) to avoid the Galactic plane.  The 2MASS input catalogue spatially overlaps the DENIS input catalogue by $\Delta\delta = 3^{\circ}$ because DENIS's sky coverage is not complete.  The long white areas within the coloured regions in Fig. \ref{f:input_aitoff} are one or more slots (12 arcmin in $\alpha$ by 30$^{\circ}$ in $\delta$), where the observed DENIS strips filling these slots are missing from DENIS DR3. The small white areas within the coloured regions are one or more missing DENIS images (12 arcmin $\times$ 12 arcmin).  The white circular region within the coloured regions is the southern equatorial pole, which is not observed by DENIS.    The width of the missing DENIS strips and images is 3.5\% of the 6dF field of view and so does not pose a fibre configuration problem.  Its effect on RAVE's random selection of targets should be negligible.  It is important to note that observations from the 2MASS input catalogue at  $|b| > 25^{\circ}$ do not include the colour cut and so include all colours.  Figure~\ref{f:input_aitoff} shows that there are only three DR4 fields that spatially overlap the DENIS input catalogue colour cut footprint and the 2MASS input catalogue footprint and so have mixed selection functions.

\begin{figure}
\begin{center}
\includegraphics[width=3.5in]{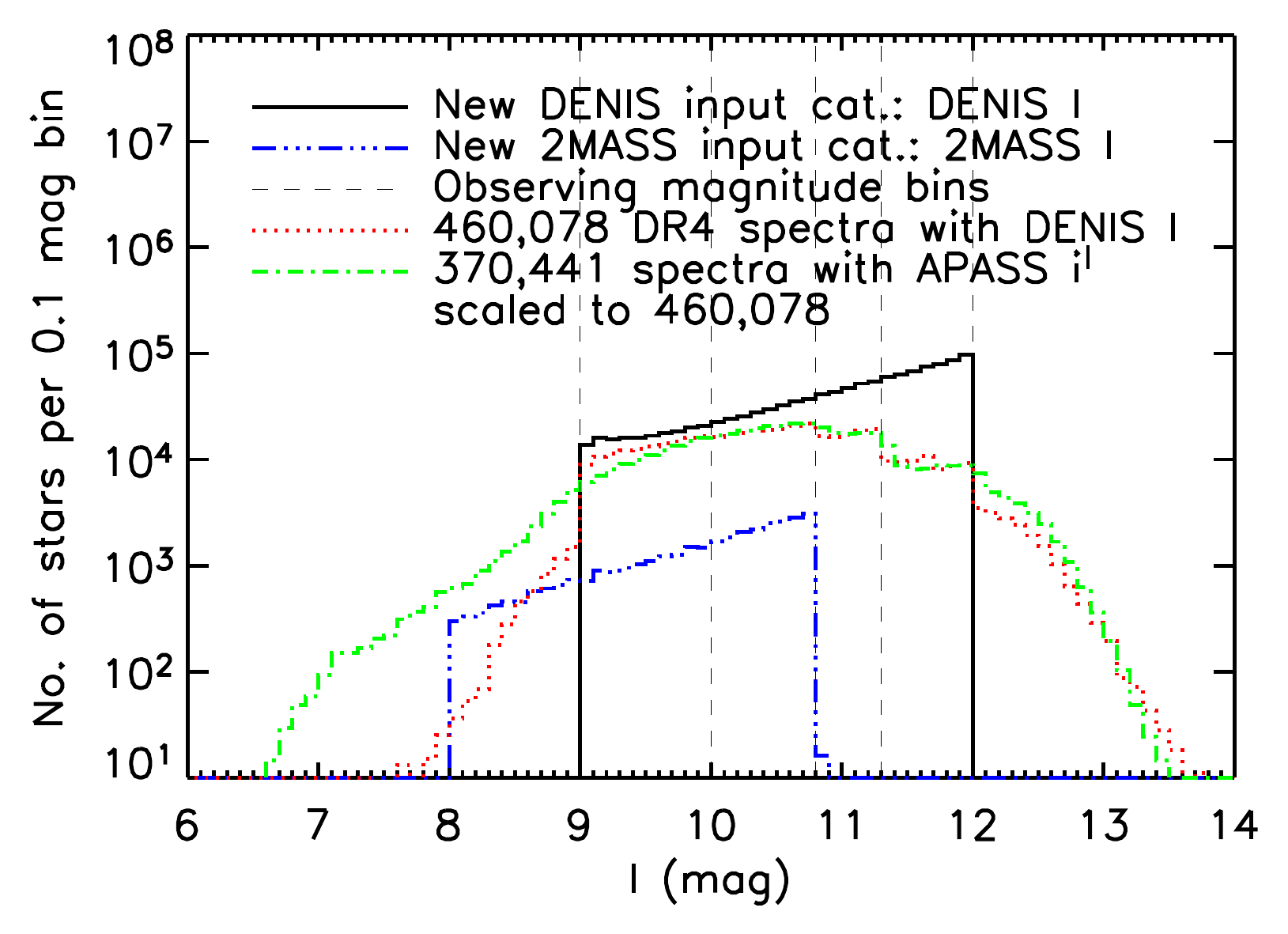}
\end{center}
\caption{Histogram comparing the new input catalogues as a function of $I$ magnitude and the selection functions of all the observed RAVE stars as a function of $I$ magnitude and as a function of APASS $i'$ magnitude, which is similar but not identical to DENIS $I$. }
\label{f:hist_input}
\end{figure}

\begin{figure*}[tbp]
\begin{center}
$\begin{array}{c}
\includegraphics{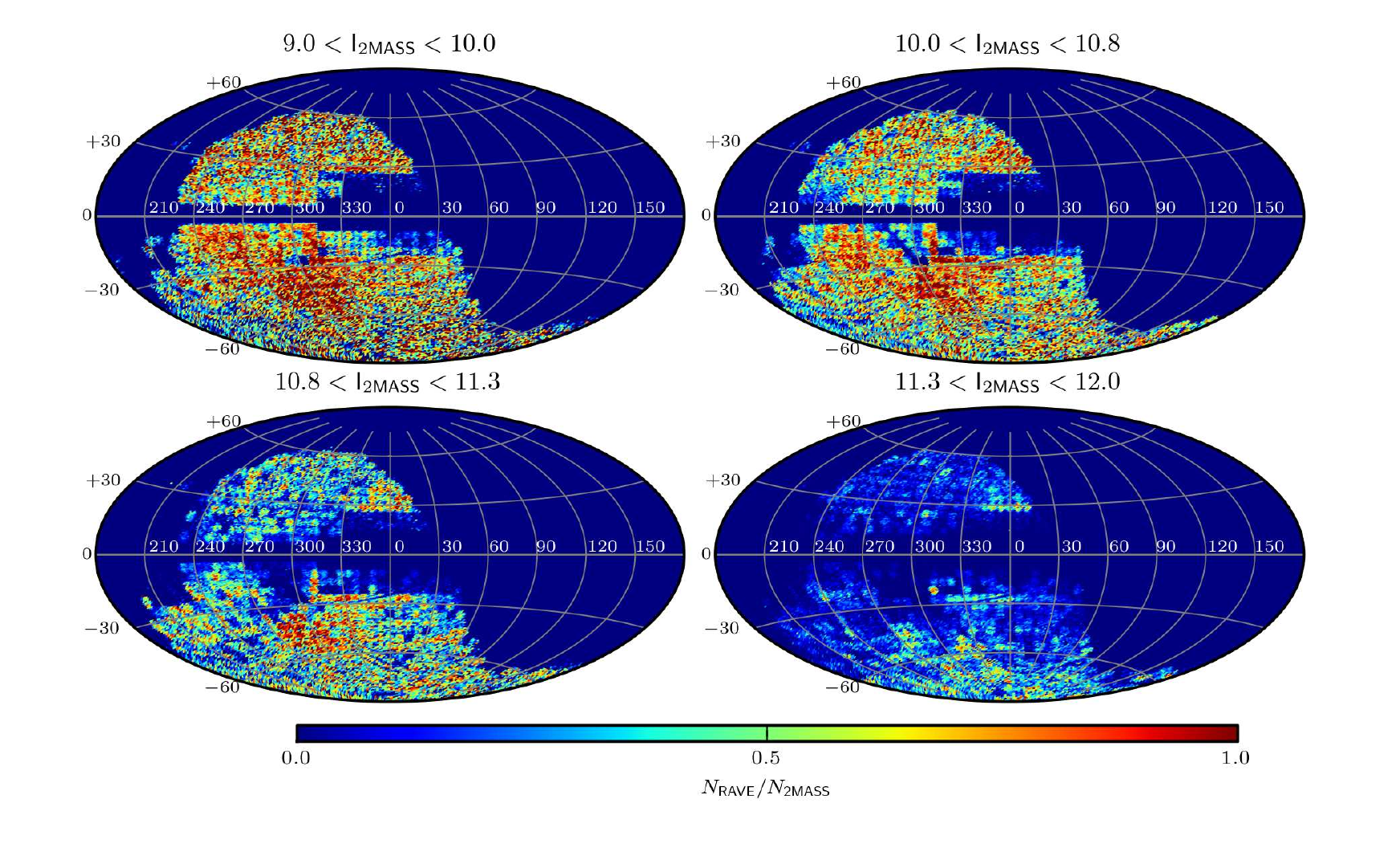} 
\end{array}$
\caption{Aitoff projection of Galactic co-ordinates of the completeness of the stars in the $I_{\rm 2MASS}$ band for which radial velocity measurements are available. Each panel shows a different magnitude bin. Grey-scale coding represents the ratio of RAVE observations to 2MASS stars. Only stars with errors in radial velocity less than 10~\kms\ are shown. }
\label{Fig:Completeness_maps}
\end{center}
\end{figure*}

Figure~\ref{f:hist_input} compares the new input catalogues and the number of observed DR4 spectra as a function of $I$ magnitude.  It emphasises that the observed DR4 spectra have not exhausted the new DENIS input catalogue overall and so are not complete with respect to DENIS overall, although individual fields may be complete.  Indeed, the new DENIS input catalogue is only complete where DENIS has sky coverage.   DR4's completeness with respect to 2MASS is shown in Figure~\ref{Fig:Completeness_maps}.  Figure~\ref{f:hist_input} also shows that the 2MASS input catalogue extends the bright limit of the survey to $I$ = 8 mag.  Each 6dF field set-up only selects new input catalogue targets from one of the following four magnitude bins: $9.0 < I < 10.0$ ($8.0 < I < 10.0$ for the new 2MASS input catalogue), $10.0 < I < 10.8$,  $10.8 < I < 11.3$ and $11.3 < I < 12$ mag, which are visible in Figure~\ref{f:hist_input}.   This minimises the magnitude range within any one 6dF field set-up to be within a bin, meaning exposure times can be scaled more appropriately for all the targets in the same field.  Each field set-up is a random selection of unobserved targets within these bins (apart from designed repeat observations).  Any spectrum within a 6dF field set-up can be adjacent to any other in the same set-up on the CCD but the bins limit the magnitude difference, which also minimises fibre cross-talk.  

Figure~\ref{f:hist_input} compares the selection functions of the new input catalogues and the observed DR4 stars.  The input catalogues have step functions at their bright and faint ends.  However, the observed DR4 stars do not have step functions but extend to brighter and fainter than the input catalogues.  Because DENIS saturates at $I < 9.8$~mag, the selection function of RAVE's new input catalogue and observations need to be compared to an $I$-band survey that doesn't saturate in the RAVE magnitude range.  
The recent advent of the {\it AAVSO Photometric All-Sky Survey} (APASS - see Sect.~\ref{sect:APASS}) means that this can now be explored.  
Future data releases will be supplemented with APASS photometry.  
At the time of writing, APASS $i'$ (internal release DR7) was available for 370,441 spectra and is plotted in Figure~\ref{f:hist_input}, scaled to the number of DR4 spectra. 
 It shows that the distribution of APASS $i'$ approximately agrees with the distribution of DENIS $I$ at the faint end but extends to much brighter magnitudes at the bright end.  This is because DENIS saturates at $I < 9.8$~mag, which means some sources are actually $I < 9$~mag.

\begin{figure}
\begin{center}
\includegraphics[width=3.5in]{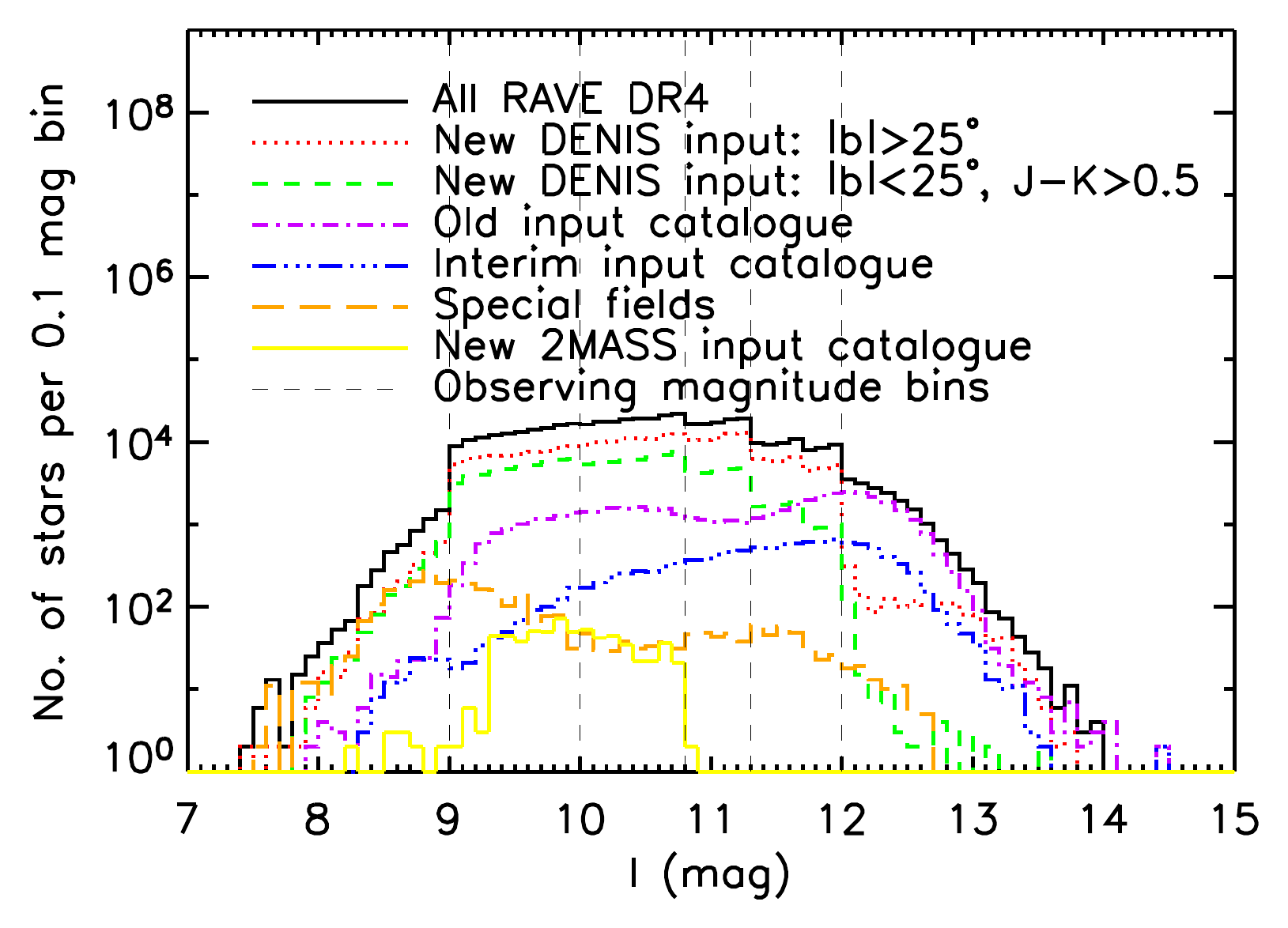}
\end{center}
\caption{Histogram decomposing DR4 into its constituent input catalogues as a function of $I$ magnitude.  Note that this is either DENIS $I$ or 2MASS constructed DENIS $I$ and so are not reliable at $I < 9.8$~mag due to DENIS $I$ saturation.}
\label{f:hist_input_zoom}
\end{figure}

The overall selection function of DR4 is more clearly seen in Fig. \ref{f:hist_input_zoom}.  It has a complex shape due to the observing strategy and because DR4 includes all RAVE observations up to the end of 2012, which have been selected from four different RAVE input catalogues.  Numerically, DR4 is dominated by stars selected from the new DENIS input catalogue ($|b| > 25^{\circ}$).  This means the observing magnitude bins are visible in the DR4 selection function.  It also means that overall the new input catalogue observations have filled in the incompleteness of the old input catalogue (mainly due to Tycho-2) thus reducing the presence of its subtle selection biases in DR4, compared to DR1, DR2 and DR3.  Therefore, overall RAVE DR4 is more representative of the underlying Galactic stellar populations than previous DRs, with the exception of $|b| < 25^{\circ}$ which deliberately targets giants with $J - K > 0.5$ mag.  Nevertheless, on a field by field basis, Galactic science still requires care to account for the various selection biases introduced by the inhomogeneous photometry used to derive some of the input samples (detailed in the DR1 paper).  Care is required if stars are selected from DR4 using their observation date as the selection function changes as a function of time: old and interim input catalogues obsdate $\le$ 20060312; new DENIS input catalogue obsdate $>$ 20060312; new 2MASS input catalogue obsdate $>$ 20120128.  There may also be a stellar position bias because the 6dF fibre positioner software avoids putting targets too close to each other and also avoids crossing of fibres.

Figure~\ref{f:hist_input_zoom} shows that the faint tail of the DR4 magnitude distribution is dominated by the old input catalogue.  This tail is fainter than the nominal selection of $I < 12$ mag due to SuperCOSMOS photographic $I$ magnitudes being saturated.  The twin peaks of the old input catalogue are due to the merging of Tycho-2 and SuperCOSMOS.  The interim input catalogue also includes Tycho-2 and SuperCOSMOS but weights their relative numbers to achieve a single peak in its magnitude distribution.  The special fields are identified by being away from the main data.  However, there are many calibration and science stars within the main footprint of the survey and so are classed as stars observed from the new input catalogue.  These dominate the bright tail of the DR4 magnitude distribution in Figure \ref{f:hist_input_zoom} and also contribute to the faint tail.  The old and interim input catalogues contribute to the bright tail because their bright magnitudes were constructed from Tycho-2 $B_{T}$ and $V_{T}$.

To summarise, DR4 has no kinematic bias and no overall photometric bias.  The notable exception is the colour criterion ($J - K > 0.5$~mag) to deliberately target giants at $|b| < 25^{\circ}$ (except where $\delta > 0^{\circ}$).  No other colour cuts exist in the data.  However, on a field by field basis, subtle selection biases may still be present.


\section{The new pipeline for the stellar parameters}
\label{sect:pipeline}

The wavelength region $\lambda \lambda 8410-8795$\AA~ is often used for Galactic archaeology purposes, as for instance with the multi-fibre spectrographs of ESO FLAMES-Giraffe at the LR8 and HR21 setups, and the Gaia-RVS. Indeed, it is a spectral region with relatively few telluric absorptions, that exhibits many iron and $\alpha$-element lines, in particular the prominent \ion{Ca}{2} triplet ($\lambda=8498.02$\AA, 8542.09\AA,  8662.14\AA). This feature is still visible at low signal-to-noise-ratio (SNR) and low metallicity, and as a consequence, it allows  relatively easily to have radial velocity measurements and metallicity estimations for any type of spectrum.  

\subsection{Spectral degeneracies in the infrared ionised Calcium triplet wavelength range}

The recent work of \citet[noted K11 hereafter]{Kordopatis11a} on spectra with a resolution $R \le 10~000 $ has shown that the \ion{Ca}{2} wavelength range suffers from spectral degeneracies that, if not appreciated,  can introduce serious biases in spectroscopic surveys that use automated parameterisation pipelines.  These degeneracies are mostly important for cool main-sequence stars, and stars along the giant branch. 
On the one hand,  the spectral signatures that are used to determine the surface gravities for the main-sequence stars are insensitive to small \logg~variations, hence reducing the accuracy of the measurement of that parameter. On the other hand, the spectra of stars along the giant branch can be identical for different sets of parameters. In this case, the degeneracy is due to the fact that the spectral signatures sensitive to variations of effective temperature, surface gravity and metallicity are the same. The degeneracy works as follows: the spectrum corresponding to a given \teff, \logg~ and \meta,  is almost identical to a spectrum  corresponding to a lower (higher) temperature, lower (higher) surface gravity and lower (higher) metallicity.

Automated parameterisation pipelines try to find the model template that minimises the distance function, defined as the difference between the observation and a set of reference  (model) spectra. In the case where not enough information is available in the data, the distance function can become non-convex,  and secondary minima can appear, in which the parameterisation algorithms can falsely converge. If these secondary minima are close in the parameter space, the resulting parameter estimation will have a large scatter around the true value, whereas if they are distant in the parameter space, the algorithms might converge randomly to one or the other according to where the noise is placed in the spectrum. 

In \citetalias{Kordopatis11a} it has been shown that for low and medium resolution spectra around the IR \ion{Ca}{2}, decision-tree methods manage to better find the absolute minimum of the distance function, compared to other algorithms, like the projection methods (e.g.: principal component analysis) or the ones trying to solve an optimisation problem (e.g.: minimum $\chi^2$), in particular when the SNR is low. 
For that reason, the pipeline presented in \citetalias{Kordopatis11a} iteratively renormalises the spectra and obtains the atmospheric parameters using a combination of a decision-tree algorithm called DEGAS \citep{Bijaoui12} and a projection method called MATISSE \citep{Matisse} which allows to better interpolate between the grid points. 

Both DEGAS and MATISSE have a learning phase based on a nominal  library of synthetic spectra {  ({\it i.e.} the templates)}, described in Sect.~\ref{sect:synthetic_grid}. 
Here we describe briefly the pipeline and the two algorithms, but we refer the reader to \cite{Matisse, Bijaoui12, Kordopatis11a} for further details.

\subsection{DEGAS: an oblique k-d decision-tree method for the low SNR spectra, and for re-normalisations}
\label{sect:DEGAS}
 
In the limit of the sampling precision of a learning grid ({\it i.e.} the parameter steps),  parameter estimation is a pattern
recognition problem.  The grid of synthetic spectra can be treated as a
known set of patterns among which  the observed spectra should be identified. 
The DEGAS ({\it DEcision tree alGorithm for AStrophysics}) algorithm is an
oblique 3-d decision tree (in our case  $T_\mathrm{eff}$, $\log~g$ and [M/H]), for which the decisions result from the
projection of the observations onto $N$ node vectors noted {$D_n(\lambda)$} ($n=1 \cdots N$).
These node vectors are associated with a subset of  {  spectra of the nominal library},  in the sense that the result of the projection of an observed spectrum on that node will select half of the subset which most closely resembles the observation. 

The recognition rules of DEGAS, at each node, are established during the learning
phase as follows:

\begin{enumerate}
\item The mean vector $M(\lambda)$ of the flux values per pixel of all the reference spectra in the subset is computed.

\item For each reference spectrum  $S_j(\lambda)$ associated with the node, the scalar product $c_j = \sum_\lambda S_j(\lambda) \cdot M(\lambda)$ is calculated. Let \~c be the median value of $c_j$. 

\item The reference spectra  are bisected in two new subsets, $T_1$ and $T_2$, according to the following
criteria:\\
$S_j$ belongs to the subset $T_1$ if $c_j \leqslant $ \~c \\
$S_j$ belongs to the subset $T_2$ if $c_j$ $>$ \~c.

\item 
The difference vector $ D(\lambda) =  M_1(\lambda) -  M_2(\lambda)$ is determined, where $M_1(\lambda)$  and $ M_2(\lambda)$ are the mean vectors of the flux values of each subset.

\item If the correlation coefficient between $M(\lambda)$ and $D(\lambda)$ is too
  small (typically, smaller than 0.999), the initial subset of reference spectra is re-separated  by the hyperplane defined by
  $D(\lambda)$, iterating until convergence (going back to step 2, replacing $M(\lambda)$ by $D(\lambda)$).

\end{enumerate}

When the previous procedure has converged for a particular node {\it n}, the final adopted  projection node vector {$D_n(\lambda)$} is determined, that will display the features that allow the separation of the subset of learning spectra at that node. The final median value $\tilde{c_n}$ of $c_j = \sum_\lambda S_j(\lambda) \cdot  D_n(\lambda)$ that will allow us to make the decisions is also set.

In this way, the recognition tree is built, having $\mathrm{log_2}(N)$
levels, where $N$ is the number of spectra of the learning grid. At
the lowest level nodes of the tree, only one training
spectrum remains associated with each node.  During the application
phase, the target data $O_i(\lambda)$ passes through all the levels of
the recognition tree, and a template is associated with it.

Of course, noise can induce  misclassifications. 
The exploration of the branches, even if the decision threshold  $\tilde{c_n}$ does not allow it, is accomplished thanks to an activation function on the directions of the tree. 
Let us consider:
\begin{equation}
u_i=\frac{c_i-\tilde{c_n}}{\sigma_{c_i}}
\end{equation}
where $c_i=\sum_\lambda O_i(\lambda) \cdot  D_n(\lambda)$, $\sigma_{c_i}=S_f \cdot \sqrt{\Sigma_\lambda D_n(\lambda)^2}$ and $S_f$ is an arbitrary constant chosen in order to explore optimally the branches. 
If $u_i \leqslant -k$ we decide that the correct
direction is 1. If $u_i \geqslant k$ the direction 2 is chosen.  If
$-k < u_i < k$ both directions are considered.  

After the scanning of all the nodes, a subset of synthetic templates
is selected, and their distances (in terms of the difference with the observed spectrum) are
computed.  Then, the parameters of the observed spectrum are determined by computing a weighted mean on the selected spectra, taking into
account these distances, setting:
\begin{equation}
W_{i}^{n} =\left (1- \left [\sum_\lambda O_i(\lambda)  -  S_n(\lambda)  \right ]^2 \right )^{p}.
\label{eq:dichodif_weight_distances} 
\end{equation}
The value $p$ of the exponent is rather arbitrary. Following \citetalias{Kordopatis11a}, where it was found that $p=64$  gave the best results, the same value was adopted in what
follows.\\

DEGAS has a key role in the pipeline used for the RAVE DR4 analyses. Because of its pattern recognition approach, it is exploring more efficiently the parameter space than other mathematical methods, and as a consequence, is less affected by secondary minima in the distance function. 
 Hence, given a roughly normalised spectrum at the rest frame, DEGAS is used in order to achieve a good normalisation for the data, using the synthetic spectrum corresponding to the intermediate solution found. 

Once DEGAS has converged on a set of parameters, the new parameters are used to re-normalise the spectrum given the intermediate solution template. 
The re-normalisation process might need several iterations until the shape of the continuum stays unchanged from one normalisation to another. The number of needed iterations may vary from three to ten. 
In the case where the distance function has many secondary minima (low SNR spectra and/or low metallicity stars) the results of DEGAS at that stage are the most accurate among the other methods. Nevertheless, it has been shown in \citetalias{Kordopatis11a} that MATISSE manages to better interpolate between the grid points when the astrophysical  information is sufficient in the spectra. 
Thus, once the shape of the continuum has converged, the mean SNR per pixel is estimated in the same way as in \cite{Zwitter08}, and  in the case of high SNR spectra MATISSE is run to get more precise atmospheric parameters.  The SNR threshold at which MATISSE is run has been established in \citetalias{Kordopatis11a} to be SNR$=30$~pixel$^{-1}$.

\subsection{MATISSE: a projection method for the high SNR regimes}
\label{sect:Matisse}
The MATISSE algorithm ({\it MATrix Inversion for Spectral SynthEsis}) is a
local multi-linear regression method.  It estimates a $\hat{\theta_i}$ stellar atmospheric parameter
($i \equiv $$T_\mathrm{eff}$, log~g, [M/H]) by projecting the observed spectrum
$O(\lambda)$ on a particular vector $B_\theta(\lambda)$ associated to
a theoretical $\theta_i$ parameter, as follows:

\begin{equation}
\hat{\theta_i}=\sum_\lambda B_{\theta_i}(\lambda) \cdot O(\lambda)
\label{eq:general_matisse} 
\end{equation}

These vectors, called $B_\theta(\lambda)$ functions hereafter, are
computed during a learning phase. They
relate, in a quantitative way, the pixel-to-pixel flux variations in a
spectrum to a given variation of the $\theta_i$ parameter. In the case
where the $B_\theta(\lambda)$ are orthogonal, the effects due to each
parameter affect the spectrum in an independent way, and hence the
atmospheric parameters are derived accurately.  When this is not the
case (as in most applications), possible degeneracies in the parameter space can occur, causing
a correlation of the parameter errors. 

 The $B_\theta(\lambda)$
functions are computed within a given range of parameters, from an optimal multi-linear combination of
theoretical, synthetic spectra $S(\lambda)$, as follows:

\begin{equation}
B_{\theta_i}(\lambda)=\sum_j \alpha_{ij} \cdot S_j(\lambda)
\label{eq:bfunctions_matisse} 
\end{equation}
where the $\alpha_{ij}$ factor is the weight associated with each
synthetic spectrum $S_j(\lambda)$, in order to retrieve the
$\hat{\theta_i}$ parameter. The weights are computed during the
learning phase by applying Eq. \ref{eq:general_matisse} to a subset of
synthetic spectra. Thus, combining Eq. \ref{eq:general_matisse} and
\ref{eq:bfunctions_matisse}, one obtains:

\begin{equation}
 \Theta_i = C \alpha_i 
\label{eq:correlation_matrix_matisse} 
\end{equation}
where $C=[c_{jj'}]$ is the correlation matrix and $\Theta_i$ the
vector of the parameters $\theta_i$ for all the considered
spectra. The weights $\alpha_{ij}$ are then obtained by inverting the correlation matrix $C$. As explained in  \citetalias{Kordopatis11a}, a direct inversion of $C$ would take into account all the available spectral signatures, including the smallest ones. 
Nevertheless, in the case of noisy spectra some lines become insignificant and should hence be discarded  from the analysis. In order to achieve that, we adopt here the same approach as in \citetalias{Kordopatis11a} which used the Landweber algorithm to iteratively invert $C$. By stopping the inversion procedure at different convergence values, smaller weights are then applied to the most insignificant lines which allows the solution to be less affected by secondary minima in the distance function.  Extensive analysis of the parameter space for spectra of different SNR values allowed adoption of a different set of $B_\theta(\lambda)$ functions for different SNR values,  and hence achieved  enhanced accuracy in the parameter determination (see \citetalias{Kordopatis11a} for further details). 


The convergence of the algorithm works as follows: if the projection of the observed spectrum on a set of $B_\theta(\lambda)$ gives results that are not within the parameter ranges for which these $B_\theta(\lambda)$ have been computed, then a new set of $B_\theta(\lambda)$ is used, centred on the previous solution. This step is repeated until the results stay within the parameter range of applicability of the projection vectors. In the case where the distance function is convex, less than five iterations are usually  needed to reach the absolute minimum, but in some cases of  degeneracy in the distance function, the algorithm might not converge (these spectra are then flagged and should not be used from the catalogue, see Sect.~\ref{sect:rejection_criteria}). \\

As noted previously, MATISSE has the advantage of interpolating accurately between the learning grid points, achieving a good parameter estimation at the high SNR regimes. Nevertheless, local projection methods such as MATISSE have, in general,  two main disadvantages. The  first consists in not exploring entirely the parameter space, and hence being easily trapped in secondary minima of the distance function if there is a lot of noise in the spectrum. The second disadvantage is that it can extrapolate results outside the boundaries of the learning grid. 
These two undesired effects are attenuated  with the parallel use of DEGAS. Indeed, as described in Sect.~\ref{sect:DEGAS}, DEGAS is first used to converge towards a sub-region of the parameter space. Then, for the lowest SNR spectra or when the derived MATISSE parameters are outside the grid's limits, the results of DEGAS are adopted, and MATISSE is not implemented\footnote{the stars for which DEGAS parameters have been adopted after MATISSE has given parameters outside the grid boundaries are flagged as well, and should be used with caution.}.

\subsection{The grid of synthetic spectra for the learning phase of the pipeline}
\label{sect:synthetic_grid}
The very high resolution grid computed in  \citetalias{Kordopatis11a} has been convolved with a Gaussian kernel, in order to obtain a new synthetic library at  $R=7500$
\footnote{ We note that the effective resolution of RAVE can in reality vary from $6500<R<8500$, the changes being a function of both time and position on the CCD.  Nevertheless, simple tests, degrading the $S^4N$ library (see Sect.~\ref{sect:additionnal_clusters})  to $R\sim6500$, 7500 and 8500 and then analysis as if the spectra  were at $R\sim7500$, show that the effect on the parameter estimation was of the second order.  We hence did not take into account these resolution changes for DR4. However, future data releases will implement these second order effects. },
  with a constant wavelength step of 0.35\AA~and covering the wavelength range  of  [8450.80--8746.55]\AA. In addition, the cores of the \ion{Ca}{2} lines have been removed from the synthetic spectra, corresponding to one pixel for the first \ion{Ca}{2} line and 2 pixels for the other two lines. This removal is justified since the cores of the lines are formed in the upper stellar atmospheric layers where {  some modelling assumptions like} the Local Thermodynamical Equilibrium (LTE)  {  or hydrostatic equilibrium might not be valid hypotheses anymore}.The flux disagreement with real spectra for these pixels can induce the algorithms to converge to false minima and hence the cores must be discarded (see \citetalias{Kordopatis11a} for further details). 

We recall that the synthetic library has been computed using the 1D, LTE and hydrostatic equilibrium MARCS model atmospheres \citep{Gustafsson08},  in combination with the Turbospectrum code \citep{Alvarez98}. The atomic line list has been calibrated on high SNR and high resolution spectra of the Sun and Arcturus \citep{Sun_Brault,Hinkle03}, assuming the Solar abundances of \cite{Grevesse08}, except for CNO where we used the values of \cite{Asplund05}.  The molecular line list includes ZrO, TiO, VO, CN, C2, CH, SiH, CaH, FeH and MgH lines with their corresponding isotopic variations (kindly provided by B. Plez).

The reference grid spans effective temperatures from 4000~K to 8000~K with a constant step of 250~K and  surface gravities from 0.0~dex to 5.0~dex with a constant step of 0.5~dex. In addition, the library spans with a constant step of 200~K  effective temperatures from 3000~K to 4000~K, and  surface gravities from 0.0~dex to 5.5~dex. 
As far as the metallicities are concerned, the number of grid models has increased compared to \citetalias{Kordopatis11a}. Instead of having a variable metallicity step, ranging from 0.25~dex for the most metal-rich stars, to 1~dex for the stars with \meta$<-3$~dex, the new grid has a constant metallicity step of 0.25~dex for all the metallicities ranging from $[-5.0;+1.0]$~dex. These new spectra, whose atmospheric models did not exist in the MARCS database, have been linearly interpolated from the existing synthetic spectra. \\

One of the noticeable differences between previous data releases and  DR4 is that, in this work, only three parameters are independent in the grid: the effective temperature: \teff, the logarithm of the surface gravity: \logg , and the overall metallicity: \meta. This restriction decreases the number of the secondary minima of the distance function, hence increasing the accuracy of the parameter derivation. It should be noted though that the {  microturbulent} velocity ($\xi$) is not constant within the grid, but is changed in lock-step  with the gravity of the stars. Dwarfs (\logg$\geq3.5$~dex) have a {  microturbulent} velocity of  $\xi=1$~\kms, whereas giants have $\xi=2$~\kms. Abundances of the $\alpha$-elements\footnote{The chemical species considered as $\alpha-$elements are O, Ne, Mg, Si, S, Ar, Ca and Ti.}  are also changed systematically with metallicity,  being scaled on the iron abundance, [Fe/H], following the standard $\alpha-$enhancement found for the metal-poor stars  of the Milky Way (thick disc and halo): 
\begin{itemize}
\item $[\alpha$/Fe]=0.0 dex for  [Fe/H] $\geq$0.0 dex
\item $[\alpha$/Fe]=$-0.4 \times $[Fe/H] dex for $-1\leq$[Fe/H]$\leq 0$~dex
\item $[\alpha$/Fe]=+0.4 dex for [Fe/H] $\leq$ $-$1.0  dex.
\end{itemize}

In addition, spectra for which the parameter combinations did not correspond to realistic astrophysical stars have been removed from the learning grid and hence from the solution space as well. To minimise the importance of our astrophysical priors in the derived parameters, we removed only the templates with \logg $= 5$~dex and \teff $> 6250$~K, those with \teff $\leq 4250$~K and $4 \leq$ \logg $\leq 3$~dex, as well as all stars with \meta $ \leq -3$~dex, \teff$\leq 4000$~K and \logg $\leq$4~dex.  These criteria correspond to excluding very young stars with extremely metal-poor {  abundances} (age $<$0.5~Gyr and [Fe/H]$<-2.5$~dex), as well as old stars that are extremely metal-rich (age $>$14~Gyr and [Fe/H]$>0.75$~dex).
The final grid contains 3580 spectra of 839 pixels each. \\

Based on this grid, a subset of reference models can be selected, according to the additional information that is available for each data set to be treated. In the case of RAVE, the 2MASS photometric information that is available for the observed targets  is used to exclude some parameter combinations from the solution space corresponding to derived temperature ranges which are grossly inconsistent with the photometric colour.
In practice, the RAVE spectra have been separated into four different colour ranges. Then, according to their 2MASS $(J-K_s)$ colour, a set of solutions has been imposed as soft photometric priors for every analysed spectrum, defining:
\begin{itemize}
\item $(J-K_s)> 0.75 \Rightarrow $ \teff $< 4500$~K
\item $0.4 < (J-K_s) < 0.75 \Rightarrow  3750 < $\teff $<6000$~K
\item $(J-K_s)< 0.4 \Rightarrow $ \teff $>5250$~K.
\end{itemize}
The few stars for which  no 2MASS photometry was available (less than 2\%) form a fourth category for which there is no prior on the solution space.

The above mentioned effective temperature ranges have been determined by requesting a colour-magnitude diagram of the Galaxy from the web interface\footnote{\url{http://stev.oapd.inaf.it/cmd}} of the Padova database with the 2MASS photometric system. Then, for the above three colour ranges we inferred the full range of effective temperatures of the simulated stars, and increased these limits  by $\pm 500$~K.  
We note that the effective temperature bins have been made deliberately large, because neither the photometric errors nor the extinction have been taken into account when separating the spectra into colour bins.


\subsection{Computation of the internal uncertainties}
\label{sect:Internal_errors}

The total uncertainty of the pipeline for a given star is the quadratic sum of its internal and external errors. The internal uncertainties relate the capacity of an algorithm to treat spectral degeneracies and  SNR, whereas external uncertainties concern the difference between the template synthetic spectra  and the true stellar spectra (see Sect.~\ref{sect:total_error}). 

Following \citetalias{Kordopatis11a}  the internal uncertainties of the algorithm have been estimated by computing a set of spectra of realistic Galactic populations. Based on the Besan\c{c}on Galactic model, a simulated catalogue of stars towards three different Galactic directions (Galactic centre, north Galactic pole and intermediate Galactic latitudes) has been constructed, from which  $10^4$ stars have been randomly selected to be our realistic Galactic sample. In addition, in order to explore different metallicity regimes, each star has been replicated  in the catalogue, with its partner having reduced (by $-0.75$~dex) metallicity.
The  $2\times 10^4$ synthetic spectra corresponding to these parameter combinations have been computed thanks to the interpolation capabilities of MATISSE, and four different values of white Gaussian noise  were added to them (SNR=100, 50, 20, 10~pixel$^{-1}$).   
Given these final $8 \times 10^4$~spectra, the pipeline was run in order to retrieve the associated errors.

In order to simulate the way the RAVE spectra are  analysed, the pipeline was run twice. Once without any photometric prior (see Table~\ref{tab:Internal_errors_noColorcuts}), and once by imposing soft priors  (see Table~\ref{tab:Internal_errors_Colorcuts}), based on their temperatures. These priors have been selected in order to be similar with the ones that are applied in the analysis of the RAVE spectra (see Sect.~\ref{sect:synthetic_grid}). 
The error values for different stellar types, presented in Tables~\ref{tab:Internal_errors_noColorcuts} and ~\ref{tab:Internal_errors_Colorcuts},  have been computed as the $70^{\rm th}$ percentile of the error distribution. Indicative atmospheric parameter  uncertainties for typical thin disc dwarfs, thick disc dwarfs and halo giants are also given in the last three lines of these tables.  A comparison of the uncertainty values with or without photometric priors shows that when the spectral degeneracy is important, the applied soft priors improve significantly the associated uncertainties (see for example the hot, metal-poor dwarfs).  In addition, it has been noticed, as expected, that the use of the soft photometric priors  improve  the $90^{\rm th}$  percentile of most of the stellar categories considered in these tables.  \\

The way the internal errors are associated with a specific parameter estimation is as follows: once the pipeline has converged towards a set of parameters, the final SNR is computed as in \cite{Zwitter08}, utilising the associated solution template. According to the SNR, the stellar type, the luminosity class, its metallicity, and the use or not of 2MASS photometric prior, the equivalent internal error estimations in one of the Tables~\ref{tab:Internal_errors_noColorcuts} or ~\ref{tab:Internal_errors_Colorcuts} are adopted. 
We note that this approach does not optimally take into account the properties of the distance function and hence the degeneracies. Nevertheless, we find that in the case of spectral degeneracy, the associated errors are larger, being consistent with what is expected. 


\begin{table*}[tdp]
\caption{Internal errors after re-normalisations  {\bf without} photometric priors}
\begin{center}
\footnotesize
\begin{tabular}{l||cccc||cccc||cccc}
& \multicolumn{4}{c||}{\teff\ (K)} &  \multicolumn{4}{|c||}{$\log~g$ (dex)} &  \multicolumn{4}{|c}{[M/H] (dex)}\\ \hline
SNR  (per pixel) & 100 & 50 & 20 & 10 & 100 & 50 & 20 & 10 & 100 & 50 & 20 & 10 \\ \hline
KII-IV, [M/H]$>$-0.5  dex &    72  &    76  &   117  &   201  &  0.12  &  0.13  &  0.20  &  0.48  &  0.08  &  0.08  &  0.10  &  0.23 \\
KII-IV, -1$<$[M/H]$<$-0.5 dex  &    62  &    85  &   133  &   302  &  0.14  &  0.20  &  0.35  &  0.72  &  0.08  &  0.09  &  0.16  &  0.30 \\
KII-IV, -2$<$[M/H]$<$-1 dex  &    75  &    96  &   178  &   330  &  0.20  &  0.30  &  0.57  &  0.97  &  0.09  &  0.11  &  0.19  &  0.33 \\
KII-IV, [M/H]$<$-2 dex  &    78  &   105  &   184  &   382  &  0.31  &  0.40  &  0.76  &  1.26  &  0.09  &  0.10  &  0.20  &  0.37 \\
GII-IV, [M/H]$>$-0.5 dex &    78  &   111  &   233  &   402  &  0.09  &  0.20  &  0.40  &  0.69  &  0.07  &  0.09  &  0.15  &  0.35 \\
GII-IV, -1$<$[M/H]$<$-0.5 dex  &    81  &   110  &   241  &   426  &  0.15  &  0.25  &  0.54  &  0.98  &  0.08  &  0.10  &  0.17  &  0.44 \\
GII-IV, -2$<$[M/H]$<$-1  dex &    98  &   164  &   282  &   472  &  0.25  &  0.46  &  0.74  &  1.08  &  0.10  &  0.13  &  0.23  &  0.43 \\
GII-IV, [M/H]$<$-2 dex  &   187  &   248  &   375  &   553  &  0.37  &  0.61  &  0.99  &  1.07  &  0.17  &  0.26  &  0.49  &  0.60 \\
    FII-IV all [M/H]  &    79  &    73  &   138  &   140  &  0.14  &  0.13  &  0.14  &  0.15  &  0.09  &  0.09  &  0.10  &  0.27 \\
\hline
KV, [M/H]$>$-0.5 dex  &    66  &    69  &    92  &   171  &  0.11  &  0.14  &  0.22  &  0.34  &  0.08  &  0.09  &  0.09  &  0.21 \\
KV, -1$<$[M/H]$<$-0.  &    75  &    85  &   112  &   225  &  0.15  &  0.17  &  0.24  &  0.38  &  0.09  &  0.10  &  0.13  &  0.30 \\
KV, -2$<$[M/H]$<$-1   &    83  &    98  &   173  &   328  &  0.16  &  0.19  &  0.25  &  0.51  &  0.09  &  0.11  &  0.14  &  0.35 \\
  KV, [M/H]$<$-2 dex  &    93  &   133  &   278  &   518  &  0.11  &  0.17  &  0.47  &  1.03  &  0.06  &  0.06  &  0.15  &  0.38 \\
GV, [M/H]$>$-0.5 dex  &    67  &    98  &   209  &   344  &  0.10  &  0.16  &  0.33  &  0.51  &  0.09  &  0.10  &  0.14  &  0.30 \\
GV, -1$<$[M/H]$<$-0.  &    87  &   147  &   246  &   426  &  0.14  &  0.22  &  0.36  &  0.55  &  0.09  &  0.12  &  0.21  &  0.38 \\
GV, -2$<$[M/H]$<$-1   &   119  &   181  &   358  &   669  &  0.19  &  0.32  &  0.44  &  0.71  &  0.11  &  0.14  &  0.28  &  0.54 \\
  GV, [M/H]$<$-2 dex  &   279  &   435  &   690  &   843  &  0.44  &  0.54  &  0.72  &  1.06  &  0.26  &  0.38  &  0.61  &  0.80 \\
FV, [M/H]$>$-0.5 dex  &    81  &   117  &   307  &   493  &  0.13  &  0.18  &  0.34  &  0.52  &  0.11  &  0.13  &  0.27  &  0.43 \\
FV, -1$<$[M/H]$<$-0.  &    96  &   151  &   306  &   575  &  0.14  &  0.21  &  0.33  &  0.49  &  0.11  &  0.14  &  0.26  &  0.48 \\
FV, -2$<$[M/H]$<$-1   &   155  &   257  &   563  &   999  &  0.21  &  0.30  &  0.40  &  0.69  &  0.13  &  0.19  &  0.43  &  0.83 \\
  FV, [M/H]$<$-2 dex  &   447  &   641  &  1046  &  1165  &  0.43  &  0.65  &  0.84  &  1.22  &  0.40  &  0.53  &  0.95  &  1.14 \\
\hline
    Thin Disc dwarfs  &    66  &    89  &   199  &   344  &  0.09  &  0.14  &  0.32  &  0.50  &  0.09  &  0.10  &  0.13  &  0.29 \\
   Thick Disc dwarfs  &    91  &   146  &   280  &   501  &  0.14  &  0.22  &  0.35  &  0.52  &  0.09  &  0.13  &  0.24  &  0.43 \\
         Halo giants  &    90  &   149  &   244  &   443  &  0.23  &  0.43  &  0.70  &  1.05  &  0.10  &  0.13  &  0.24  &  0.39 \\
\hline

\end{tabular}
\end{center}
\footnotetext{Luminosity classes I-II assume \logg $\le3.5$,  luminosity class V assumes \logg$> 3.5$. Spectral types are defined by \teff\ ranges as follows: \teff$<5000$~K  K type,    $5000\le$\teff$<6000$~K  G type, and  \teff$\ge6000$~K  F type stars.  }
\label{tab:Internal_errors_noColorcuts}
\end{table*}%

\begin{table*}[tdp]
\caption{Internal errors after re-normalisations  {\bf with} photometric priors}
\begin{center}
\footnotesize
\begin{tabular}{l||cccc||cccc||cccc}
& \multicolumn{4}{c||}{\teff\ (K)} &  \multicolumn{4}{|c||}{$\log~g$ (dex)} &  \multicolumn{4}{|c}{[M/H] (dex)}\\ \hline
SNR  (per pixel) & 100 & 50 & 20 & 10 & 100 & 50 & 20 & 10 & 100 & 50 & 20 & 10 \\ \hline
KII-IV, [M/H]$>$-0.5 dex &    71  &    76  &   112  &   180  &  0.12  &  0.14  &  0.24  &  0.50  &  0.08  &  0.08  &  0.11  &  0.21 \\
KII-IV, -1$<$[M/H]$<$-0.5 dex &    61  &    86  &   137  &   285  &  0.14  &  0.20  &  0.40  &  0.69  &  0.08  &  0.10  &  0.17  &  0.29 \\
KII-IV, -2$<$[M/H]$<$-1  dex&    75  &    96  &   173  &   312  &  0.20  &  0.31  &  0.59  &  1.02  &  0.09  &  0.11  &  0.21  &  0.34 \\
KII-IV, [M/H]$<$-2 dex  &    75  &   101  &   213  &   399  &  0.27  &  0.35  &  0.79  &  0.98  &  0.10  &  0.10  &  0.18  &  0.32 \\
GII-IV, [M/H]$>$-0.5  dex&    78  &   104  &   237  &   332  &  0.09  &  0.21  &  0.48  &  0.49  &  0.07  &  0.09  &  0.18  &  0.26 \\
GII-IV, -1$<$[M/H]$< $-0.5 dex &    79  &   103  &   238  &   412  &  0.15  &  0.23  &  0.52  &  0.97  &  0.08  &  0.10  &  0.17  &  0.37 \\
GII-IV, -2$<$[M/H]$<$-1  dex&    90  &   158  &   283  &   412  &  0.21  &  0.45  &  0.78  &  1.05  &  0.11  &  0.12  &  0.23  &  0.33 \\
GII-IV, [M/H]$<$-2 dex  &   203  &   265  &   378  &   479  &  0.33  &  0.61  &  0.91  &  1.14  &  0.15  &  0.25  &  0.49  &  0.54 \\
    FII-IV all [M/H]  &    79  &    83  &    93  &   138  &  0.14  &  0.13  &  0.14  &  0.26  &  0.09  &  0.09  &  0.09  &  0.31 \\
\hline
KV, [M/H]$>$-0.5 dex  &    66  &    69  &    92  &   168  &  0.11  &  0.15  &  0.22  &  0.36  &  0.08  &  0.09  &  0.09  &  0.20 \\
KV, -1$<$[M/H]$<$-0.5 dex  &    75  &    84  &   110  &   219  &  0.15  &  0.17  &  0.23  &  0.38  &  0.09  &  0.10  &  0.12  &  0.29 \\
KV, -2$<$[M/H]$<$-1 dex   &    82  &    97  &   172  &   293  &  0.16  &  0.19  &  0.23  &  0.52  &  0.08  &  0.10  &  0.14  &  0.33 \\
  KV, [M/H]$<$-2 dex  &    92  &   144  &   240  &   480  &  0.11  &  0.17  &  0.46  &  0.54  &  0.07  &  0.07  &  0.17  &  0.39 \\
GV, [M/H]$>$-0.5 dex  &    66  &    95  &   203  &   316  &  0.10  &  0.15  &  0.35  &  0.55  &  0.09  &  0.10  &  0.15  &  0.30 \\
GV, -1$<$[M/H]$<$-0.5 dex  &    85  &   144  &   238  &   360  &  0.14  &  0.22  &  0.38  &  0.55  &  0.09  &  0.12  &  0.20  &  0.33 \\
GV, -2$<$[M/H]$<$-1 dex   &   101  &   169  &   291  &   441  &  0.15  &  0.28  &  0.43  &  0.62  &  0.10  &  0.13  &  0.24  &  0.39 \\
  GV, [M/H]$<$-2 dex  &   220  &   317  &   393  &   480  &  0.35  &  0.46  &  0.53  &  0.80  &  0.21  &  0.30  &  0.39  &  0.50 \\
FV, [M/H]$>$-0.5 dex  &    66  &    94  &   257  &   464  &  0.13  &  0.18  &  0.38  &  0.56  &  0.10  &  0.11  &  0.25  &  0.41 \\
FV, -1$<$[M/H]$<$-0.5 dex &    84  &   127  &   277  &   498  &  0.14  &  0.21  &  0.35  &  0.52  &  0.09  &  0.13  &  0.25  &  0.43 \\
FV, -2$<$[M/H]$<$-1 dex   &   104  &   182  &   440  &   601  &  0.18  &  0.25  &  0.38  &  0.53  &  0.11  &  0.15  &  0.32  &  0.51 \\
  FV, [M/H]$<$-2 dex  &   331  &   503  &   617  &   678  &  0.32  &  0.51  &  0.54  &  0.66  &  0.33  &  0.48  &  0.54  &  0.69 \\
\hline
    Thin Disc dwarfs  &    61  &    85  &   183  &   330  &  0.09  &  0.14  &  0.33  &  0.53  &  0.09  &  0.10  &  0.12  &  0.30 \\
   Thick Disc dwarfs  &    84  &   133  &   256  &   406  &  0.14  &  0.21  &  0.38  &  0.54  &  0.09  &  0.12  &  0.22  &  0.37 \\
         Halo giants  &    83  &   143  &   258  &   399  &  0.21  &  0.41  &  0.74  &  1.04  &  0.10  &  0.12  &  0.24  &  0.38 \\
\hline

\end{tabular}
\end{center}
\label{tab:Internal_errors_Colorcuts}
\end{table*}%

\subsection{Description of the observed input spectra}
\label{sect:input_spectra}
The extraction and reduction procedures applied to  the observed spectra are as described  in \cite{Steinmetz06, Zwitter08}. As far as the normalisation and rest-frame corrections are concerned, we have used the results and the raw data coming from DR3, details of which can be found in \cite{Siebert11} and for  which the general properties are summarised in Sect.~\ref{sect:Vrad}. 
In addition, in order to be able to perform a pattern recognition and a pixel-to-pixel comparison with the spectra of the learning grid, the RAVE spectra have been interpolated at the wavelengths of the templates and the cores of the IR \ion{Ca}{2} triplet lines have been removed (see Sect.~\ref{sect:synthetic_grid}). 

Since the DR4 pipeline re-normalises the observed spectra, initially erroneous or inaccurate  continuum shapes do not influence the final parameter accuracy of DR4. 
One might worry about the accuracy of the radial velocity of the stars, and hence their rest-frame correction. Indeed, the radial velocities have been computed using cross-correlation with the solution template derived in  DR3, whereas in the present work, the atmospheric parameters are recomputed,  leading to different values. 
Nevertheless, the pipeline requires a radial velocity precision only better than $\sim7-10$~\kms~(see \citetalias{Kordopatis11a}) in order not to be affected by Doppler shifts in the parameter estimation. This threshold is much higher than the accuracy of the radial velocities coming from the DR3, where $95\%$ of the sample has errors of less than $\sim 4$~\kms\ and 98\% less than 7~\kms~\citep[][ and Sect.~\ref{sect:Vrad}]{Siebert11}. Thus, we can assume that the spectra are indeed at the rest frame for the purpose of our analysis.


\section{Validation of the parameterisation with external data sets}
\label{sect:Validation_pipeline}
Up to here in this paper, only the internal performances of the pipeline have been discussed. Nevertheless, any given pipeline based on a grid of synthetic spectra needs to be verified and calibrated on  observed spectra with high SNR,  and well-determined parameters. The grid of synthetic spectra that has been used for this work has been computed with an atomic line list calibrated on the  high-resolution and high-SNR spectra of the Sun and Arcturus of \cite{Sun_Brault,Hinkle03}. However, these calibrations concern only two particular stars, and further investigation needs to be done in order to correct possible biases in the parameter's estimation. 
In order to calibrate the \teff, \logg~and \meta, instead of going through the process of calibrating all the lines for many reference stars and improve the quality of the atmosphere modelling, one can also validate the pipeline's parameter results with reference parameter measurements from  the literature. 

This calibrating data set needs to cover as much as possible the parameter space investigated by the survey. Ideally the calibration of the parameters would be done using only RAVE spectra of suitable standards, but RAVE-like spectra, at the same resolution and (if possible) reduced in the same manner, can be sufficient in the case where not enough calibration spectra are available.

\subsection{The calibration data sets of observed spectra}
\label{sect:calibration_datasets}

\begin{table*}[tdp]
\caption{Calibration data sets with SNR$>$40~pixel$^{-1}$}
\begin{center}
\begin{tabular}{ccccc}
Dataset     & Object  &  N spectra & [Fe/H]  & Reference for the stellar parameters \\ \hline
IC 4651   &  open cluster  & 6 & $+0.10$   &\cite{Pasquini04}  \\
M67          & open  cluster  & 16 & $+0.05$  &\cite{Pancino10} \\
CFLIB       & dwarfs \& giants & 224 & $[-1.0,0.0]$   &PASTEL database\\
CD-38245 & \teff=4800~K, \logg=1.5 & 2 & $-4.2$ &\cite{Cayrel04}\\ 
\cite{Ruchti11}\footnotemark[1] & giants \& dwarfs & 229 & $[-2.5 ; -0.5]$   &\cite{Ruchti11}   \\
Fulbright et al.    & giants & 163  & $[-2.5 ; 0.0]$ & {  Fulbright} et al. (in prep.)\\ 
RAVE spectra & giants \& dwarfs & 169 & $[-1.5 ; 0.0]$  &PASTEL database \\ \hline
\end{tabular}
\end{center}
\footnotetext[1]{For the \cite{Ruchti11} catalogue we selected  stars with \meta$> -2.5$~dex and \logg$<$3, and stars with \meta$< -0.8$~dex and \logg$>$3. }
\label{tab:calibration_sets}
\end{table*}%

First, the RAVE database has been explored to find spectra of stars which had atmospheric parameter determinations available from high-resolution spectroscopy. For that purpose, we made extensive use of the heterogeneous PASTEL catalogue\footnote{\url{http://pastel.obs.u-bordeaux1.fr/}} to identify such targets, retrieving roughly 400 star candidates.
 Following \cite{Soubiran10}, we considered only  the reference values coming from 
\cite{Fuhrmann98b, Fuhrmann98a, Fuhrmann04,Fuhrmann08, Gratton96, Gratton03,Hekker07, Luck06, Luck07, McWilliam90, Mishenina01, Mishenina04, Mishenina06, Mishenina08, Ramirez07, Valenti05}.  {  These studies, when considered by author, all include a large number of stars (at minimum 222 stars), and are all analysed in a homogeneous way. This allows  to minimise the discrepancies between the sub-catalogues of PASTEL. }
When for a given star several measurements were available, the mean was computed and the dispersion of the parameters has been considered as the uncertainty on the reference values. We kept only those stars for which measurements were available for the three parameters from a single study, and for which the dispersions among the literature values were less than 100~K, 0.2~dex and 0.1~dex for \teff, \logg\ and \meta, respectively.
In total, 169 stars were selected that way, mainly dwarfs of intermediate metallicity.

In order to investigate the pipeline's behaviour in the low metallicity regime for giant stars, we chose to use the parameters of 229 thick disc stars analysed by \cite{Ruchti11}, as well as 163 stars observed by Fulbright et al. (in prep.).
The targets of both of these data sets are drawn from RAVE, while the stellar parameters have been obtained from an equivalent-width analysis of high resolution spectra.  
In addition, the very metal-poor giant star CD-38245 \citep[\meta$=-4.2$~dex,][]{Cayrel04}, which has been observed twice by RAVE, has been included in the list, in order to calibrate the results at the very metal-poor regime.

Metal-rich giant stars have been explored thanks to the CFLIB library \citep{Valdes04}. The entire spectral library was  downloaded from the webpage\footnote{\url{http://www.noao.edu/cflib/}} of that project, excluding only spectra which did not include the wavelength range around the IR \ion{Ca}{2} triplet. The final comparison catalogue is the same as in \citetalias{Kordopatis11a}, where once again, we used the updated values that can be found in the PASTEL database, and discarded the stars for which the dispersion in the literature values were greater than 100~K, 0.2~dex and 0.1~dex for \teff, \logg\ and \meta, respectively.

Finally, in order to have a more significant statistical sample at the high-metallicity regime, we used the 2.3m telescope at the Siding Spring Observatory (SSO) to obtain spectra of stars belonging to open clusters.  
Although the data have not been obtained with the same instrument, the same reduction pipeline has been used as for the RAVE spectra. 
For calibration purposes, we have used 16 RAVE-like SSO spectra of giant stars belonging to the open cluster M67 and 12 RAVE-like SSO spectra of giants belonging to the open cluster IC4651, with a few additional data sets used as testing sets (see Sect.~\ref{sect:additionnal_clusters}). These targets were selected given their positions, colours and radial velocities when available, prioritising bright stars in order to have high SNR spectra. For these stars, no individual atmospheric parameters were available, but their metallicity is expected  to have a small dispersion around their mean open cluster metallicity value.  

\begin{figure*}[tbp]
\begin{center}
$\begin{array}{cc}
\includegraphics[width=0.4\textwidth,angle=0]{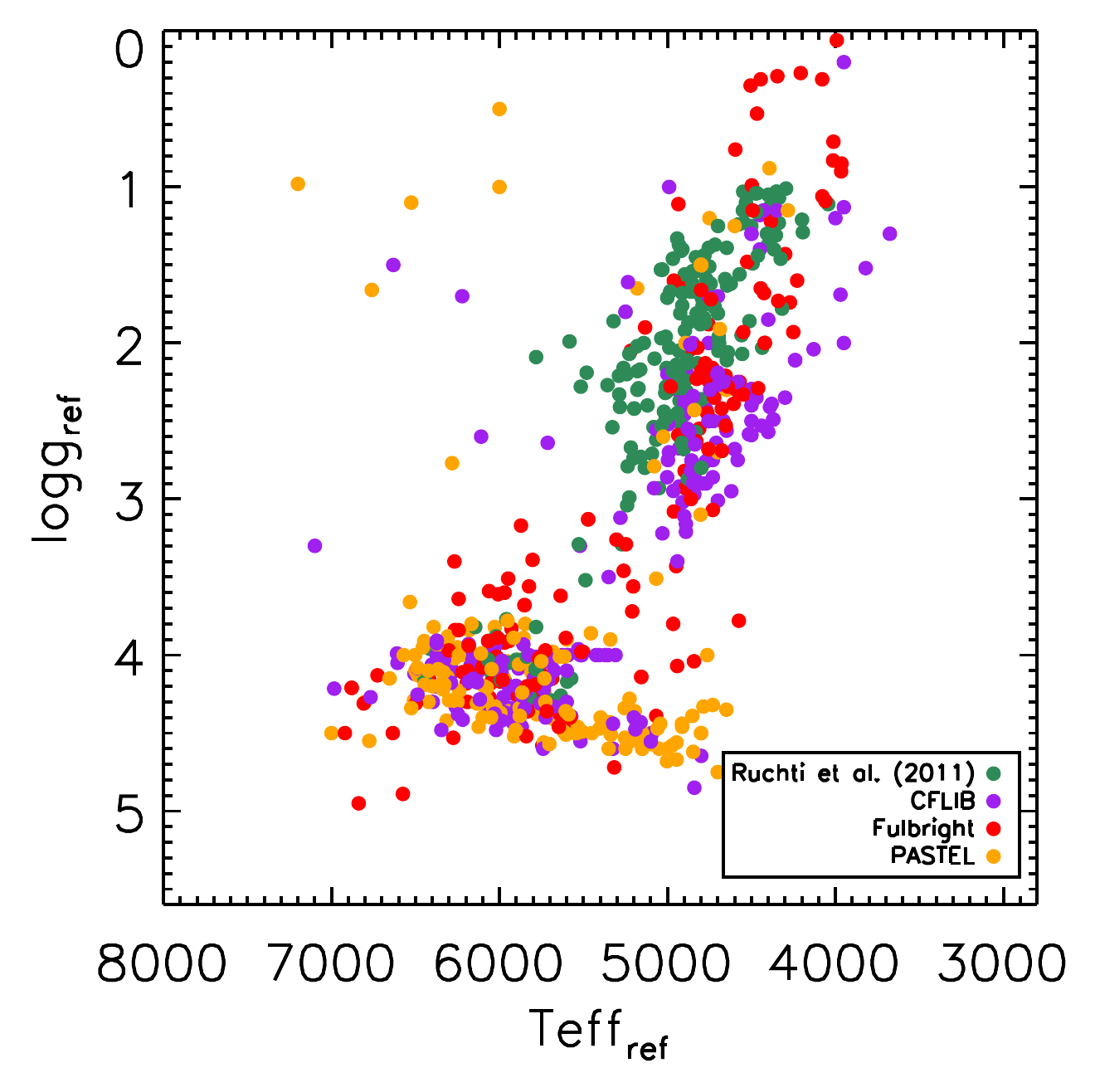} &
\includegraphics[width=0.4\textwidth,angle=0]{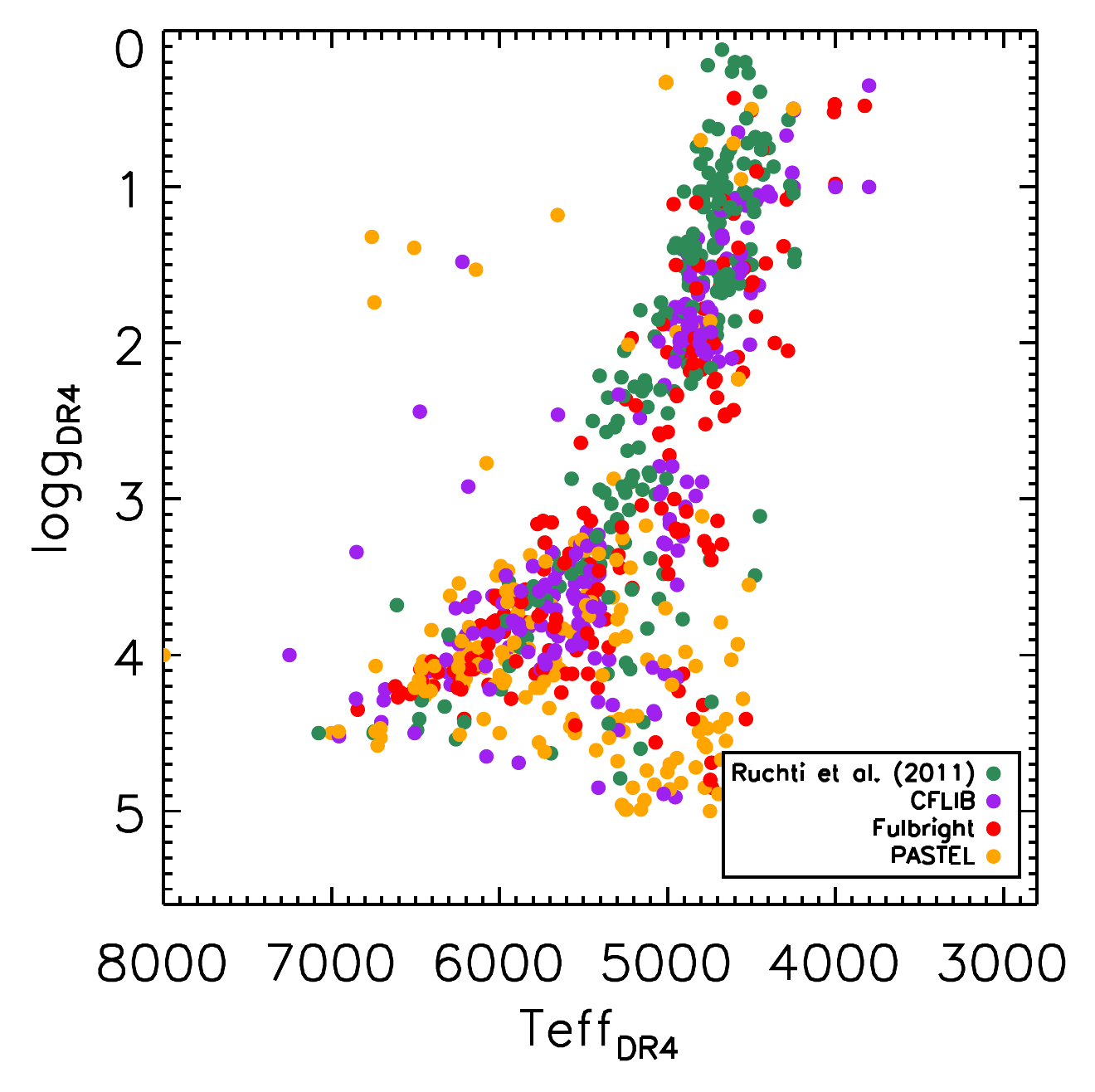}
\end{array}$
\end{center}
\caption{Surface gravity (\logg) versus effective temperature (\teff) diagram of the calibration data sets for which we have parameter estimations coming from high-resolution spectroscopy. At the left hand side are represented the values found in the literature, whereas on the right hand side are plotted the results obtained from the DR4 pipeline.} 
\label{Fig:Calib_HRD}
\end{figure*}

In total, 809 stars are used as calibrators, each having  SNR$>40$~pixel$^{-1}$.  The final number of spectra used from each data set is summarised in Table~\ref{tab:calibration_sets} and their reference and retrieved \teff--\logg~ diagrams are plotted in Fig.~\ref{Fig:Calib_HRD}.

\subsection{Validation of the effective temperatures and surface gravities }
\begin{figure*}[tbp]
\begin{center}
$\begin{array}{cc}
\includegraphics[width=0.42\textwidth,angle=0]{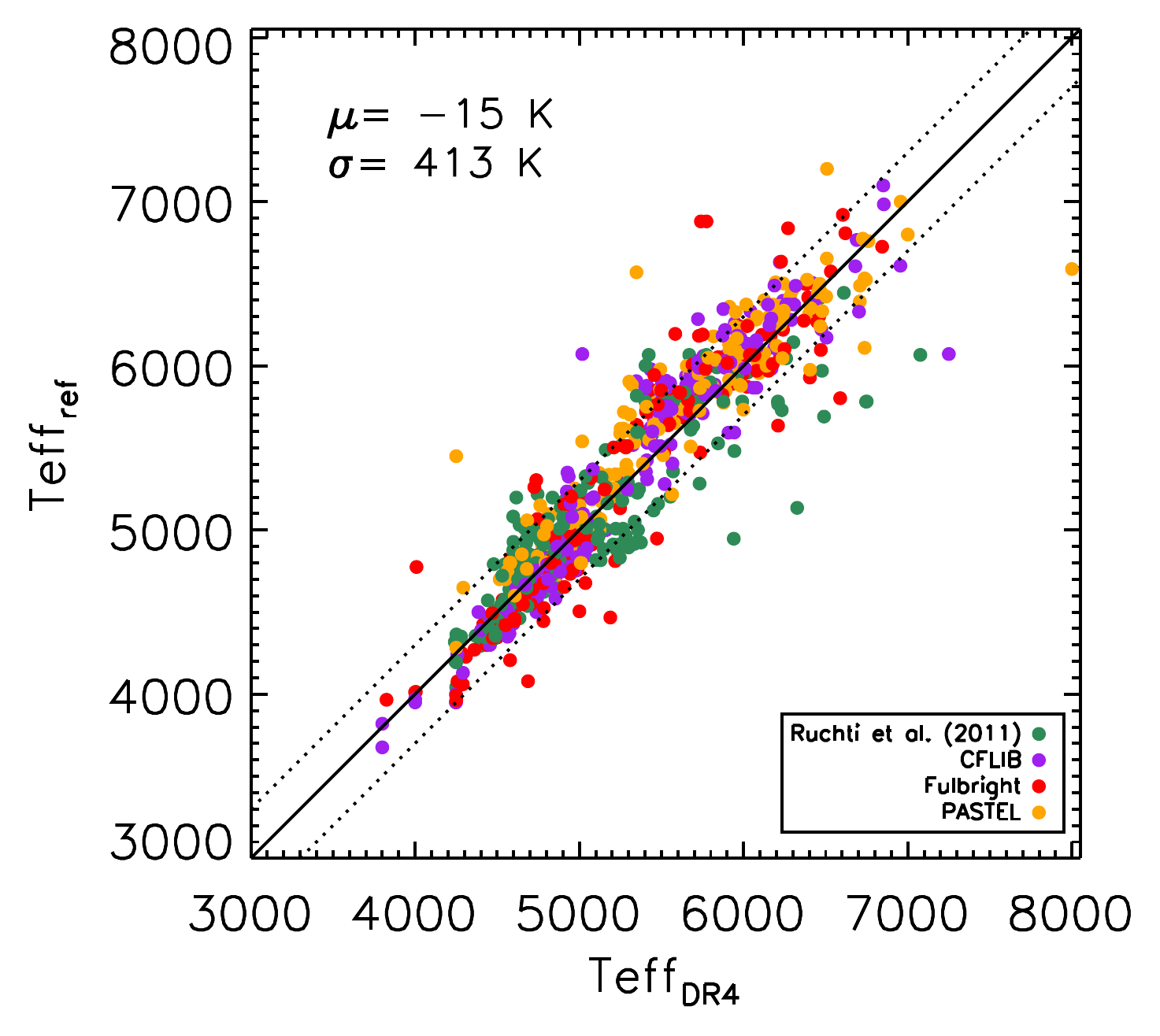} &
\includegraphics[width=0.38\textwidth,angle=0]{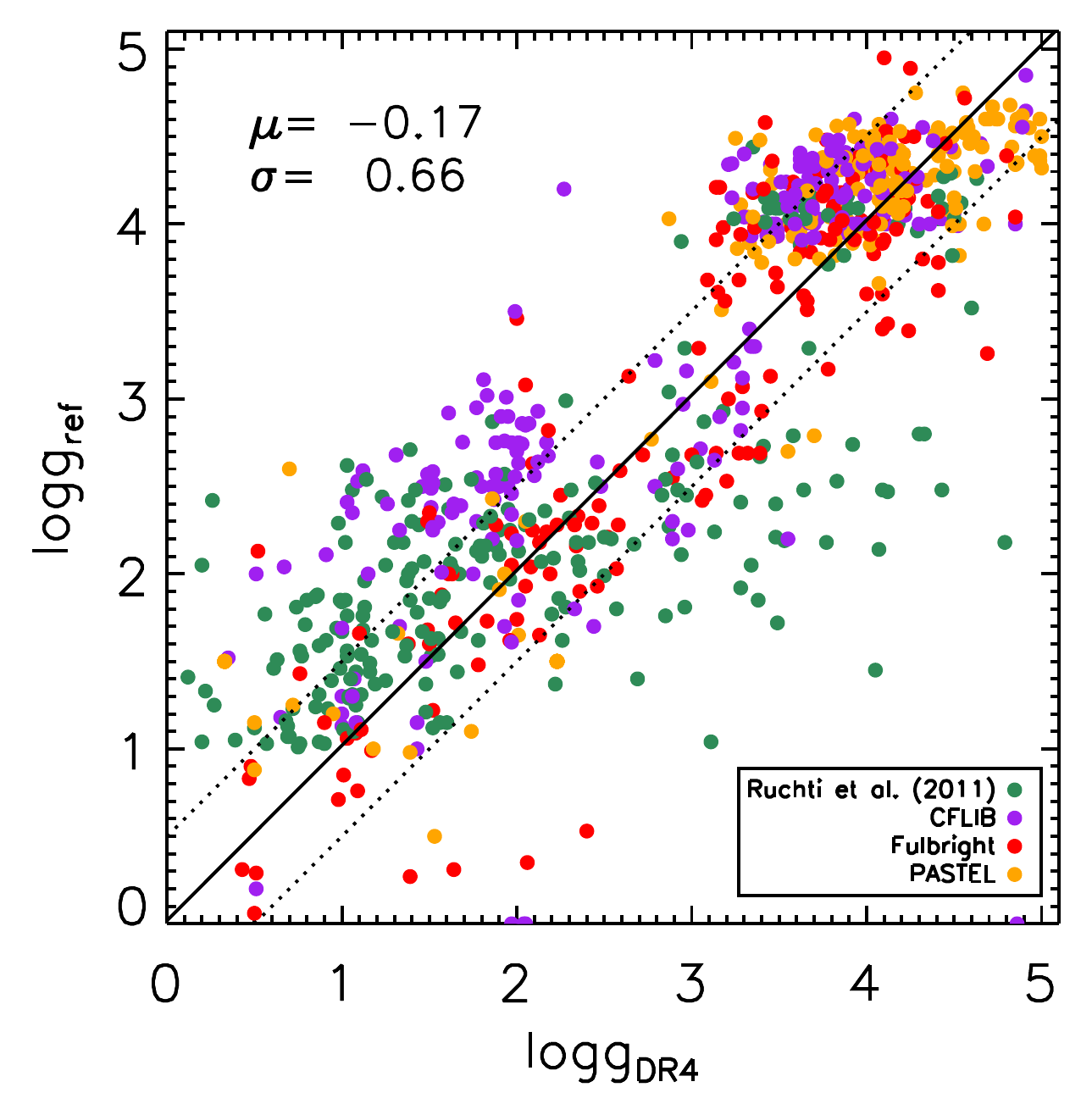}
\end{array}$
\end{center}
\caption{Comparison of the reference values found in the literature and the derived effective temperatures (left hand side) and surface gravities (right hand side). Colour coding for each data set is the same as in Fig.~\ref{Fig:Calib_HRD}. Dotted diagonal lines represent offsets from unity of $\pm 300$~K and $\pm 0.5$~dex for \teff~and \logg, respectively. The mean offsets ($\mu$) and the dispersions ($\sigma$) are indicated in the upper left corner of each plot.} 
\label{Fig:Calib_Teff_logg}
\end{figure*}

Figures~\ref{Fig:Calib_HRD} and~\ref{Fig:Calib_Teff_logg}  show the comparison between the reference values found in the literature and those found with the present pipeline for all the data sets except those for open cluster members (where no reference values were available). As far as the effective temperature is concerned,  good agreement is found, with a mean offset of 15~K and a dispersion of roughly 400~K. On the other hand, the agreement is less good for the surface gravity, with a rather big scatter  for the giant stars. This effect is a manifestation of the previously cited spectral degeneracy which is present for the low and intermediate resolution spectra around the IR \ion{Ca}{2} triplet. According to the stochastic position of the noise on the spectrum, a metal-rich turn-off star can be easily confused with a star on the sub-giant branch of lower metallicity.  
Unless precise photometric temperatures are known, this degeneracy cannot be lifted using medium resolution spectrum alone and is a true degeneracy.  Nevertheless, as  noted in the previous sections, the DR4 pipeline takes advantage of the 2MASS  photometric information (see Sect.~\ref{sect:synthetic_grid}), hence partly reducing the effect of these degeneracies. 

\begin{figure}[tbp]
\begin{center}
\includegraphics[width=0.4\textwidth,angle=0]{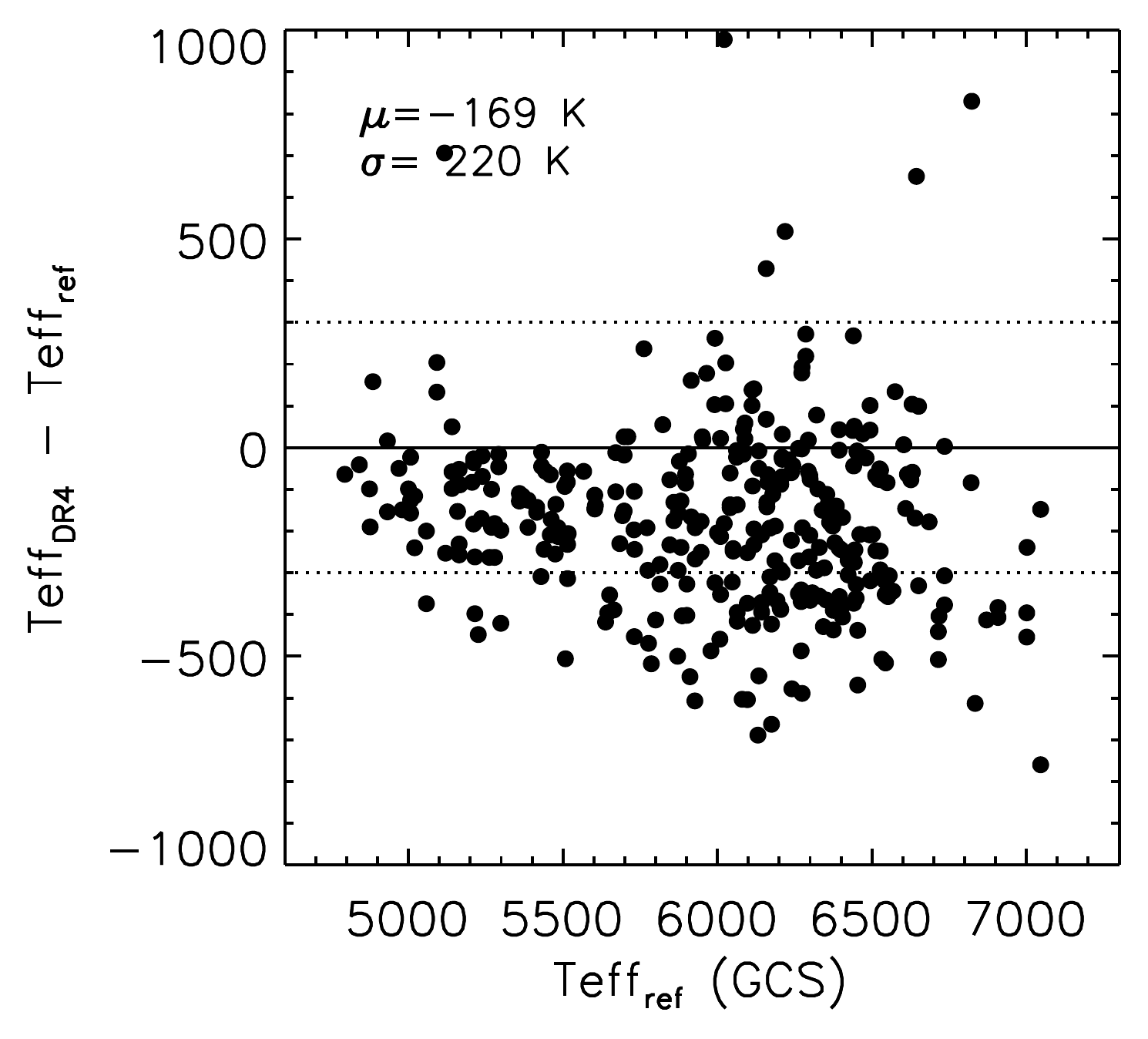}
\caption{Comparison of the effective temperatures found by the DR4 pipeline for RAVE spectra of stars that are part of the Geneva-Copenhagen Survey with the updated values of \cite{Casagrande11}. The constant offset of $\sim170$~K shows that DR4 is not on the same effective temperature scale.}
\label{Fig:GCS_Teff}
\end{center}
\end{figure}

\paragraph{Discussion on the effective temperature scale}
In order to verify the determination of the effective temperatures, we compared  the DR4 values of  RAVE spectra with SNR$>20~{\rm pixel}^{-1}$ for 327 stars in common with the photometric effective temperatures  from the \cite{Casagrande10} calibration of the Geneva-Copenhagen Survey \citep[GCS,][]{Nordstrom04}. 
Figure~\ref{Fig:GCS_Teff} shows on one hand a  small dispersion of the DR4 pipeline's effective temperatures when comparing with the values published by \cite{Casagrande11} for the GCS
 but on the other hand there is a constant  underestimation of $\sim170$~K. Nevertheless, since the GCS covers only a limited range of the parameter space (only metal-rich dwarfs), and because any such offset is not seen with the other calibration data sets, it has been decided not to apply any correction to the RAVE \teff\ scale. We note though that  the user of the DR4 effective temperatures should be aware that in order to be in agreement with the \citet{Casagrande10} effective temperature scale, for the type of stars analysed by the GCS, an offset correction should be performed.

\subsection{Overall metallicity calibration}
\begin{figure*}[tbp]
\begin{center}
$\begin{array}{cc}
\includegraphics[width=0.4\textwidth,angle=0]{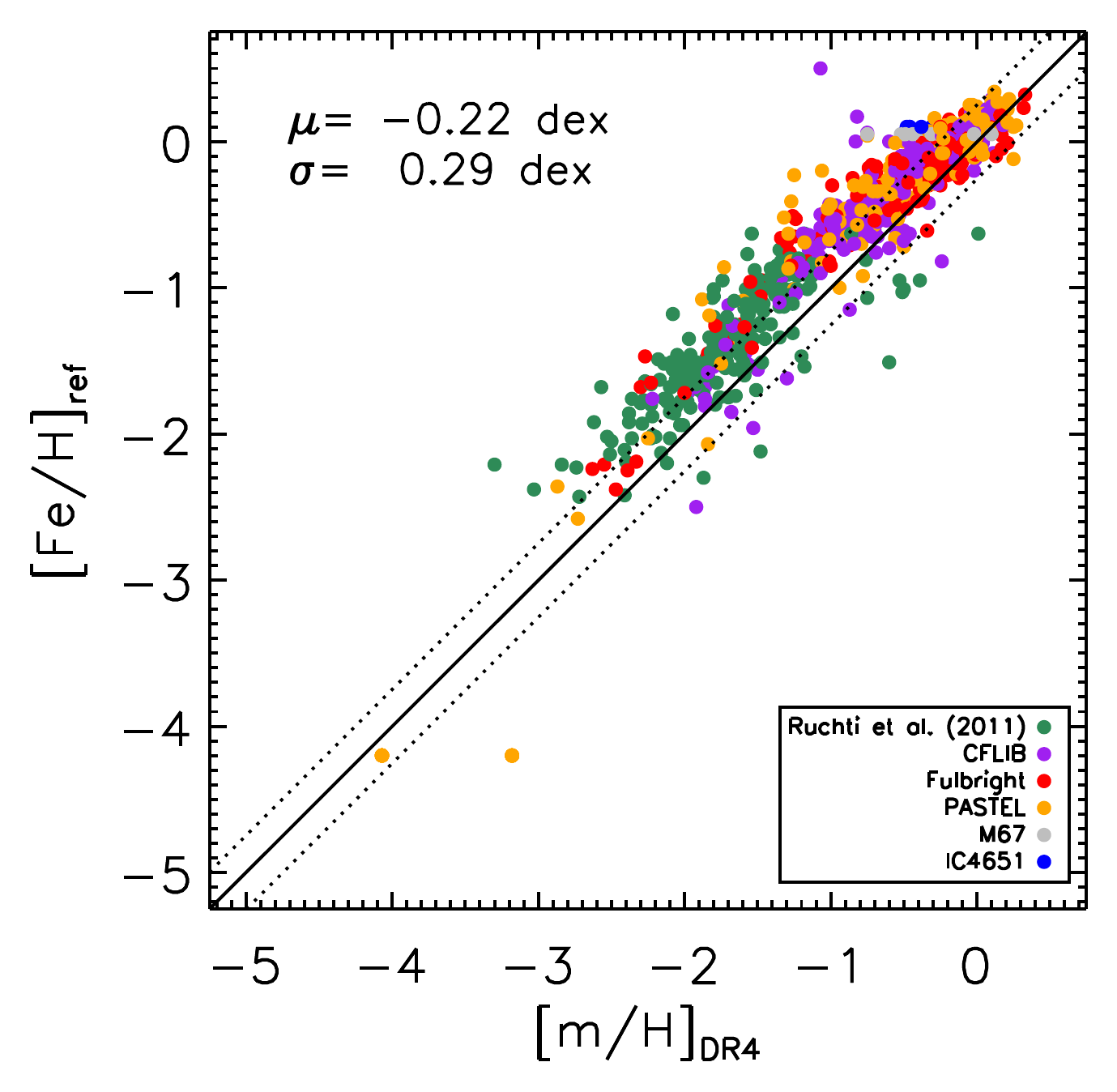} &
\includegraphics[width=0.4\textwidth,angle=0]{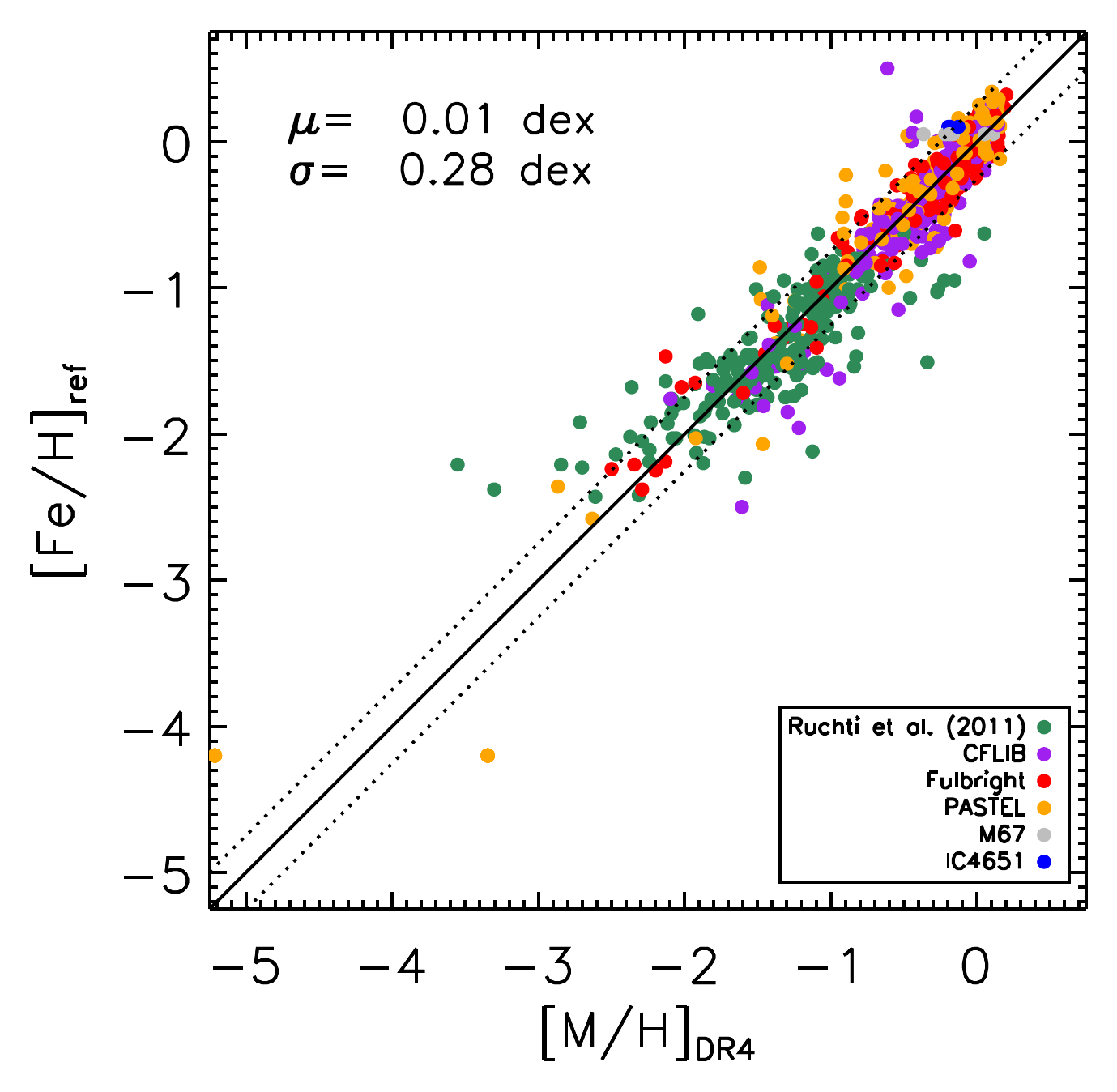}
\end{array}$
\end{center}
\caption{Comparison between the reference iron abundances found in the literature ([Fe/H]$_{\rm ref}$) and the derived overall metallicities ([m/H]$_{\rm DR4}$, left hand side) and the calibrated overall metallicities (\meta$_{\rm DR4}$, right hand side), according to Eq.~\ref{eq:bias_correction}. Colour coding for each data set is the same as in Fig.~\ref{Fig:Calib_HRD}. Dotted diagonal lines represent offsets from unity of $\pm 0.25$~dex.} 
\label{Fig:Calib_meta}
\end{figure*}

In order to investigate the calibration needs for the metallicities, we used the iron abundances ([Fe/H])  from the literature. Indeed, we recall that the iron-peak  and the  $\alpha-$element abundances  of the synthetic spectra are scaled to the iron abundance. Since the metallicity measurement is dominated by the \ion{Ca}{2} lines (which correspond to an $\alpha-$element), in our case and for standard Galactic $\alpha-$abundances ({\it i.e.} following the trend defined in Sect.~\ref{sect:synthetic_grid}) we have: [M/H]$_{\rm DR4}\approx$[Fe/H]. Nevertheless, for non-standard stars, the overall metallicity will not be equal to the iron abundance, and hence it has been  decided to keep in what follows the notation \meta~ instead of [Fe/H].  In what follows, the nomenclature of the DR3 paper is adopted, denoting the raw metallicity estimation of the DR4 pipeline as [m/H], and the calibrated (final) metallicity as [M/H]. Comparison of the reference values and those derived by the DR4 pipeline is presented in the left panel of  Fig.~\ref{Fig:Calib_meta}.


 \begin{figure*}[tbp]
\begin{center}
$\begin{array}{ccc}
\includegraphics[width=0.3\textwidth,angle=0]{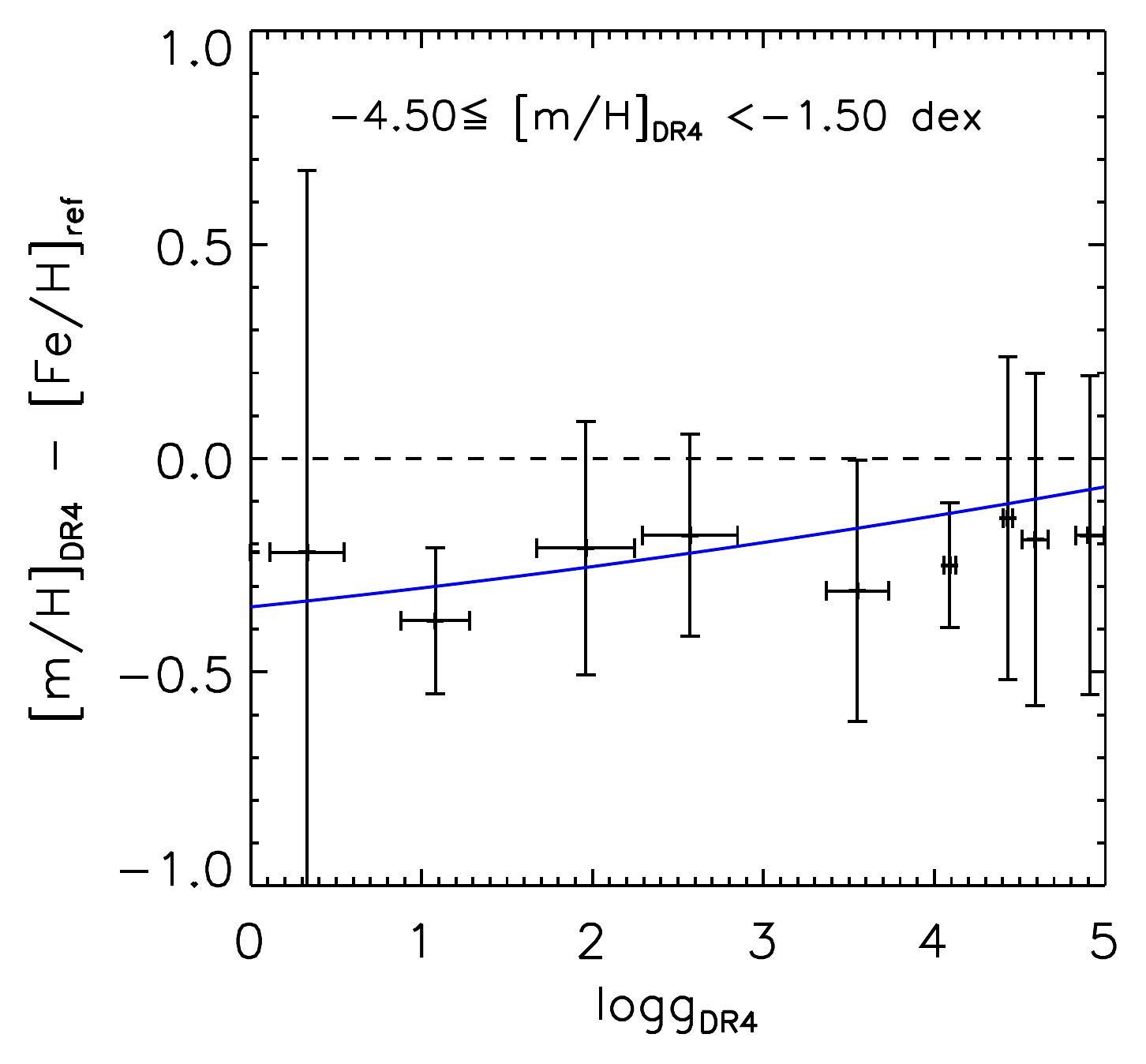} &
\includegraphics[width=0.3\textwidth,angle=0]{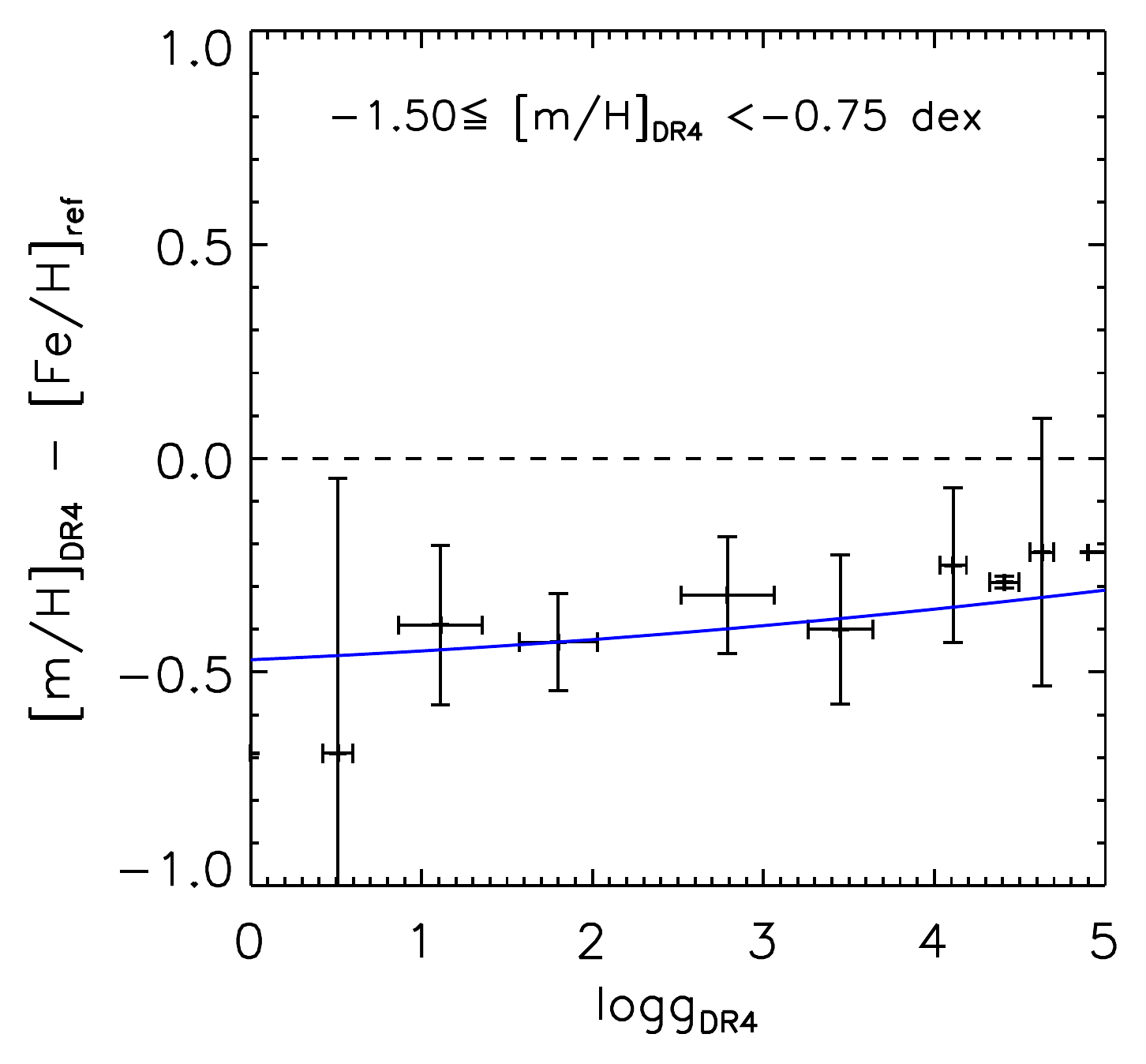} &
\includegraphics[width=0.3\textwidth,angle=0]{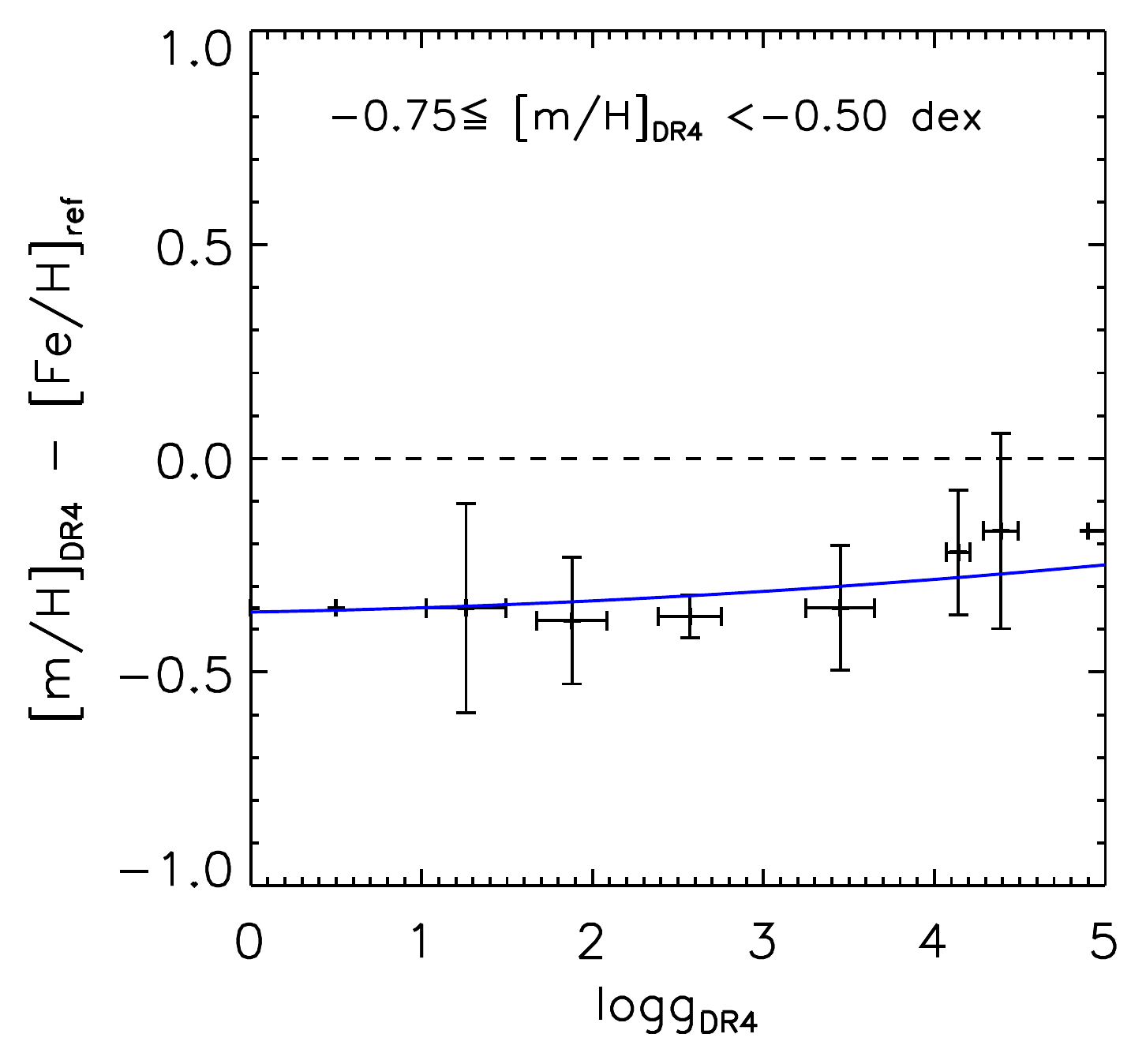} \\
\includegraphics[width=0.3\textwidth,angle=0]{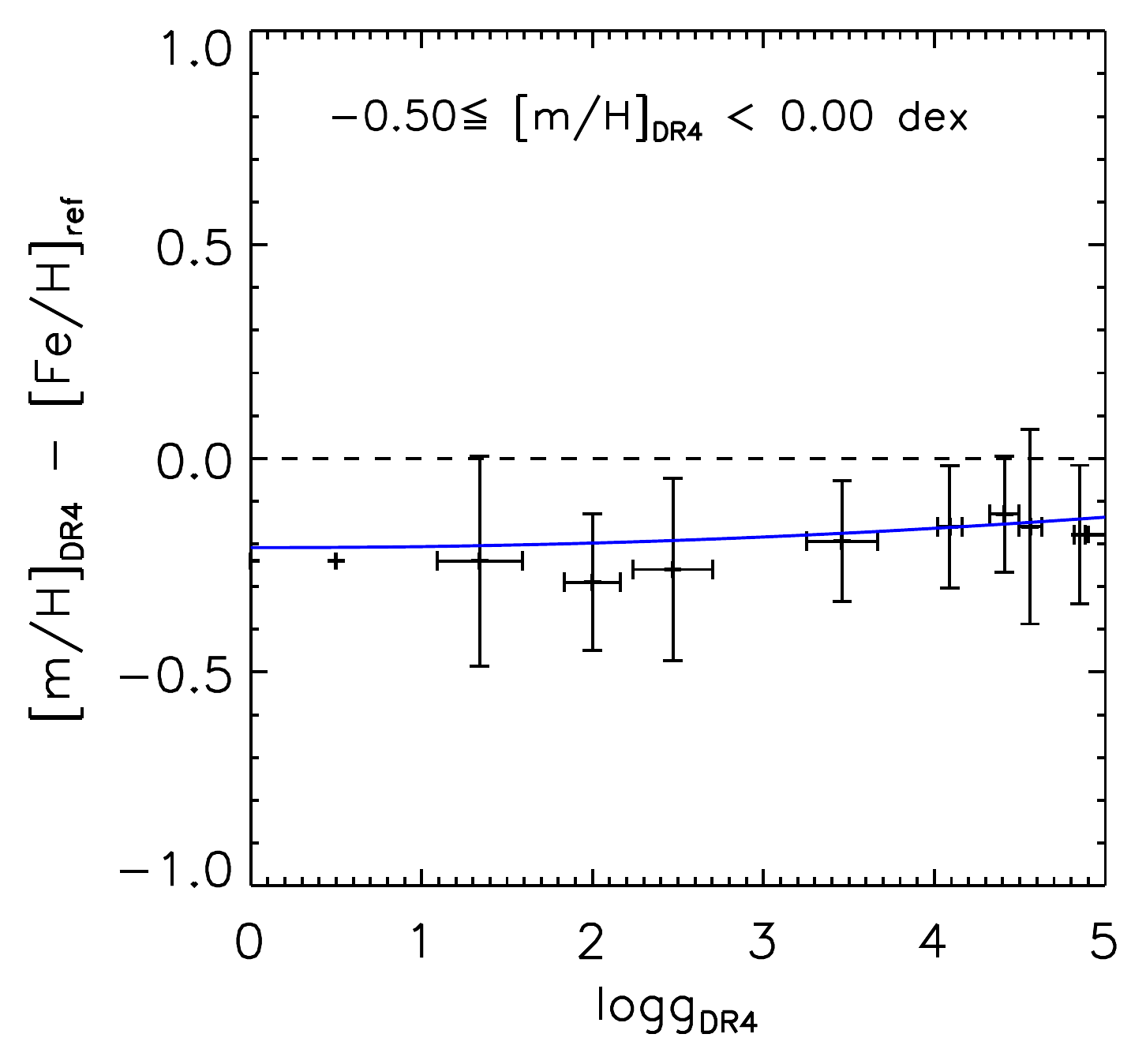} &
\includegraphics[width=0.3\textwidth,angle=0]{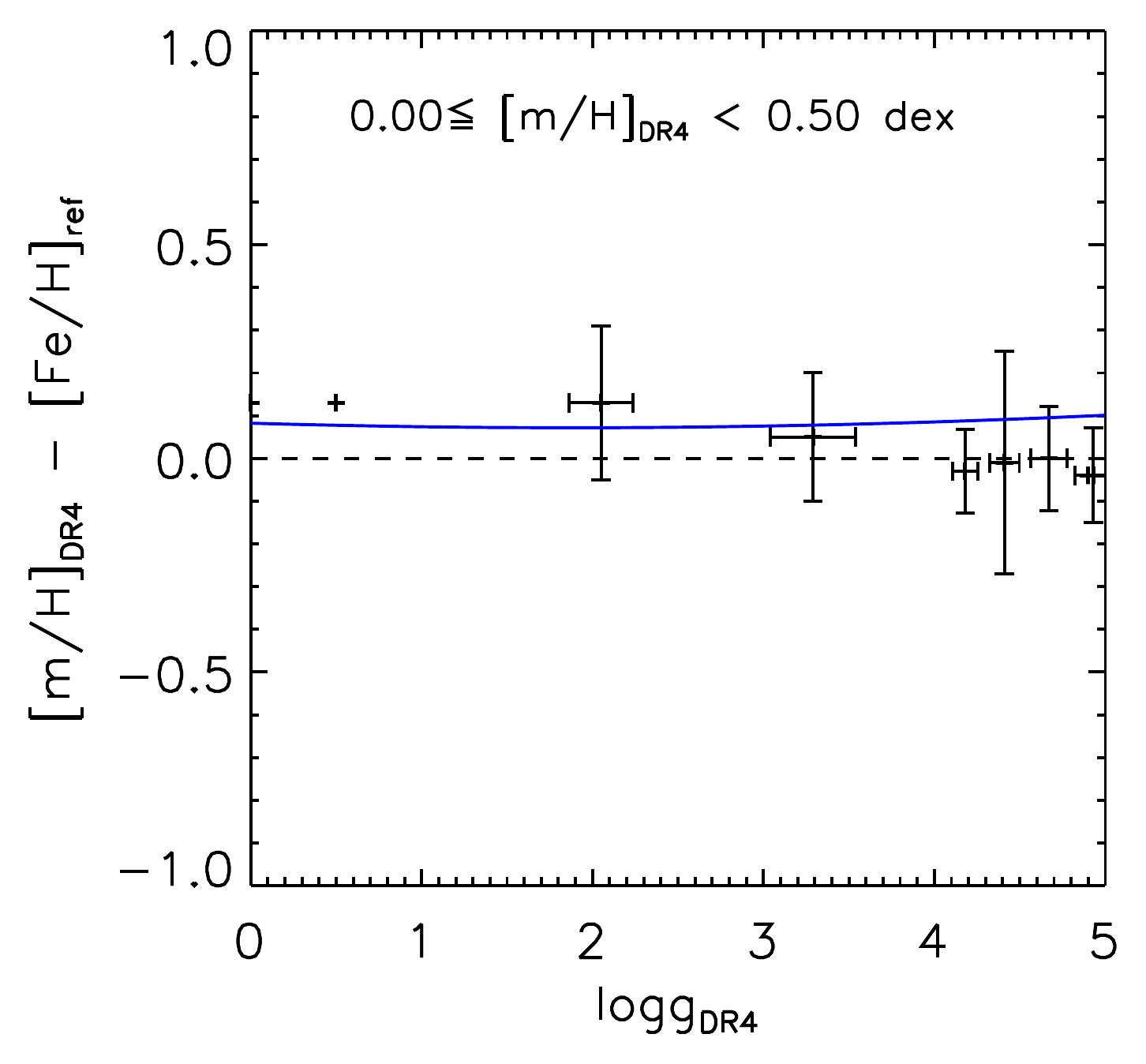} 
\end{array}$
\end{center}
\caption{Trends in the metallicity determination for different gravity bins, at different metallicity ranges. The error bars correspond to the dispersion of \logg\ and the error in metallicity inside each bin. The blue lines represent the polynomial that has been adopted in order to best describe the metallicity offsets. It corresponds to the polynomial of Eq.~\ref{eq:bias_correction}, and has been computed for the mean metallicity of each subsample.} \label{Fig:pipeline_fits}
\end{figure*}

The results of Fig.~\ref{Fig:Calib_meta} have been obtained with all the data sets of Table~\ref{tab:calibration_sets}, assuming the metallicities given in the fourth column of that table. 
From the left hand side plot of Fig.~\ref{Fig:Calib_meta} one can notice that there is an offset between the derived metallicities from the RAVE spectra and the reference iron abundances. This bias is not the same for all metallicities, and it is  more important for metal-poor stars than in the metal-rich regime. 
We investigated the correlations of the errors, and found that the main parameters driving the bias are the surface gravity and the metallicity itself. Figure~\ref{Fig:pipeline_fits} illustrates the covariance of the residual errors on the metallicity with respect to the surface gravity, for different metallicity ranges. 
In this figure, 
each point and error bar represent the median and the dispersion of the  metallicity error for the stars inside each gravity bin. This binning approach smooths the errors, minimises the impact of outliers and highlights the general trends of the biases. 
 On one hand, the results of Fig.~\ref{Fig:pipeline_fits} show for the lowest metallicities a rather constant underestimation of the metallicity  by  $0.2$~dex.
On the other hand, there are some clear trends in the more metal-rich regimes, where the giant stars exhibit higher offsets than the dwarfs. 
These trends are  too strong to be explained by a variation of {  microturbulent} velocity along the giant branch, where the expected  offsets should be less than  0.1~dex \citep[see for example ][]{Kirby09}. 


Using the binned points of Fig.~\ref{Fig:pipeline_fits}, the resulting fit of a quadratic surface for the errors in metallicity, taking into account the dependences on both the surface gravity and the metallicity,  is:
\begin{multline}
{\rm [m/H]}-{\rm [M/H]_{\mathrm{ref}}}=
-0.076-0.006*\log g + 0.003 * \log^2 g \\- 0.021 *{\rm [m/H] }* \log g +0.582 *{\rm [m/H]} + 0.205*{\rm [m/H]}^2.
\label{eq:bias_correction}
\end{multline}

Given this relation, the trend for the typical mean metallicity inside each box of Fig.~\ref{Fig:pipeline_fits} has been plotted in blue. As expected, the fits are in good agreement with the offsets, hence assimilating the metallicity calibration relation to Eq.~\ref{eq:bias_correction}.
The right panel of Fig.~\ref{Fig:Calib_meta}  shows   the improvement that has been made on the metallicity determination thanks  to the correction of Eq.~\ref{eq:bias_correction}.  We note, however, that due to the lack of reference stars with super-solar metallicities, our calibration is not optimal for [M/H]$>+$0.1~dex.  This limitation will be addressed in a future study.

\begin{figure*}[tp]
\begin{center}
\includegraphics[width=0.9\textwidth,angle=0]{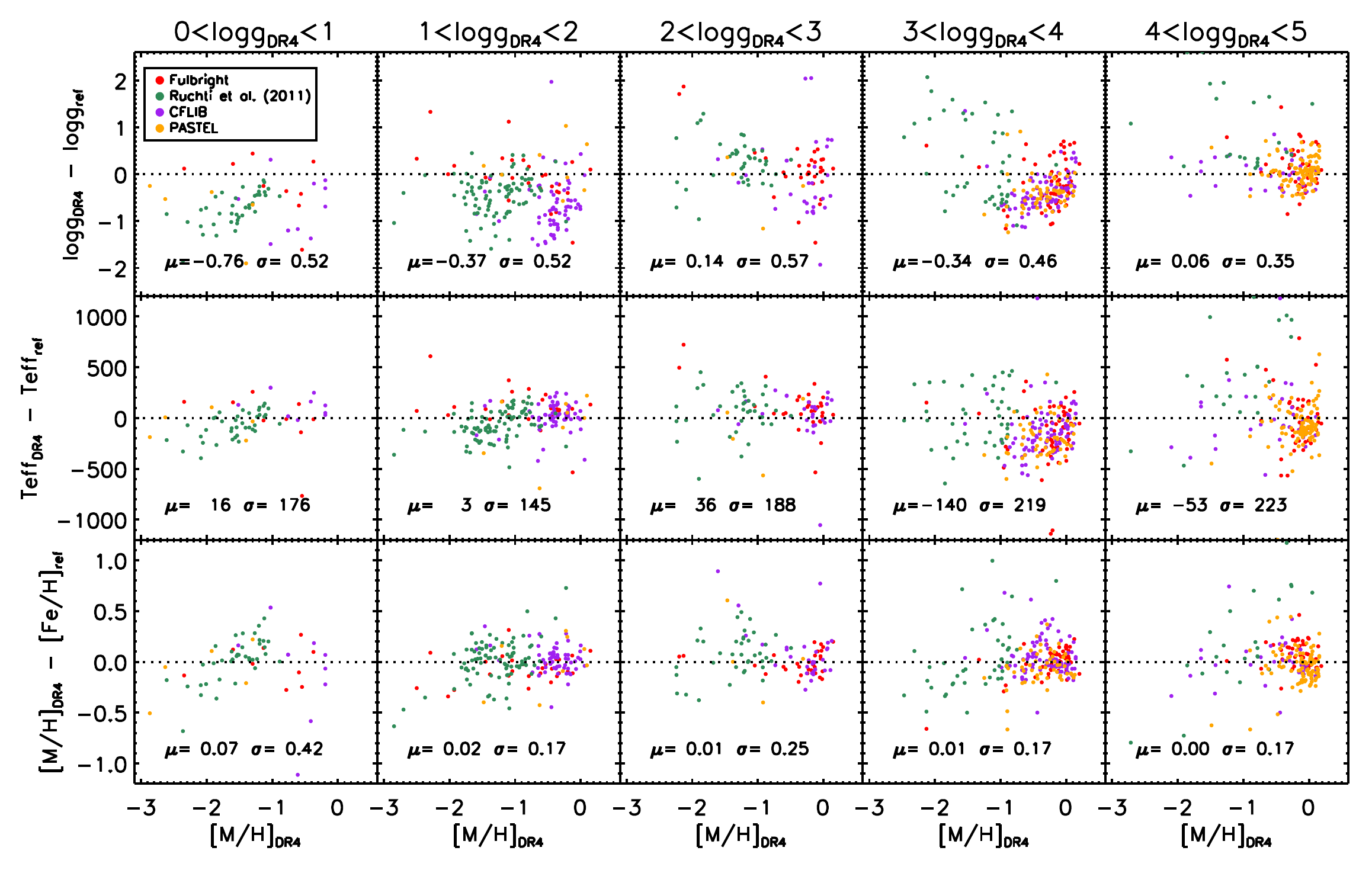}
\caption{DR4 residual plots for RAVE  (RAVE-PASTEL in yellow, Ruchti in green and Fulbright in red) and RAVE-like spectra (CFLIB, in purple). The trends in the stellar parameter systematics are shown with respect to the calibrated metallicity, divided into different (DR4) \logg\ bins of 1~dex. In each panel, the median offset and the dispersion is indicated. }
\label{Fig:Kordopatis_bias}
\end{center}
\end{figure*}
A more detailed investigation of the residuals for different gravity regimes and with respect to the calibrated  metallicity is shown in Fig.~\ref{Fig:Kordopatis_bias},  for the Ruchti, Fulbright, PASTEL and CFLIB libraries.  As expected, there is no bias for the calibrated metallicity (lower plots), nor for the other parameters, except for the surface gravity of the lowest gravity giant stars (see also Sect.~\ref{sect:rejection_criteria}). The self-consistency of the calibration is hence validated.

\subsection{A comment on  RAVE DR1, DR2, \& DR3  parameters}
Previous RAVE data releases  used the \cite{Munari05} grid of synthetic spectra and a penalised $\chi^2$ algorithm in order to determine the effective temperatures, surface gravities, overall metallicities and $\alpha-$abundances. 
The stellar rotational velocities ($V_{\rm rot}$) and the {  microturbulent} velocities ($\xi$) were also left as free parameters, although without attributing any constraint on these values in the end\footnote{for DR3 the {  microturbulent} parameter  was fixed at $\xi=2$~\kms.}. Furthermore, the 2MASS photometric information was not used to help reduce  spectral degeneracies and  the calibration data sets were not fully available.

The present DR4 pipeline reduces  the parameter space to only the three free atmospheric parameters we are trying to measure.  In addition to imposing a photometric effective temperature range, the new pipeline explores more efficiently the parameter space, thanks to the decision-tree algorithm. This makes the new results more robust and less susceptible to biases caused by spectral degeneracies.

\begin{figure*}[tp]
\begin{center}
\includegraphics[width=0.9\textwidth,angle=0]{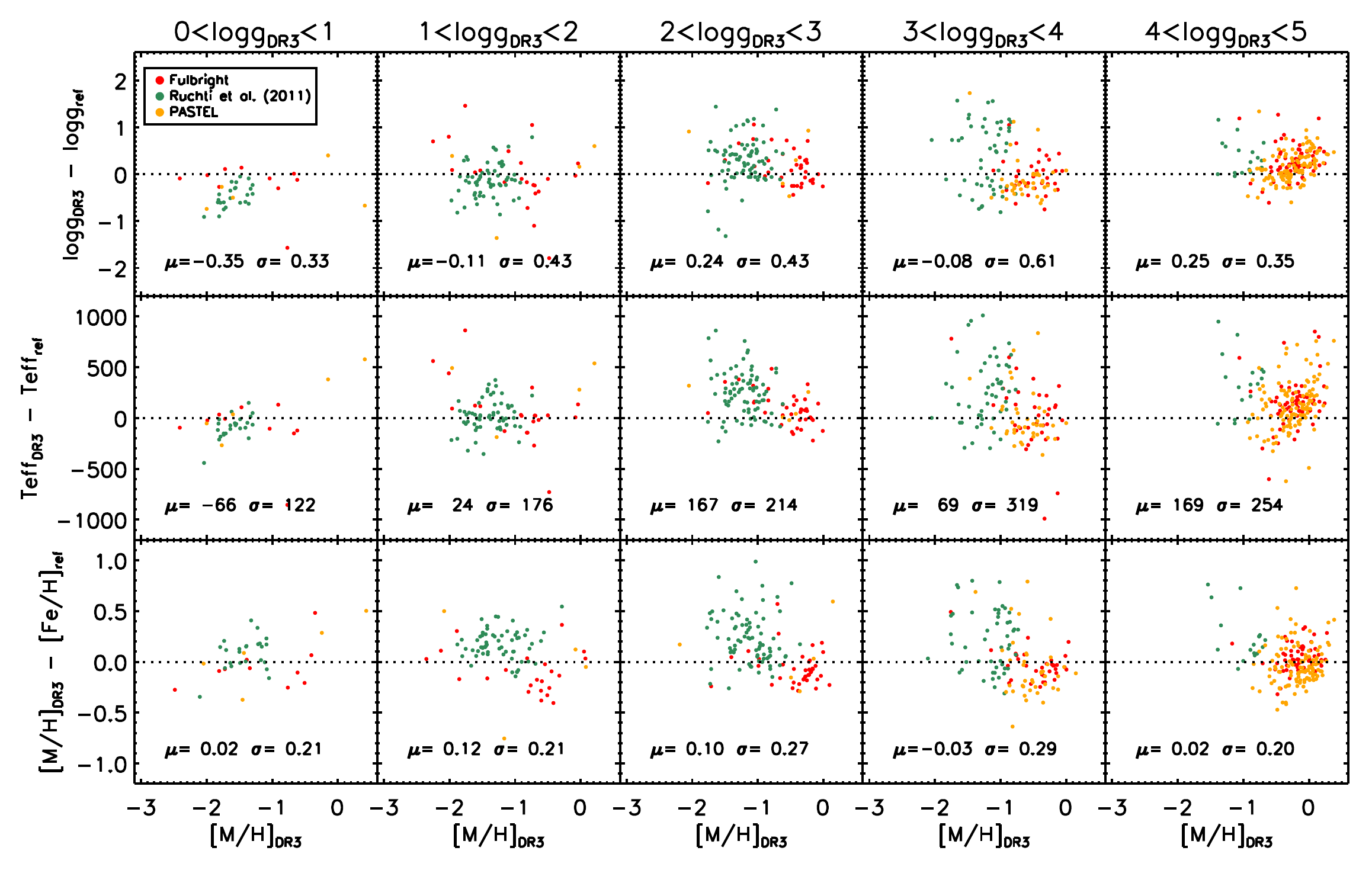}
\caption{Same as Fig.~\ref{Fig:Kordopatis_bias} but for the RAVE DR3 pipeline and without the CFLIB analysis. The calibrated metallicities correspond to those obtained using Eq.~2 of \cite{Siebert11} with parameters $c_0=0.578, c_1=1.095, c_3=1.246, c_4=-0.520$. The metallicity trends found  for the turn-off stars are representative of the uncalibratable biases present in RAVE DR3.  }
\label{Fig:RAVE_bias}
\end{center}
\end{figure*}

Efficient exploration of the low dimension parameter space is crucial for the accurate determination of  the atmospheric parameters, and calibration of  the results. 
Indeed, tests done on the above mentioned calibration data sets, using the DR3 pipeline output, showed that the metallicity biases could not be calibrated adequately, especially for the turn-off stars, where the degeneracy of the distance function is the most important (see Fig.~\ref{Fig:RAVE_bias}). 
As it will be shown in Sect.~\ref{sect:results_checks},  this lead to biases and interdependences between the DR3 parameters, and motivated the effort to develop the approach implemented here in DR4. 

\subsection{Sanity check of the metallicity  calibration on a set of observed spectra}
\label{sect:additionnal_clusters}
Our proposed metallicity  calibration relation  (see Eq.~\ref{eq:bias_correction}) has been further verified on  spectra that are not  part of the calibration process. For that purpose, we used the  327 RAVE spectra  of the GCS stars described previously, 105 non-RAVE spectra from the $S^4N$ library degraded to  RAVE resolution  \citep{Allende-Prieto04} and 65 RAVE-like spectra of open and globular cluster stars obtained by the 2.3m telescope at the SSO, listed in Table~\ref{tab:testing_sets}. 
We note though that the reference metallicity values that have been adopted for these test spectra are not as reliable as the calibration data sets. Indeed, except for the $S^4N$ library, all the other data sets do not have individual spectroscopically measured metallicities. In addition, non-member stars might be included in the cluster data sets. Finally, the mean metallicity value has been considered for the stars belonging to the globular clusters, whereas dispersions up to few tenths of a dex \citep{Gratton04} can be expected in some cases.

\begin{table*}[tdp]
\caption{Post-calibration verification data sets}
\begin{center}
\begin{tabular}{cccccc}
Dataset     & Type  &  N stars & $<$[Fe/H]$>$  &$\sigma($[Fe/H]$)$& Reference \\ \hline
M5                   &  globular cluster  & 8 & -1.28 & 0.11&   \cite{Ramirez03}\\
NGC 3680    &  open  cluster & 7 & -0.04 &    0.03 & \cite{Pace08}\\
IC4651           & open  cluster  & 5 & +0.10 &   0.05 & \cite{Pasquini04}\\
M67                & open cluster& 10 & +0.05 &   0.04   & \cite{Pancino10} \\
NGC 6752     & globular cluster & 12 & -1.42 &  0.10 & \cite{Gratton01}\\ 
NGC 2808    & globular cluster & 10  & -1.14 &  0.06  & \cite{Carretta04}  \\
NGC 6397    & globular cluster & 11 & -2.10 &    0.05 & \cite{Koch11a} \\ 
Praesepe      & open cluster & 35 & +0.14 & 0.04 & mean literature value\\
GCS               & MW dwarf stars & 327 & solar & -- & \cite{Casagrande11} \\ 
$S^4N$         & MW dwarf stars &  105 & solar & -- & \cite{Allende-Prieto04}  \\ \hline
\end{tabular}
\end{center}
\label{tab:testing_sets}
\end{table*}%

 \begin{figure*}[tbp]
\begin{center}
$\begin{array}{ccc}
\includegraphics[width=0.33\textwidth,angle=0]{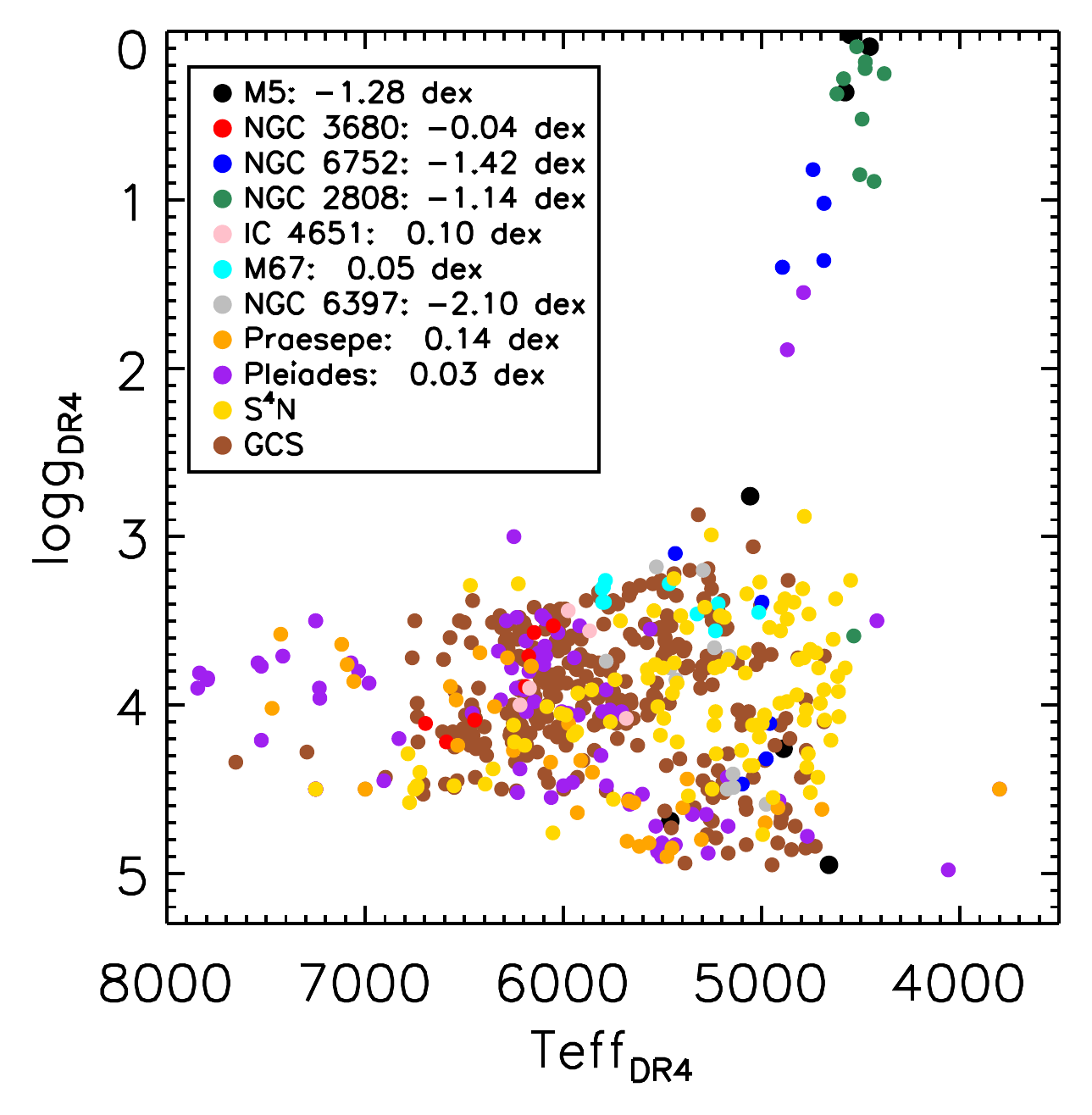} &
\includegraphics[width=0.33\textwidth,angle=0]{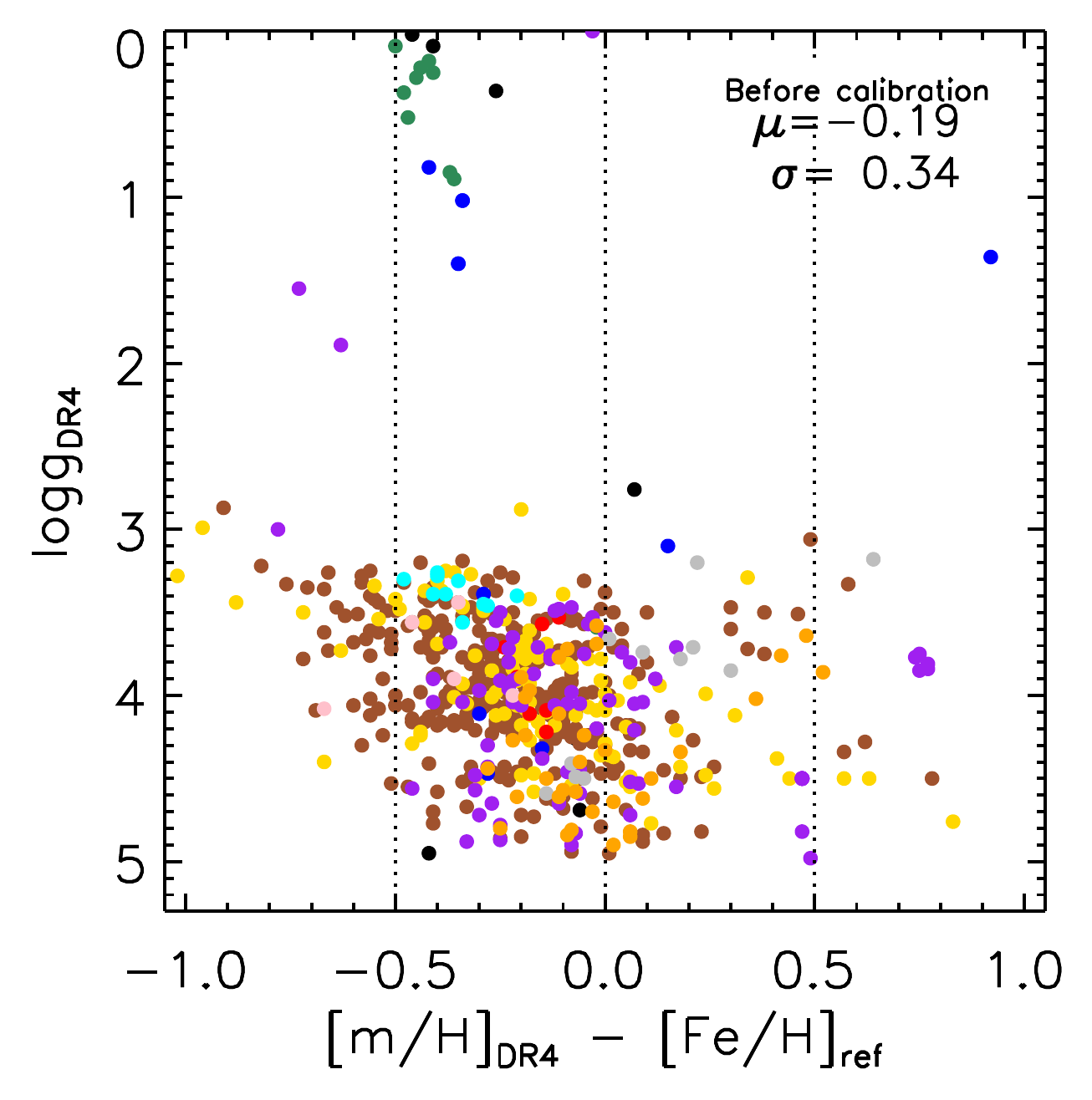} &
\includegraphics[width=0.33\textwidth,angle=0]{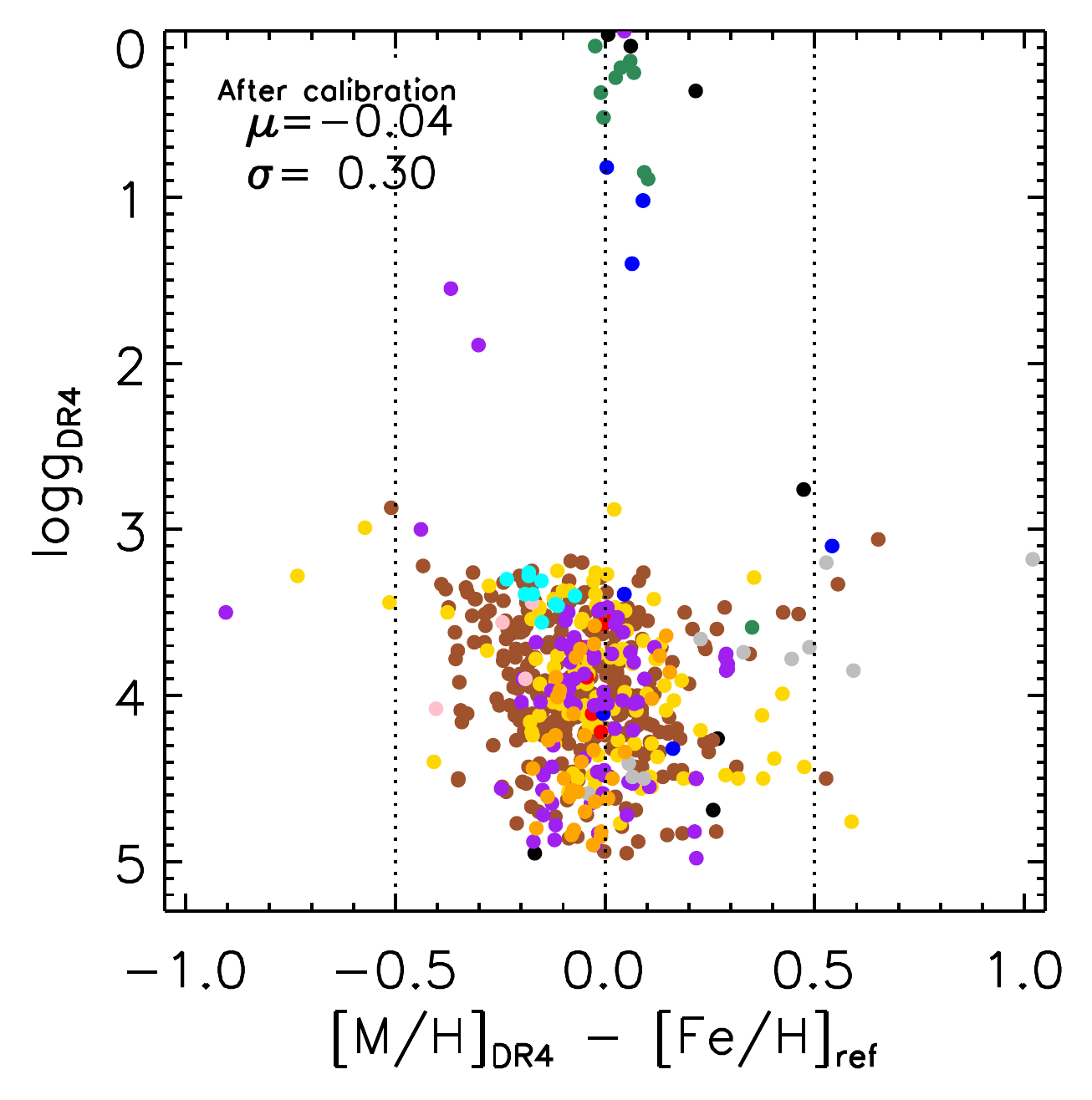} 
\end{array}$
\end{center}
\caption{Post-calibration verification data sets of open and globular cluster stars, the Geneva-Copenhagen Survey, and the $S^4N$ library.  The plot on the right shows that the offsets  are greatly improved once the \logg-dependent corrections in metallicity have been applied. The mean offsets ($\mu$) and dispersions ($\sigma$) of the residuals are noted in the upper corner of each plot.} \label{Fig:testing_sets}
\end{figure*}

The three plots of Fig.~\ref{Fig:testing_sets} show the recovered  \teff--\logg~ diagram of the total considered sample (left), the \logg~ versus residuals in [m/H] (middle) and versus residuals in calibrated [M/H] (right). Despite the relatively large dispersion due to the heterogeneous quality of the data sets, one can see that the bias is greatly reduced in all the  samples, and for all gravities, providing a sanity validation check of the calibration relation established previously.

\subsection{Computation of the total uncertainties of the pipeline}
\label{sect:total_error}
The errors described in Sect.~\ref{sect:Internal_errors} concern only the internal accuracies of the method. 
To estimate the total uncertainties of the pipeline one needs also to estimate the external errors. 

We used  all the spectra with SNR$\geq 50$~pixel$^{-1}$ of the previously described calibration data set  to estimate  the external uncertainties for different ranges of stellar parameters.  Given the total number of spectra in the data set, we divided the sample into cool (\teff$\leq 6000$~K) and hot (\teff$> 6000$~K) dwarfs (\logg$\geq 3.5$~dex) and giants  (\logg$< 3.5$~dex). 
Furthermore,  we also divided  into metal-rich (\meta$\geq -0.5$~dex) and metal-poor  (\meta$< -0.5$~dex) regimes, except for the hot giants for which not enough stars were available in the sample. 
The dispersion of the residual differences is presented in Table~\ref{tab:external_errors}, together with the number of stars that have been considered in order to compute these uncertainties. 

\begin{table}[tdp]
\caption{Estimation of the external uncertainties}
\begin{center}
\begin{tabular}{c|cccc}
\multicolumn{5}{c}{Dwarfs} \\ \hline
Parameter range       &             $N$   &    $\sigma$(\teff)     & $\sigma$(\logg) &$\sigma$(\meta) \\ \hline
\teff$> 6000$,   \meta$< -0.5$   &      28   &   314 &    0.466 &     0.269 \\
\teff$> 6000$,   \meta$\geq -0.5$   &  104    &  173   &  0.276 &     0.119  \\
\teff$\leq 6000$,   \meta$< -0.5$    &     97   &   253   &  0.470 &     0.197  \\
\teff$\leq 6000$,   \meta$\geq -0.5$   &   138    &  145 &     0.384 &     0.111 \\ \hline

\multicolumn{5}{c}{ } \\ 
\multicolumn{5}{c}{Giants} \\ \hline
Parameter range         &             $N$   &    $\sigma$(\teff)     & $\sigma$(\logg) &$\sigma$(\meta) \\ \hline
\teff$> 6000$             & 8     & 263 &    0.423 &     0.300 \\
\teff$\leq 6000$,   \meta$< -0.5$   &     273 &     191 &     0.725 &     0.217 \\
\teff$\leq 6000$,   \meta$\geq -0.5$     &   136   &   89 &     0.605 &    0.144 \\ \hline
\end{tabular}
\end{center}
\label{tab:external_errors}
\normalsize
\end{table}%

Using the values presented in Table~\ref{tab:external_errors}, the total uncertainties of the pipeline parameter determinations are then estimated by adding in quadrature the external errors with the internal errors given in  Tables~\ref{tab:Internal_errors_noColorcuts} and \ref{tab:Internal_errors_Colorcuts}.

\section{Computation of the chemical abundances}
\label{sect:Chemi_pipe}
{ 
The atmospheric parameters inferred in the previous sections {  are} used as an input in order to determine abundances of individual elements.  For that purpose, we use an improved version of the RAVE chemical pipeline described in detail in \citet[afterwards B11]{Boeche11}.  Below, we recall the general features of that pipeline, and present the current improvements.

The chemical pipeline relies on an equivalent widths (EWs) library which contains the expected EWs of the lines visible in the RAVE wavelength range (604 atomic and molecule lines). These EWs are computed for a grid of stellar parameters values covering the range [4000,7000]~K in \teff, [0.0,5.0]~dex in \logg\ and $[-2.5,+0.5]$~dex in \meta\ and five levels of abundances in the range $[-0.4,+0.4]$~dex relative to the metallicity, in steps of 0.2~dex \citep[adopting the Solar abundances of][]{Grevesse98}.
The chemical pipeline constructs on-the-fly spectrum models by adopting the effective temperatures and surface gravities obtained by the DR4 pipeline (see Sect.~\ref{sect:pipeline}). It then searches for the best fitting model by minimising the $\chi^2$ between the models and the observations. 

For a given normalised, RV corrected and wavelength calibrated spectrum, the chemical pipeline determines the elemental abundances, following the steps described below:  
\begin{enumerate}
 \item[{\it i)}]
Upload the EWs for the lines at the estimated DR4 \teff, \logg, \meta\  and for the five different abundance levels.
\item[{\it ii)}]
Keep only the lines which, at the given stellar parameters, have large enough EWs to be visible above the noise. Mathematically, the condition to satisfy is the following: 
$$
\mathrm{EW (m\AA)} > \frac{\sqrt{2\pi}\sigma_{res}}{SNR}\cdot 1000
$$
where $\sigma_{res}=0.56$\AA\ is the standard deviation of the RAVE Gaussian line profile. In practice, any absorption line whose intensity is larger than 1$\sigma$ of the noise fulfills this condition.

\item[{\it iii)}]
Fit the strong  \ion{Ca}{2} and  \ion{H}{1}  lines and correct the continuum (see Sect.~\ref{Sect:continuum_correction}).
\item[{\it iv)}]
Construct the Curve of Growth (COG) of the lines by fitting a polynomial function through the five EW-abundance points.
\item[{\it v)}]
Create the model by assuming a Voigt profile for each line and summing these profiles together (see Sect.~\ref{Sect:line_profile}).
\item[{\it vi)}]
Vary the chemical elemental abundances to obtain different models by changing the EWs of the lines according to their COG.
\item[{\it vii)}]
Finally, minimise the $\chi^2$ between the models and the observed spectrum to find the best-matching model.
\end{enumerate}
Further details on the line list and the way the EW library has been constructed can be found in \citetalias{Boeche11}. In the following subsections we describe the changes that have been brought to \citetalias{Boeche11}. These concern a better consideration of the opacity of neighbouring lines, an implementation of a pseudo-Voigt profile to model the lines   and an improved continuum re-normalisation.


\subsection{Equivalent-width corrections for the opacity of the neighbouring lines} 

The EW library is  built using the driver {\em ewfind} of the spectrum synthesis code MOOG \citep{Sneden73} which computes the EW of every line as if they were isolated. Nevertheless, line blends when not carefully taken into account can lead to abundance over-estimations. 
In the case of lines instrumentally (but not physically) blended, the observed blend has a total EW that is the sum of the EWs of the two isolated lines, and thus no problem arises. 
However, when two lines are physically blended ({\it i.e.} not instrumentally), the quantity of radiation absorbed by one line is affected by the opacity of the neighbouring line, and the total EW of the blend is smaller than the sum of the two isolated EWs. In this case the blend in the constructed model is too strong, leading to abundance overestimations.
 In order to avoid such overestimation, we corrected the EWs of the blended lines in the EW library with the following procedure:

\begin{enumerate}
\item Consider the line $l_0$ having EW$_0$, blended with some lines $l_i$
with EW$_i$. Compute the ratio  $EW_r= EW_0 / {\sum EW_i}$ with $EW_0$ and
$EW_i$ computed as if they were isolated
\item Synthesise the blend composed by $l_0$ and all $l_i$, and measure the overall
$EW_{tot}$
\item Compute the corrected $EW$ of the line $l_0$ as $EW_0^{corr}=EW_r\cdot EW_{tot}$.
\end{enumerate}
Two lines are considered blended if they are closer than 0.2~\AA.
In addition, we applied this correction to lines which are blended with one or more lines having EW$>$10m\AA. Lines with EWs smaller than 10m\AA\ would affect the EW of the neighbouring lines by less than 0.7\%, which can be considered negligible. 
Although $EW_0^{corr}$ is only an approximation, the constructed blends with such corrected EWs match the synthesised blend better than 1\% of the normalised flux. 
This correction replaces the previous one adopted in the \citetalias{Boeche11} chemical pipeline.}

\subsection{Improved line profile}
\label{Sect:line_profile}
{  
Most of the absorption lines in the RAVE wavelength range and resolution have an intrinsic width smaller than the RAVE instrumental profile. Therefore, their line profile is dominated by the instrumental one which is Gaussian. Nevertheless, this is not the case for the strongest lines, where the broad wings extend beyond the instrumental profile. In that case, the line is better approximated by a Voigt profile.

Compared to \citetalias{Boeche11}, the new chemical pipeline drops the simplistic Gaussian assumption and now uses an improved line profile. Because of the difficulties of implementing a real Voigt profile} 
we use the approximation  implemented by \citet{Bruce00}:
\begin{equation}\label{eq_voigt}
V(x)=EW\cdot [rL(x)+(1-r)G(x)]
\end{equation}
where $L$ and $G$ are the Lorentzian and Gaussian functions, respectively,
and EW is expressed in \AA.
The $r$ parameter rules the linear combination between $L$ and $G$, so that
when $r=0$, $V(x)$ is a pure Gaussian and when $r=1$, $V(x)$ is a pure
Lorentzian.
The Full Width Half Maximum (FWHM) of $L$ and $G$ is forced to be identical,
and varies as a function of the EW with the following relation:
\begin{equation}
FWHM=FWHM_{best}+EW/4
\end{equation}
where $FWHM_{best}$ is the best matching FWHM found by the minimisation
routine during the best matching model searching.
Unlike \citet{Bruce00}, we make the parameter $r$ dependent from the EW: 
\begin{equation}
r=0.5\cdot exp\left (\frac{-1}{(3EW)^2+0.001} \right)
\end{equation}
so that, for small EW the line profile is Gaussian, and for large EW the line profile 
approximates a Voigt profile.

\cite{Kielkopf73} showed that the difference between the real and
the pseudo-Voigt profile described by Eq.~\ref{eq_voigt} is always smaller 
than 1.2\% for $EW=0.5$\AA, which corresponds to an error smaller than 0.72\% in EW \citep{Bruce00}.\\

\subsection{Improved continuum re-normalisation}
\label{Sect:continuum_correction}

{  
In order to remove some fringing effects that sometimes affect the initial input RAVE spectra (the same ones as used by the DR4 pipeline, see Sect.~\ref{sect:input_spectra}), the chemical pipeline has its own internal re-normalisation algorithm. It can be summarised as follows (a more detailed discussion can be found in Sec.~2.5 and Fig.~3 of \citetalias{Boeche11}): 
\begin{enumerate}
 \item[{\it i)}]
A preliminary metallicity estimation is performed and the modelled metallic lines are subtracted from the observed spectrum. 
 \item[{\it ii)}]
The strong lines belonging to the \ion{Ca}{2} and  \ion{H}{1} are fitted with a Lorentzian profile and subtracted from the observation.
 \item[{\it iii)}]
The continuum profile is then defined by a box-car smoothing of the residuals obtained after the previous two steps. 
 \item[{\it iv)}]
The strong \ion{Ca}{2} and  \ion{H}{1}  lines are added to the continuum profile to obtain the new "continuum". The chemical pipeline does not measure the broad lines of the \ion{Ca}{2} and \ion{H}{1}, and their profiles are considered part of the continuum level.  Therefore, by adding them to the classical continuum, they are excluded from the chemical analysis. It is  by comparison with this level of "continumm" that the metallic lines are measured.
\end{enumerate}

This re-normalisation permits  better continuum placement around the absorption lines for a better elemental abundance estimation. In particular, the new adopted Voigt profile (see Sect.~\ref{Sect:line_profile}) contributes to improve the continuum placement thanks to the superior fit of the line's wings, which can now be properly subtracted during the re-normalisation procedure. 

The present chemical pipeline applies the continuum placement
like the chemical pipeline outlined in \citetalias{Boeche11} (see their Sect.~2.5), }
with the difference that it is applied twice for spectra with SNR$\geq$40~pixel$^{-1}$ and only once for SNR$<$40~pixel$^{-1}$. Indeed,  thanks to the  pseudo-Voigt profile, the continuum placement process becomes more stable at
SNR$\geq$40~pixel$^{-1}$, and when applied iteratively the continuum estimation converges after 2 iterations.
On the other hand, for SNR$<$40~pixel$^{-1}$ the continuum estimation cannot converge to the right level.
The noise spikes (mistaken as metallic lines by the code) lead to a
too high continuum placement and, consequently, to a too high metallicity estimation.
Thus, the re-normalisation is applied only once for low SNR spectra.

\subsection{Precision and accuracy of the RAVE chemical elemental abundances}
In order to evaluate the precision and accuracy of the 
new chemical pipeline, we ran tests on synthetic and real spectra with
known chemical abundances and compared the results with the expected
abundances. The samples of synthetic and real spectra
used are the same as those employed in \citetalias{Boeche11}. This allows us to have a clear view of the achieved improvement between the two pipelines. 

Unlike the work presented in \citetalias{Boeche11}, we present here the tests and
results for only 6 elements (aluminium, magnesium, silicon, titanium, nickel and iron). We rejected the calcium abundance as not being reliable (see Sect.~\ref{sect:chemical_reliability}). 

\subsubsection{Internal errors: tests on synthetic spectra}
\begin{figure*}[tbp]
\begin{center}
$\begin{array}{cc}
\includegraphics[width=8cm]{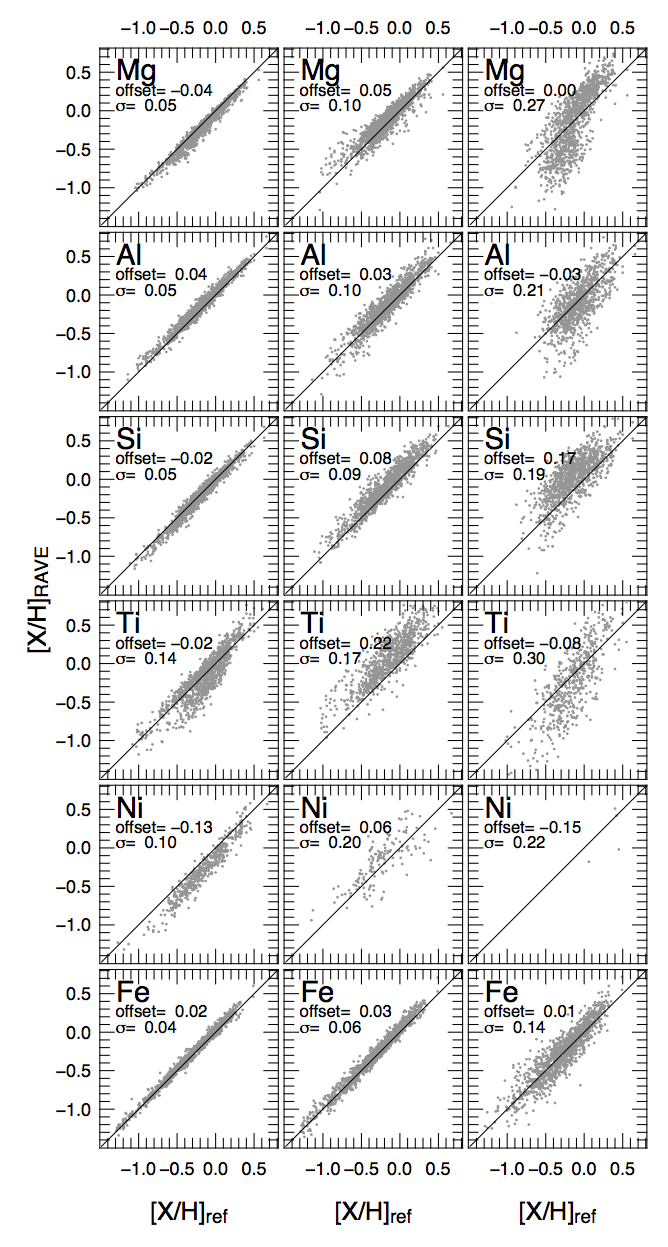}&
\includegraphics[width=8cm]{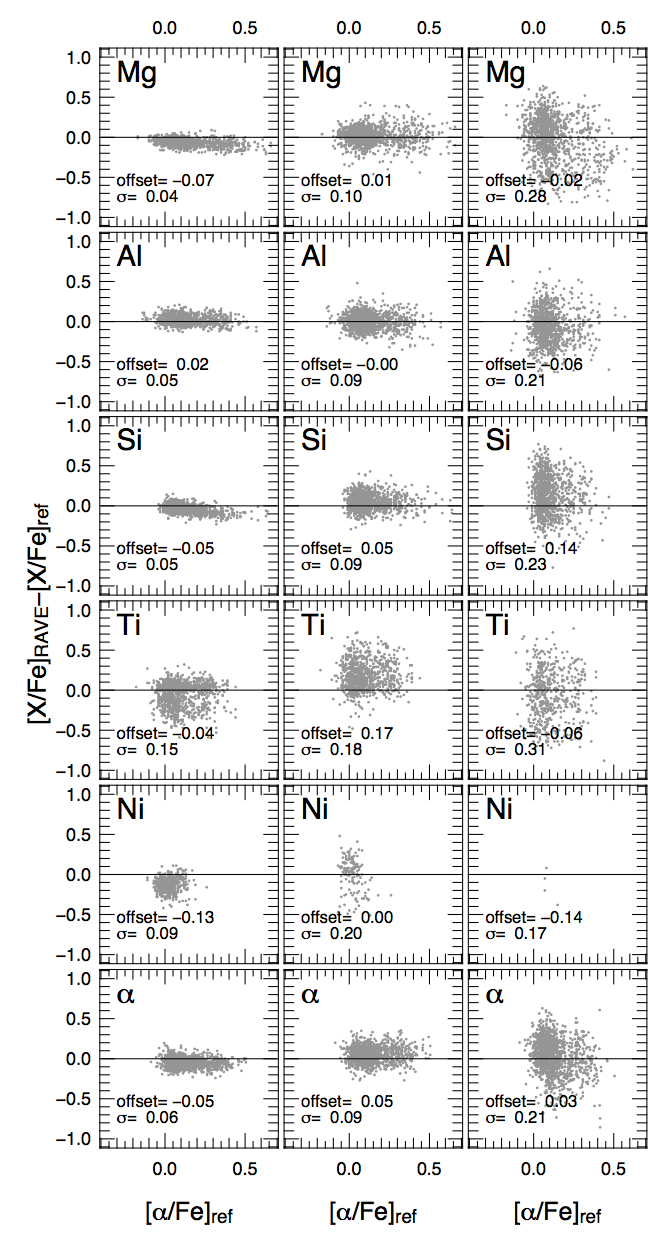}
\end{array}$
\caption{{\bf Left}: expected elemental abundances [X/H] (x-axis) versus measured elemental abundances
(y-axis) for the sample of
synthetic spectra at SNR=100, 40, 20~pixel$^{-1}$ (for the left, middle and right column,
respectively) and assuming no errors in stellar parameters.
{\bf Right}: as in left panels
but for the expected enhancement [X/Fe] (x-axis) and the residuals
measured-minus-expected (y-axis). Offsets and standard deviations are
reported in the panels. $\alpha-$abundances are computed as in \citetalias{Boeche11}, {\it i.e.} the mean of Mg and Si abundances. }
\label{X_H_comparison}
\end{center}
\end{figure*}

\begin{figure*}[tbp]
\begin{center}
$\begin{array}{cc}
\includegraphics[width=8cm]{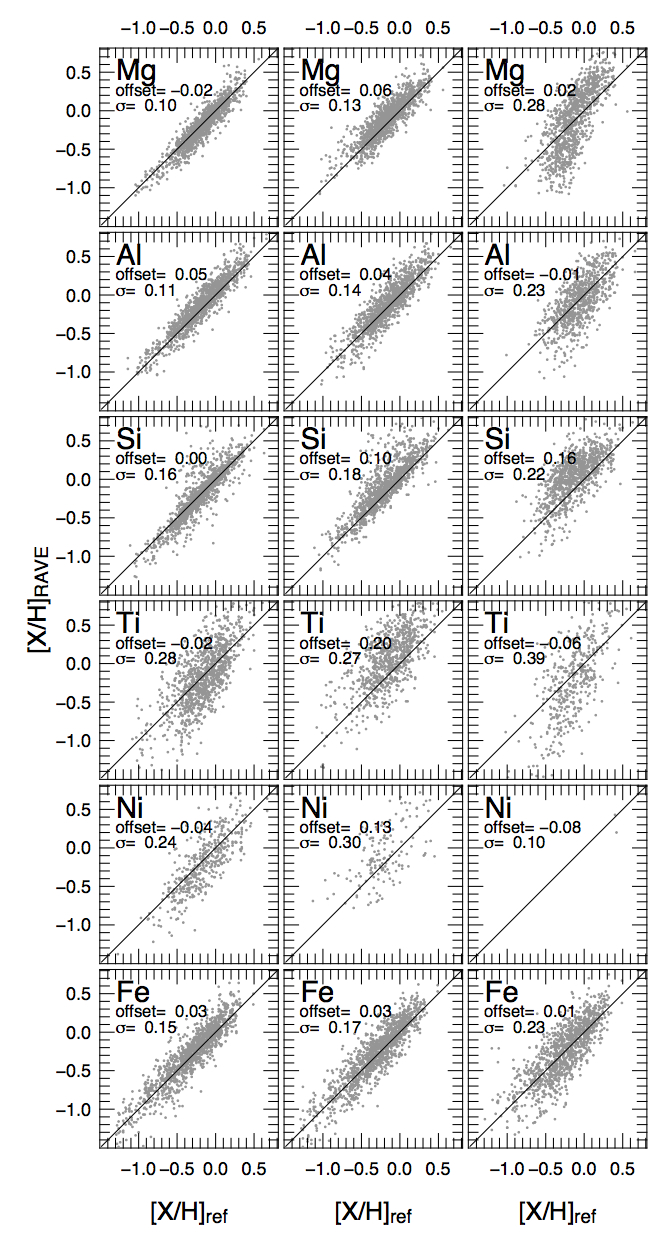}&
\includegraphics[width=8cm]{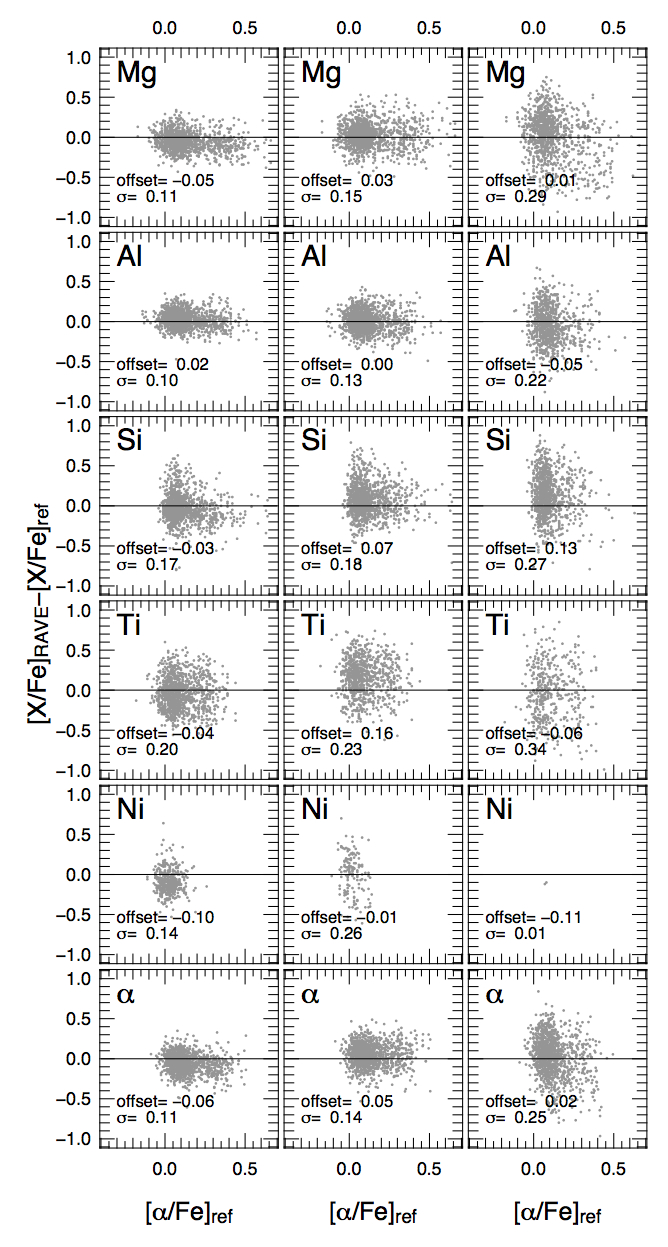}
\end{array}$
\caption{As in Fig.~\ref{X_H_comparison} but with noisy stellar parameters
to simulate the errors appropriate for the RAVE pipeline.}
\label{X_H_comparison_err}
\end{center}
\end{figure*}

The tests have been performed on a sample of 1353 synthetic spectra. 
The values for the effective temperature and the surface gravity for these spectra have been taken from a mock sample 
of RAVE observations created  using the Besan\c{c}on model, 
whereas the adopted chemical abundances have been taken 
from the \cite{Soubiran05}   catalogue whose star metallicities span from $-1.5$~dex to $+0.4$~dex (for further details on how  
the sample has been constructed, see \citetalias{Boeche11}).  This ensures plausible
stellar parameters and chemical abundance distributions of the synthetic
spectra.

We evaluated the precision and accuracy of the results at SNR=100, 40, 20~pixel$^{-1}$.
In Fig.~\ref{X_H_comparison} and \ref{X_H_comparison_err} we report the detailed  results.

\paragraph{Results at SNR=100~pixel$^{-1}$}
While the \citetalias{Boeche11} chemical pipeline gave slightly underestimated abundances,
the present one reduces or removes such underestimation for most of the
elements at SNR=100~pixel$^{-1}$. [Ni/H] is
under-estimated by $\sim$0.1~dex  whereas the [Ti/H] estimate is good 
for giants but should be rejected for dwarfs (for which Ti lines are too weak
for a good estimation). 

\paragraph{Results at SNR=40~pixel$^{-1}$}
All the elements have reliable abundances, except for [Si/H] and [Ti/H],
which look overestimated by $\sim$+0.1, and $\sim$+0.2~dex, respectively. 

\paragraph{Results at SNR=20~pixel$^{-1}$} 
[Fe/H], [Si/H] and [Al/H] are reliable, with uncertainties of $\sim 0.15 - 0.20$~dex.  [Mg/H] and
[Ti/H] show significant systematics, and [Ni/H] cannot be 
measured because its lines are too weak.\\

For Ni and Ti the selection effect due to the SNR is particularly evident.
Moving to lower SNR the number of spectra with Ti and Ni estimations
decreases, because the lines of Ti (in dwarfs stars) and Ni are
weak in the RAVE wavelength range and they do not overcome the noise at low SNR.
This selection bias is further discussed in Sec.~\ref{sect:chemical_reliability}.\\

\begin{figure*}[t]
\begin{center}
\includegraphics[width=12cm]{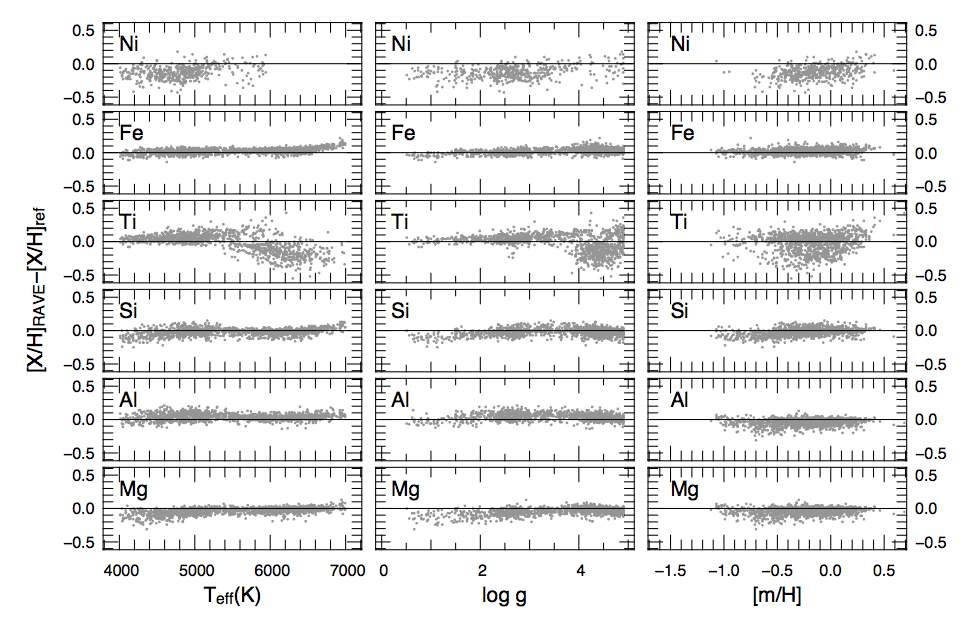}
\caption{Correlation between the elemental abundance residuals and the stellar
parameters at SNR=100~pixel$^{-1}$.}
\label{resid_abd_params}
\end{center}
\end{figure*}

In general, the new chemical pipeline suffers smaller systematics with
respect to the old one. Underestimations are reduced and abundances of
important elements (Fe, Si, Al and Mg) do not correlate  with the effective temperature (see Fig.~\ref{resid_abd_params}) as they did with the previous pipeline. 
[Ti/H] appears reliable only for cool giants.

We further tested the robustness of our results by repeating the abundance
measurements after adjusting randomly the initial \teff, \logg\ and \meta\ by values 
normally distributed around their true values with
$\sigma_{\rm T_{eff}}=250$~K,
$\sigma_{\log g}=0.5$~dex and $\sigma_{\rm [M/H]}=0.2$~dex, respectively.
The results are shown in Fig.~\ref{X_H_comparison_err}.
The shifts in stellar parameters  (representing the input errors) simply inflate the errors in abundances
seen in the test without stellar parameters errors, without introducing any new  systematics.

\subsubsection{External errors: tests on real spectra}

\begin{figure*}[tbh]
\begin{center}
\includegraphics[width=14cm]{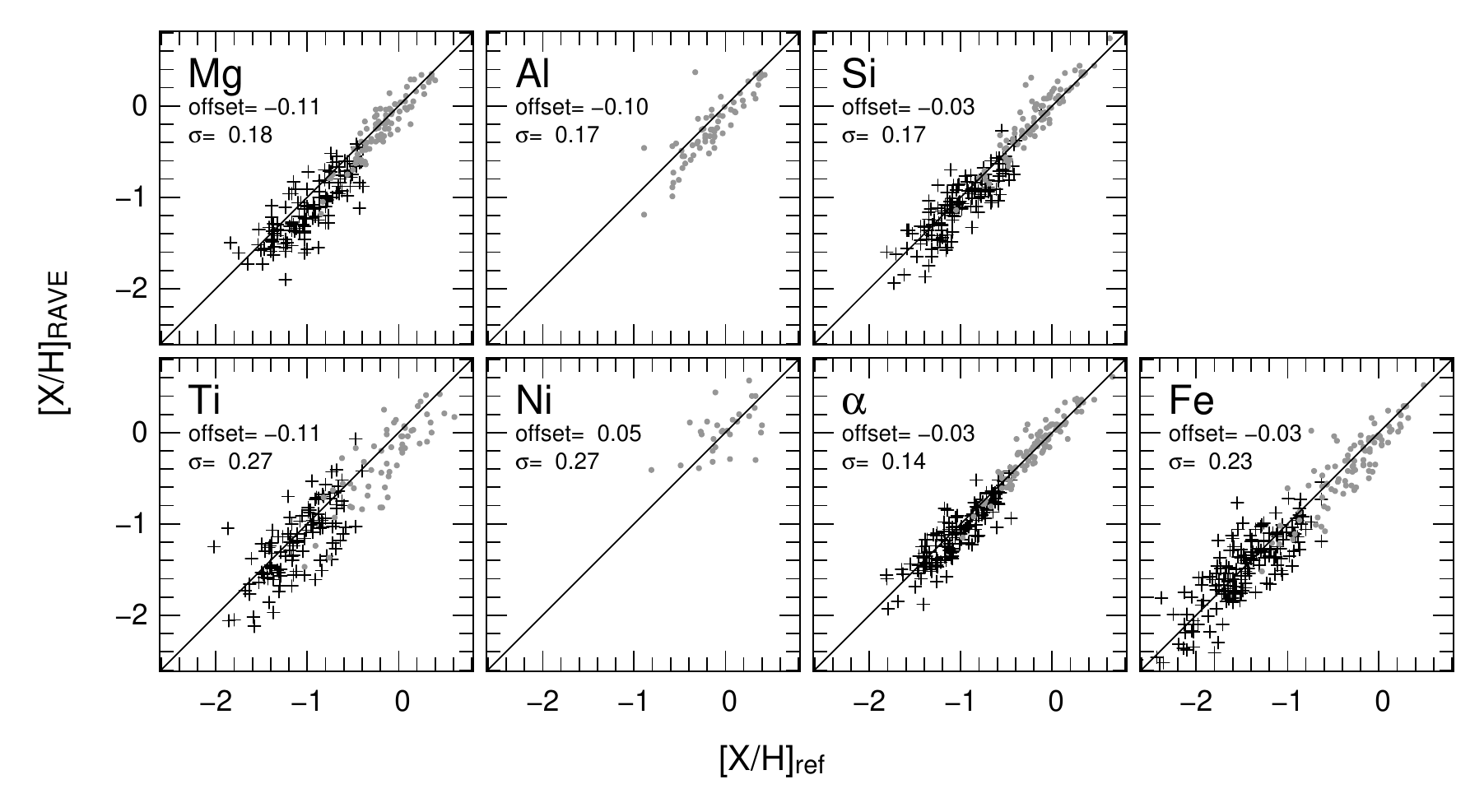}
\caption{Comparison between the reference high-resolution elemental abundances
($x$-axis) and RAVE elemental abundances ($y$-axis)
for the SG05 (98 dwarf stars, gray dots) and the R11 samples (233 spectra of 203
giant stars, black ``+") measured by adopting the stellar parameters
provided by the RAVE pipeline.}
\label{Ruchti_Soub_XH_rave}
\end{center}
\end{figure*}

\begin{figure}[tbh]
\begin{center}
\includegraphics[width=\columnwidth]{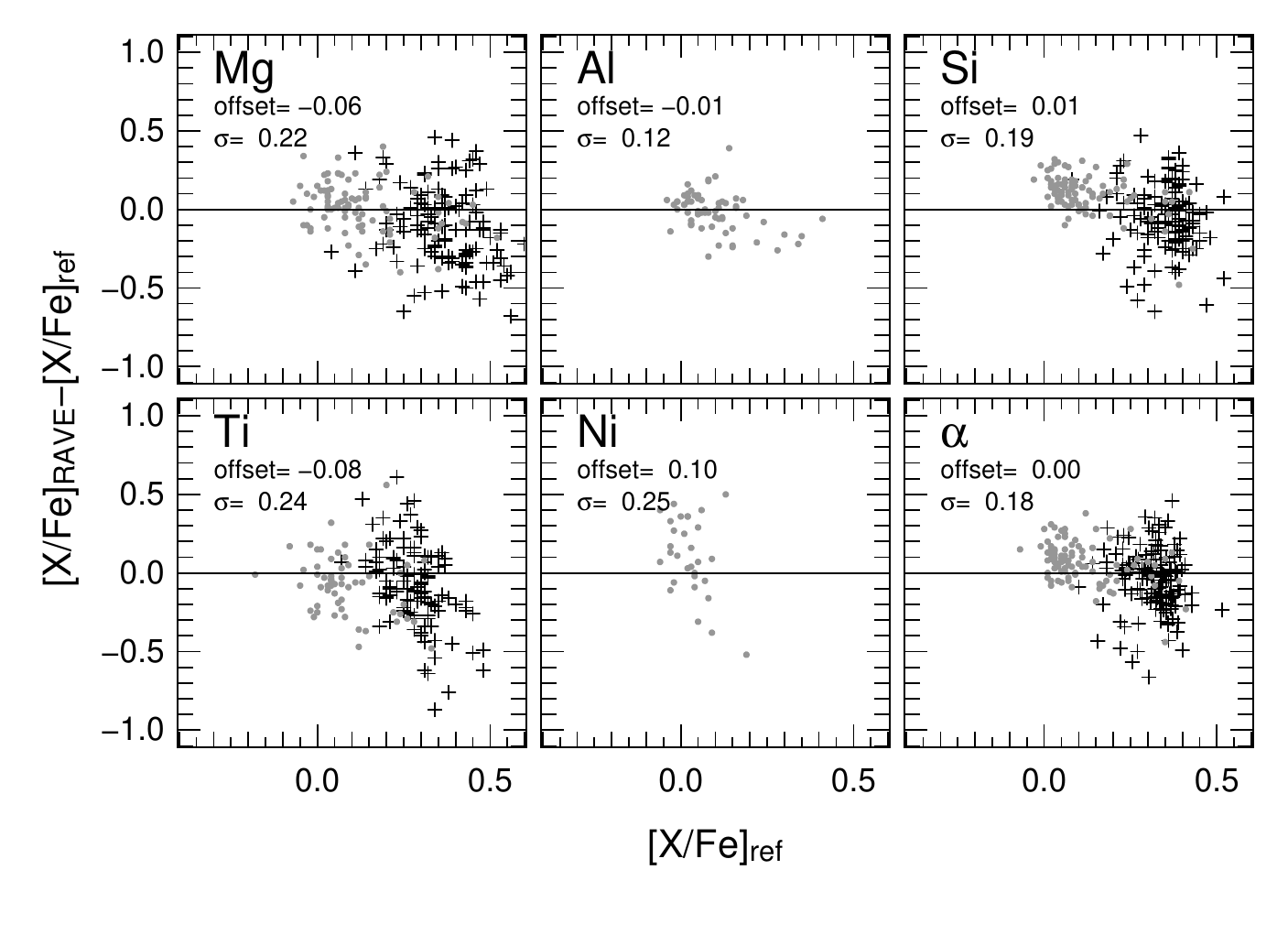}
\caption{Comparison between expected relative elemental abundance
($x$-axis) and residual abundances RAVE-minus-reference ($y$-axis).
Stellar parameters adopted and symbols are as in Fig.~\ref{Ruchti_Soub_XH_rave}.}
\label{Ruchti_Soub_XFe_rave}
\end{center}
\end{figure}

The overall uncertainties have been estimated by computing the elemental abundances  of 98 RAVE spectra of dwarf stars from \citet[hereafter SG05]{Soubiran05}  and 233 RAVE spectra of 203 giant stars from  \citet[hereafter R11]{Ruchti11}. 
Most of the SG05 stars have RAVE spectra with SNR$>100$~pixel$^{-1}$  whereas the R11 stars have RAVE spectra with 
SNR ranging between 30~pixel$^{-1}$ and 90~pixel$^{-1}$.  Hence the results are representative of the medium-high SNR regime.
Figure~\ref{Ruchti_Soub_XH_rave} and Fig.~\ref{Ruchti_Soub_XFe_rave} show the results obtained for the six elements in common with SG05 and the four  in common with R11.

Adopting the RAVE DR4 stellar parameters, the RAVE chemical pipeline delivers
slightly underestimated abundances for Mg, Al and Ti ($\sim-$0.1~dex).
There is a general improvement in precision for most of the elements with
respect to the \citetalias{Boeche11} pipeline (dispersions smaller than $\sim$0.05-0.07~dex for Mg,
Ti, Fe) with no visible systematic offsets.
The estimated errors in abundance depend on the element and
range from 0.17~dex for Mg, Al and Ti to 0.3~dex for Ti and Ni. The error for
Fe is estimated as 0.23~dex.
We  note that the errors reported here are conservative estimations
of the RAVE abundance errors, because we are comparing our results with
other more precise, but still uncertain measurements, and we have not
corrected the variance for the second contribution.
For illustration, assuming an uncertainty in the reference
abundances of ~0.1dex, our estimated RAVE errors decrease by 0.03-0.05~dex.

\section{Fourth public data release: Catalogue presentation}
\label{sect:dr4}
The fourth public data release of the RAVE data (RAVE DR4) includes the observations obtained from the 3rd of April 2004 to the 20th of December 2012. In total, \totalstarnumber\  stars have been observed, collecting \totalspectranumber\ spectra.  The catalogue is accessible online, and it contains also  radial velocities, proper motions,  photometric information, stellar morphological flags \citep[coming from][]{Matijevic12}, line-of-sight distances, ages and interstellar extinction for each star. In addition, the parameters obtained with the previous DR3 pipeline are also published, to assist readers of papers published based on those parameters, though we strongly recommend the use of the parameters  obtained with the latest DR4 pipeline in all future analyses. The DR4 catalogue  can be queried or retrieved from the Vizier database at the {\it Centre de Donn\'ees Astronomiques de Strasbourg} (CDS), as well as from the RAVE collaboration Web site (\url{www.rave-survey.org}).

\begin{figure}[tbp]
\begin{center}
\includegraphics[width=0.45\textwidth]{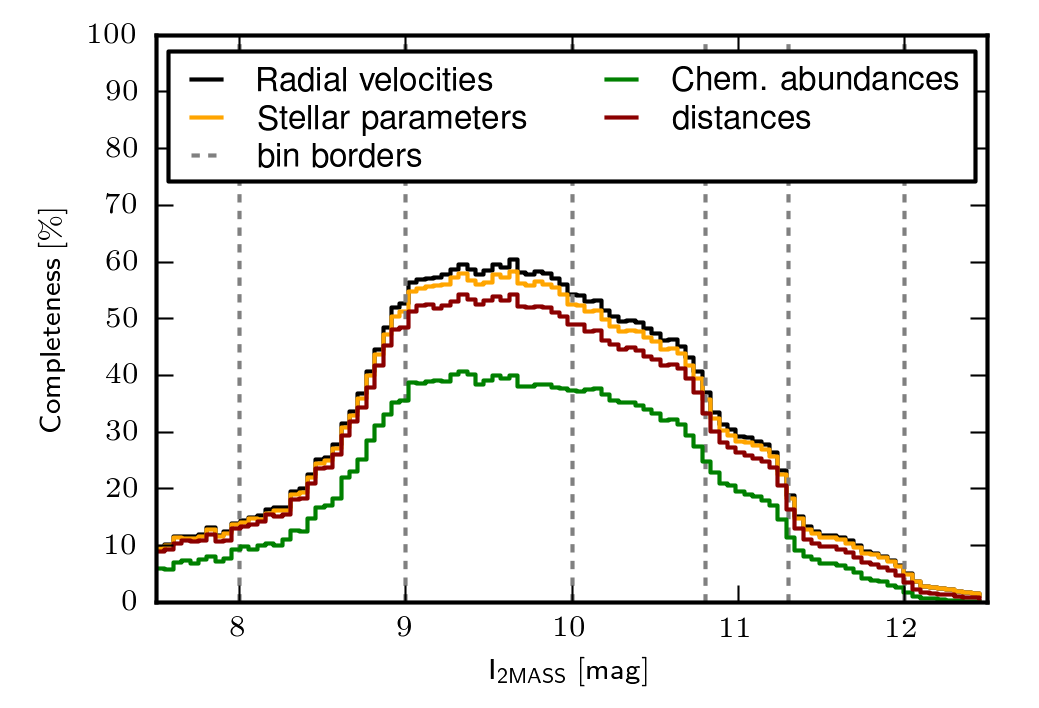}
\caption{Fractional completeness of the RAVE DR4 sample with respect to the $I_{\rm 2MASS}$ stars for the published radial velocities, stellar atmospheric parameters, chemical abundances and line-of-sight distances.}
\label{Fig:Completeness_hist}
\end{center}
\end{figure}

The completeness of the published catalogue for the radial velocities, atmospheric parameters, distances and chemical abundances, with respect to the $I_{\rm 2MASS}$ catalogue can be seen in Fig.~\ref{Fig:Completeness_hist}.
In addition,  Aitoff maps for the completeness of the catalogue at four different magnitude bins as a function of the stellar positions on the sky are shown in Fig.~\ref{Fig:Completeness_maps}.

Below we discuss which criteria to apply in order to obtain a high quality and reliable sample  of stars from the catalogue of the atmospheric parameters and the elemental abundances.  Brief discussions about proper motions, radial velocities, distances as well as the new APASS photometry are also included in what follows, but we refer the reader to \citet{Siebert11},  \citet{Zwitter10}, \citet{Binney13} and Munari et al. (2013a,b) for full details. \\

\subsection{Criteria for reliable sub-sample selection considering the atmospheric parameters}
\label{sect:rejection_criteria}
The following criteria need to be understood as the confidence limits for selection based on observational  (signal-to-noise ratio) and pipeline limitations (mainly the boundaries of the grid).

We selected all the stars which had a SNR$>20$~pixel$^{-1}$, had errors in the radial velocity estimation of  less than 8~\kms (measured by RAVE DR3, see Sect.~\ref{sect:Vrad}), had  derived  \logg$>$0.5~dex, determined \teff$>3800$~K and calibrated metallicity [M/H]$>-5$~dex  (measured by the DR4 pipeline), and for which the DR4 algorithm had converged towards a stable solution\footnote{MATISSE iterates up to 10 times until the result of the projection of the spectrum on the projection functions $B_\theta(\lambda)$ is within the parameter range defined by the $B_\theta(\lambda)$ \cite[see][ and Sect.~\ref{sect:Matisse}]{Matisse, Kordopatis11a}}. 
In total, roughly 19\% ($\sim 8.7 \times 10^4$) of the spectra have been rejected after these quality criteria. We are working towards the next  DR5 RAVE data release, with work which we hope will improve the parameters for at least some of these currently non-reliable stars.  

The cut on the error on the radial velocity ($\Delta$\vrad$<8$~\kms, 12~974 spectra) has been defined based on the results of \cite{Kordopatis11a}, where it has been shown that for Doppler shifts  larger than approximately half a pixel, the results of the pipeline were seriously degraded.
Nevertheless, a criterion based on the Tonry-Davis correlation coefficient ($R$) might be preferred in some cases, since some stars can have large errors but good $R$ (due, for example,  to strong hydrogen lines), and vice-versa (small errors but $R<5$).

The removal of the stars with gravities lower or equal to 0.5~dex (25~882 spectra), \teff\ lower than 3800~K (20~143 spectra) and/or calibrated \meta\ lower than $-5$~dex (1~282 spectra) has been decided because the results are considered both unrealistic (the synthetic spectra computed with the MARCS atmospheric models at such \logg\ have not been carefully compared to real spectra)   and less reliable (e.g.: missing models in the reference grid). 
Finally, the cut on the convergence of the DR4 algorithm (14~454 spectra) is made in order to minimise cases badly affected by  the spectral degeneracies. Indeed, these degeneracies can cause,   in some cases, an  impossibility for the algorithm to converge  due to a negative gradient in the distance function between the spectrum and the templates.  
MATISSE can in some cases oscillate between two solutions ($\sim 11\%$ of the published sample). We decided to keep these solutions because, in general, they are close in the parameter space. Nevertheless, in case the user decides not to use them, we have flagged these stars in the {\it algo\_conv} parameter which is also published with this data release (see the appendix).

An additional cut, based on the velocity width parameter of the spectral lines, $V_{rot}$,  
 has been applied, since our algorithm cannot treat fast rotators.  We discarded empirically  stars at the high velocity tail of the distribution ($V_{\rm rot} > 100$~\kms, 11~735 in total). We recall that the estimation of the $V_{rot}$ is made through the DR3 pipeline, as a free parameter, at the same moment as the first estimation of \teff, \logg~ and \meta~ is made. Nevertheless, the rather low resolving power of RAVE spectra ($\sim 1.2$~\AA\ or 30~\kms) does not allow the determination of rotational velocities for slow rotators which represent the vast majority of RAVE stars. Hence this parameter is not published, but true fast rotators will be discussed in a separate paper.

Finally, we note that targets at low Galactic latitudes should also be treated with caution, since the possibly high interstellar extinctions in these directions are not taken into account in the photometric constraints imposed by the DR4 pipeline.

\subsection{Criteria for reliable sub-sample selection considering the chemical abundances}
\label{sec_catalog}
From the whole RAVE internal data set, we measured chemical abundances 
only for spectra with the following features:
\begin{itemize}
\item Effective temperature $4000\le$\temp(K)$\le7000$~K
\item Signal-to-noise SNR$>$20~pixel$^{-1}$
\item Rotational velocity V$_{rot}<$50~km~s$^{-1}$.
\end{itemize}
Such limitations are due to the following facts. First, the EW library (and the \citetalias{Boeche11} line list on which it is based) is reliable only in
this effective temperature range. In addition, the  line measurement and stellar parameters are reliable only for signal-to-noise larger than 20~pixel$^{-1}$. Finally,  the absorption lines can be reliably measured only if their FWHM does not significantly exceed the RAVE instrumental FWHM ($\sim1.2$\AA) which corresponds to a rotational velocity of 30~km~s$^{-1}$.
Such criteria leave 313,874 spectra selected from the RAVE database.

Besides the chemical abundances of this selected sample, we provide some extra 
statistical quantities and flags to be employed for further quality selection:
\begin{enumerate}
\item {\em $\chi^2$ between best matching model and observed spectrum}: the
lower the values the better the expected abundance precision. We suggest a user
reject spectra with $\chi^2>2000$.

\item {\em the value $frac$} which represents
the fraction of the observed spectrum which satisfactorily matches the model. We
suggest a user reject spectra with $frac<0.7$ (see \citetalias{Boeche11} for further details).


\item {\em classification flags by \cite{Matijevic12}}: we
suggest a user selects spectra classified as ``normal" by Matijevic et al. in
order to avoid peculiar objects on which the chemical pipeline fails.

\item {\em $Algo\_Conv$ value}: this value indicates if the DR4 pipeline has converged or if the stellar parameters were either outside the grid boundaries or MATISSE was oscillating between two values.
The higher quality data have  $Algo\_Conv=$0.
\end{enumerate}

\begin{figure*}[t]
\begin{center}
\includegraphics[width=12cm]{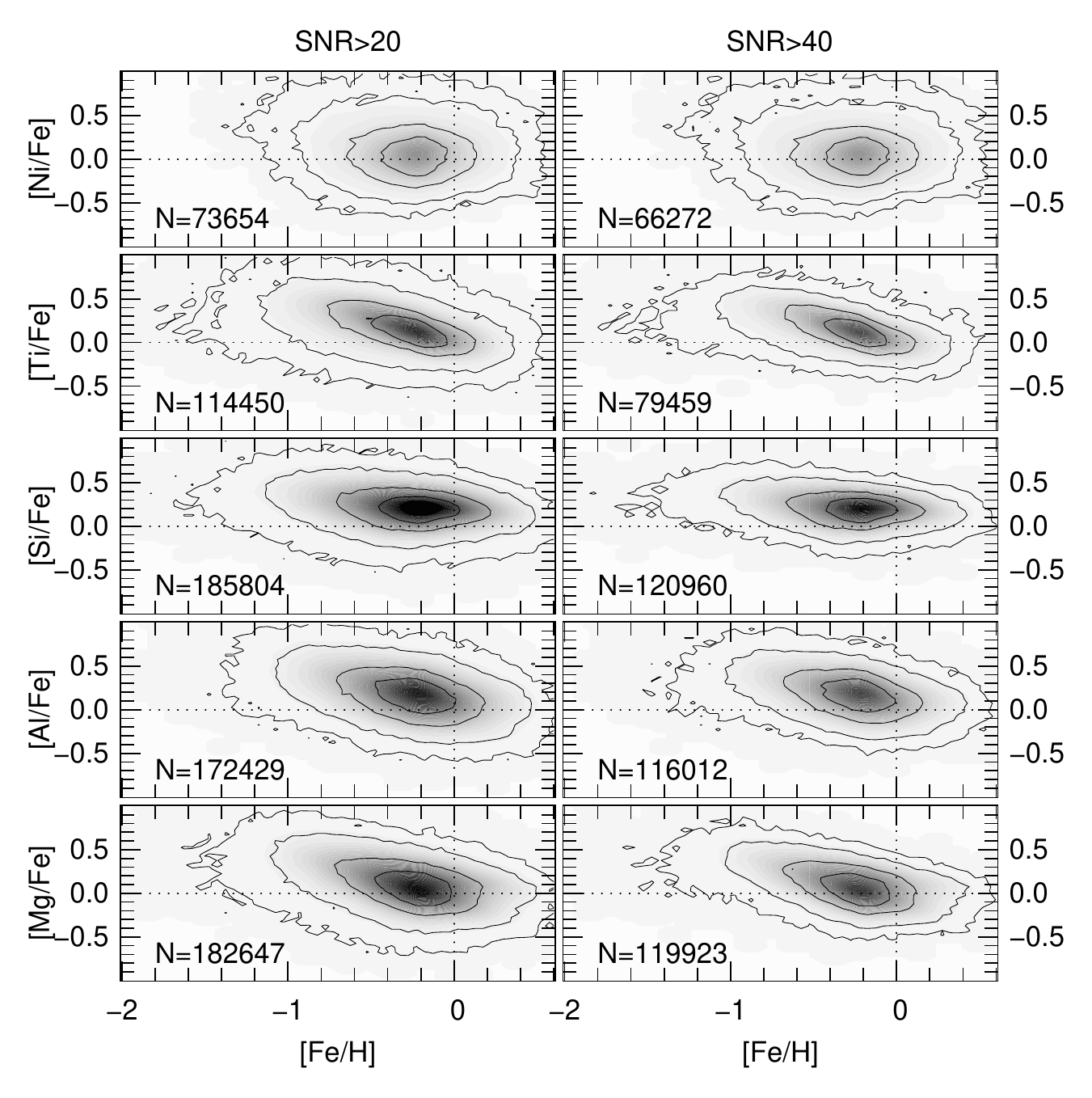}
\caption{Distribution on the chemical plane of spectra after the
application of the quality indicators reported in
Sec.~\ref{sec_catalog}. The isocontours hold 34.0\%, 68.0\%, 95.0\% and
99.5\% of the sample.}
\label{XFe_DR4}
\end{center}
\end{figure*}

The application of these quality flags is left to the user.
The number of spectra which meet all such quality flags is 187,305.
In Fig.~\ref{XFe_DR4} we show the distribution of the chemical abundances, given the above mentioned criteria, for  SNR$>$20~pixel$^{-1}$ and SNR$>$40~pixel$^{-1}$.

\subsection{Results and comparisons with DR3}
\label{sect:results_checks}
A description of Galactic properties based on the published parameters of this catalogue are beyond the scope of this paper. Nevertheless, as a sanity check, we explore in this section the general properties of the catalogue, by analysing the correlation of the parameters and the change of the metallicity properties according to the SNR, the effective temperature  or the surface gravity. 
By comparing the behaviours of the DR3 and the DR4 pipelines, we show that although the differences between the atmospheric parameters of the two methods are relatively subtle, DR4 better reproduces the expected behaviour for different subpopulations of stars and thus is the method of choice for most Galaxy evolution studies.

\begin{figure}[tbph]
\begin{center}
\includegraphics[width=0.4\textwidth,angle=0]{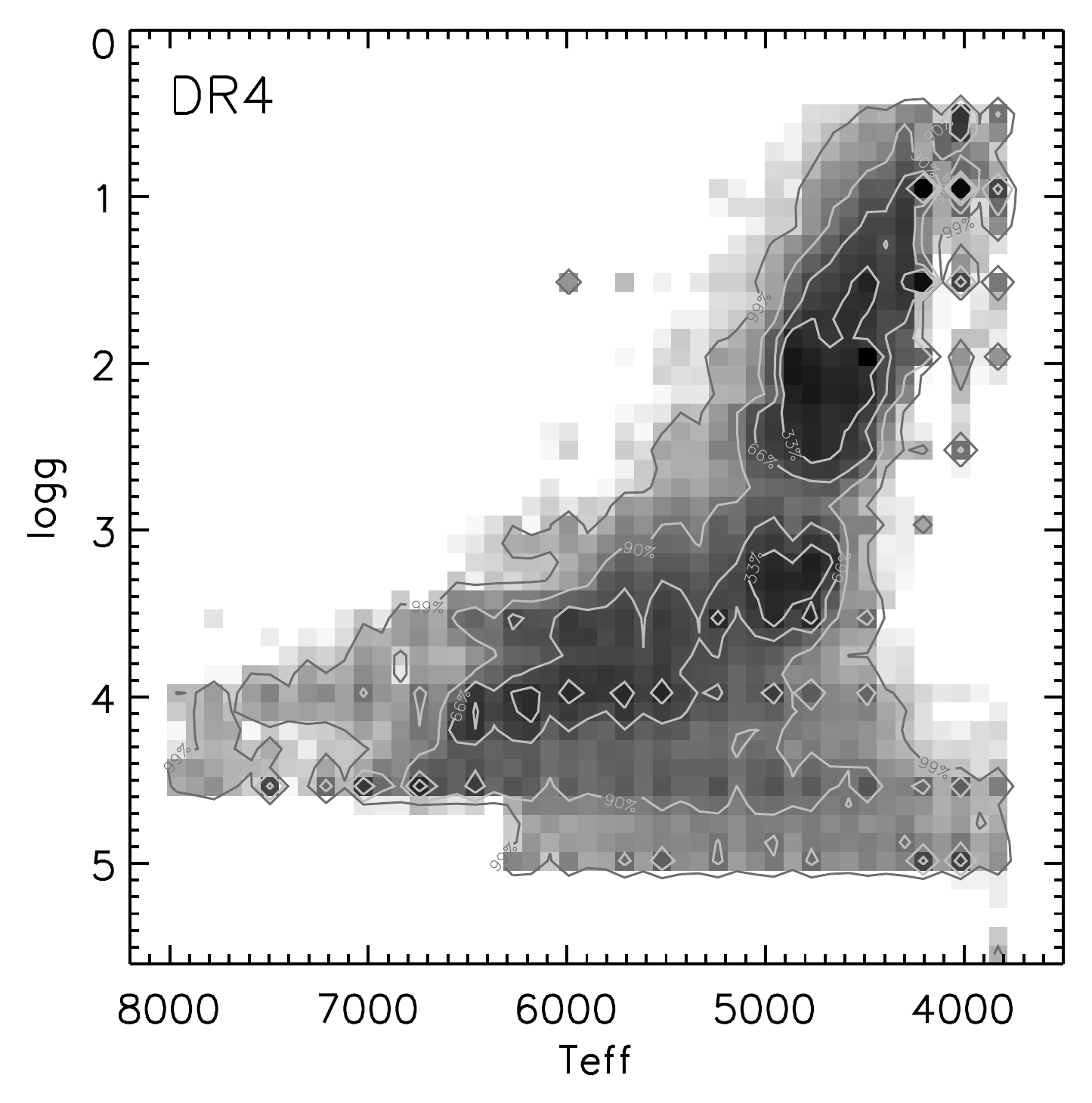}\\
\includegraphics[width=0.4\textwidth,angle=0]{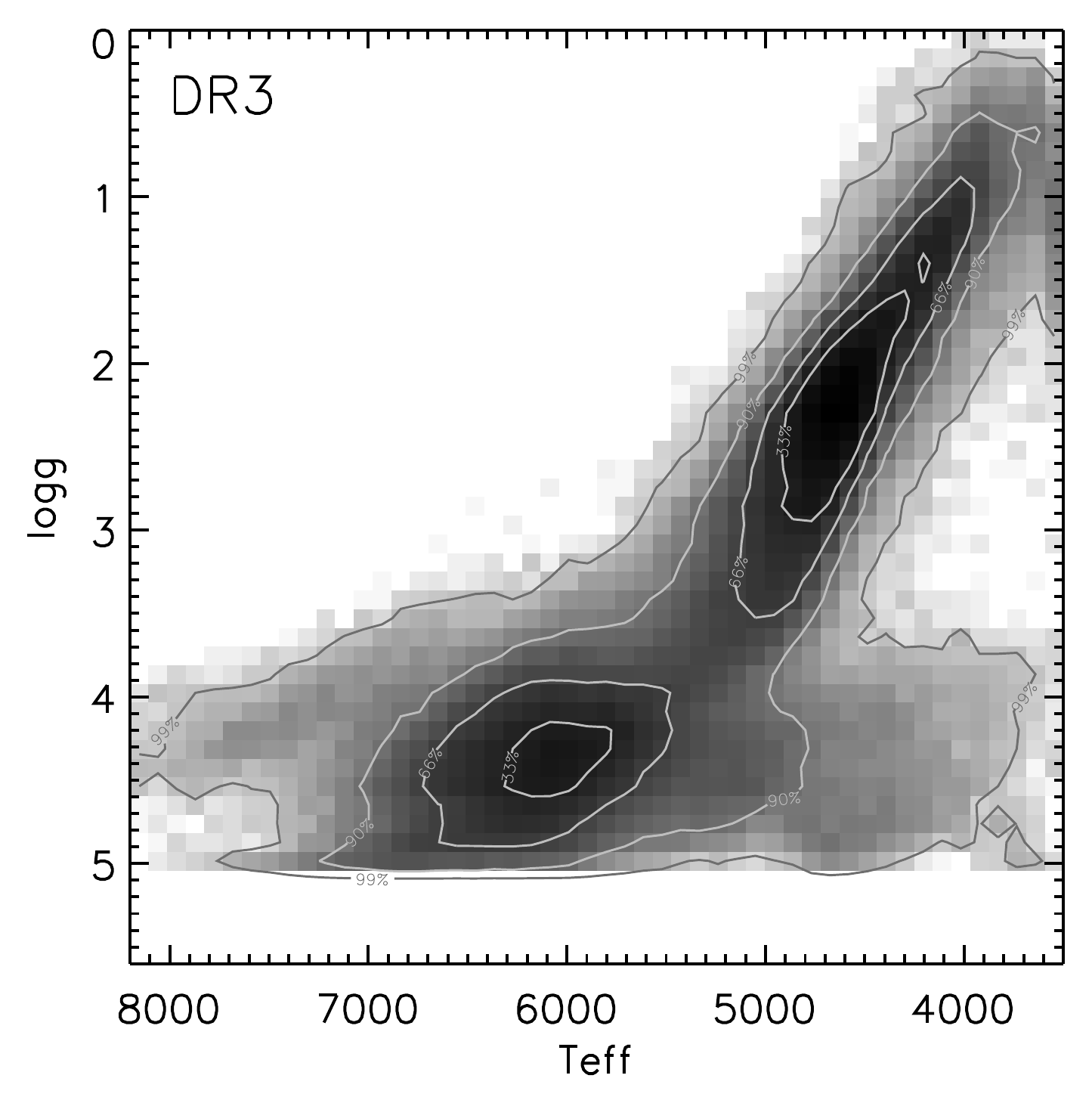}
\caption{\teff-\logg\ diagram for the selected RAVE stars (top: DR4 pipeline, bottom: DR3 pipeline), meeting our criteria defined in Sect.~\ref{sect:rejection_criteria}. In total the parameters of  $\sim 4 \times 10^5$~spectra are represented in these diagrams. The contour lines contain 33\%, 66\%, 90\% and 99\% of the total considered sample.}
\label{Fig:HR_Kordo_RAVEDB}
\end{center}
\end{figure}

Figure~\ref{Fig:HR_Kordo_RAVEDB} compares the resulting \teff--\logg\ diagrams of the DR4 and the DR3 pipelines as selected according to the criteria of Sect.~\ref{sect:rejection_criteria}. 
One can notice that besides the well understood and described discretisation due to the DEGAS algorithm, there are some additional subtle differences in the parameters of the two pipelines. In particular, hot dwarfs, as well as turn-off stars have now smaller surface gravities and the main-sequence is better defined. Finally, giants have slightly higher effective temperatures.

\begin{figure*}[tbp]
\begin{center}
$\begin{array}{cc}
\includegraphics[width=0.35\textwidth,angle=0]{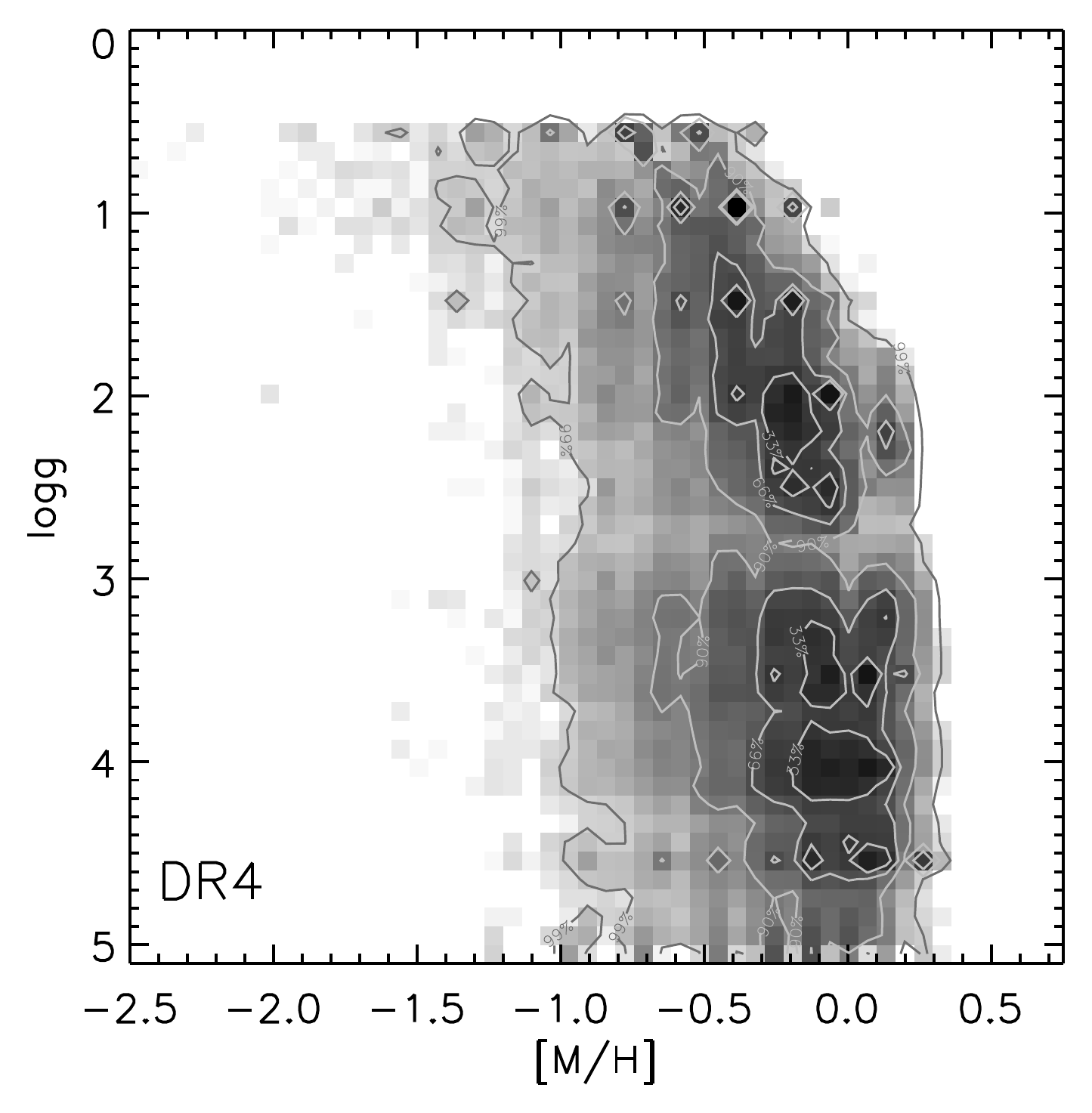} &
\includegraphics[width=0.35\textwidth,angle=0]{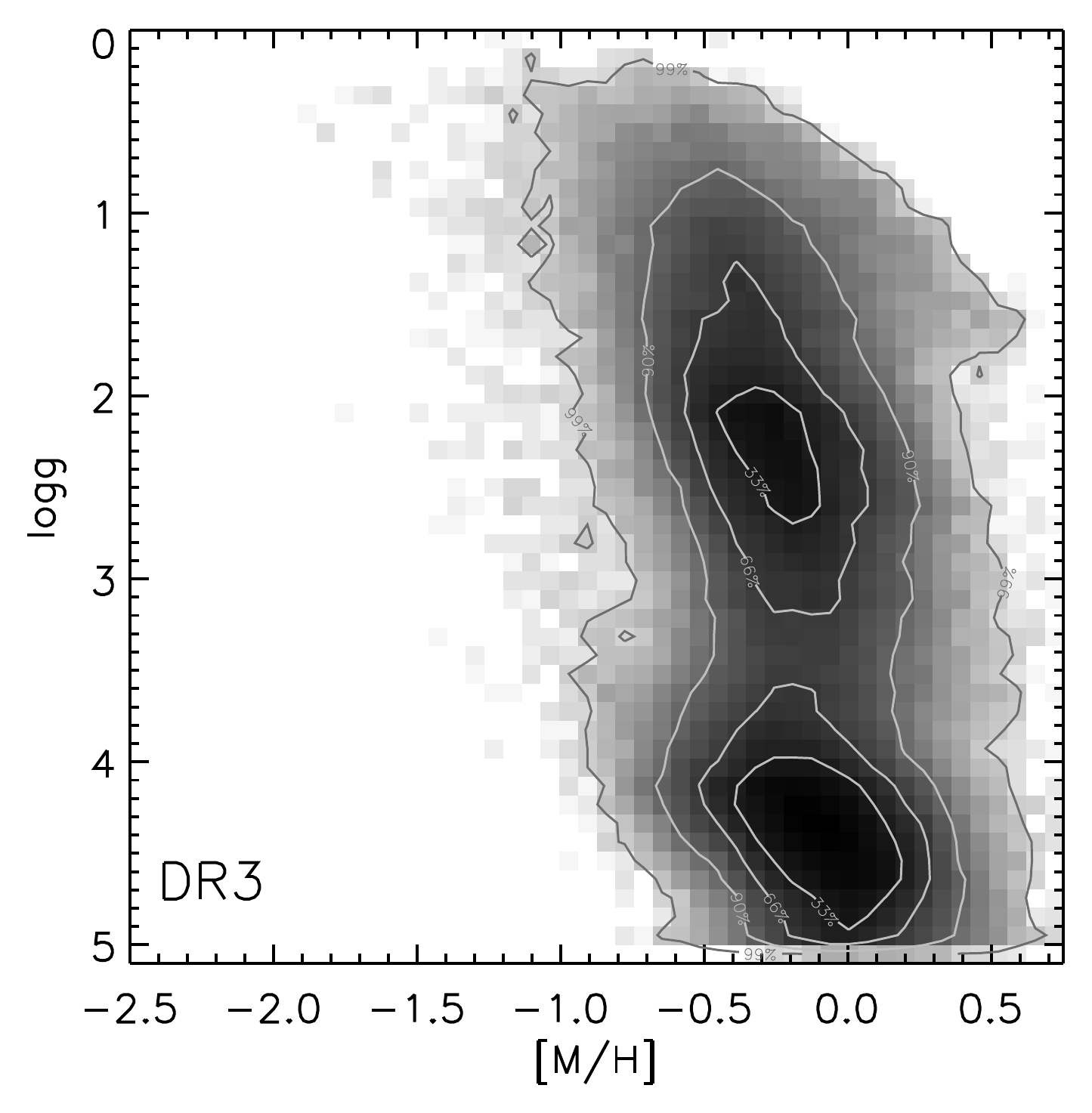} \\
\includegraphics[width=0.35\textwidth,angle=0]{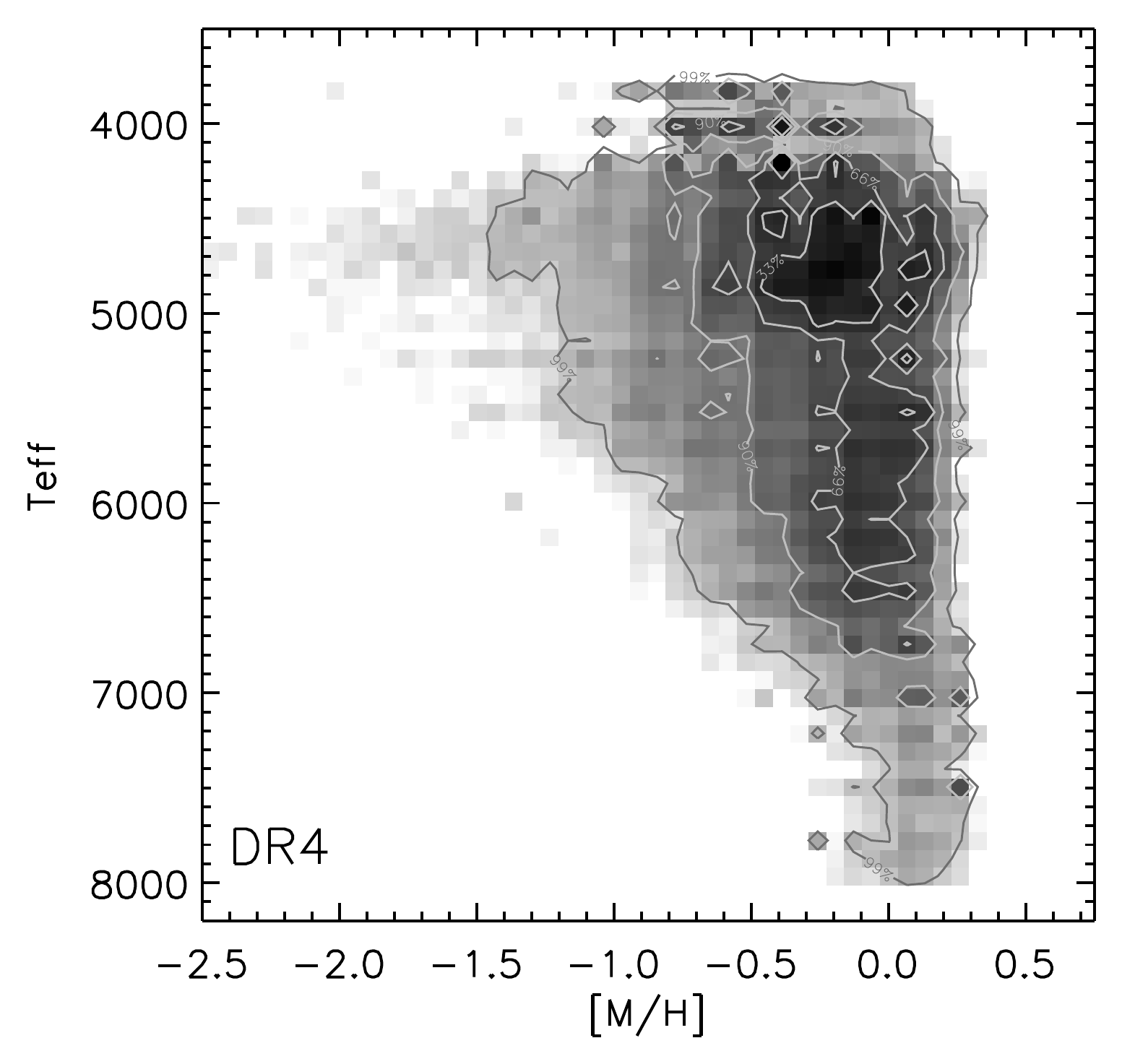} &
\includegraphics[width=0.35\textwidth,angle=0]{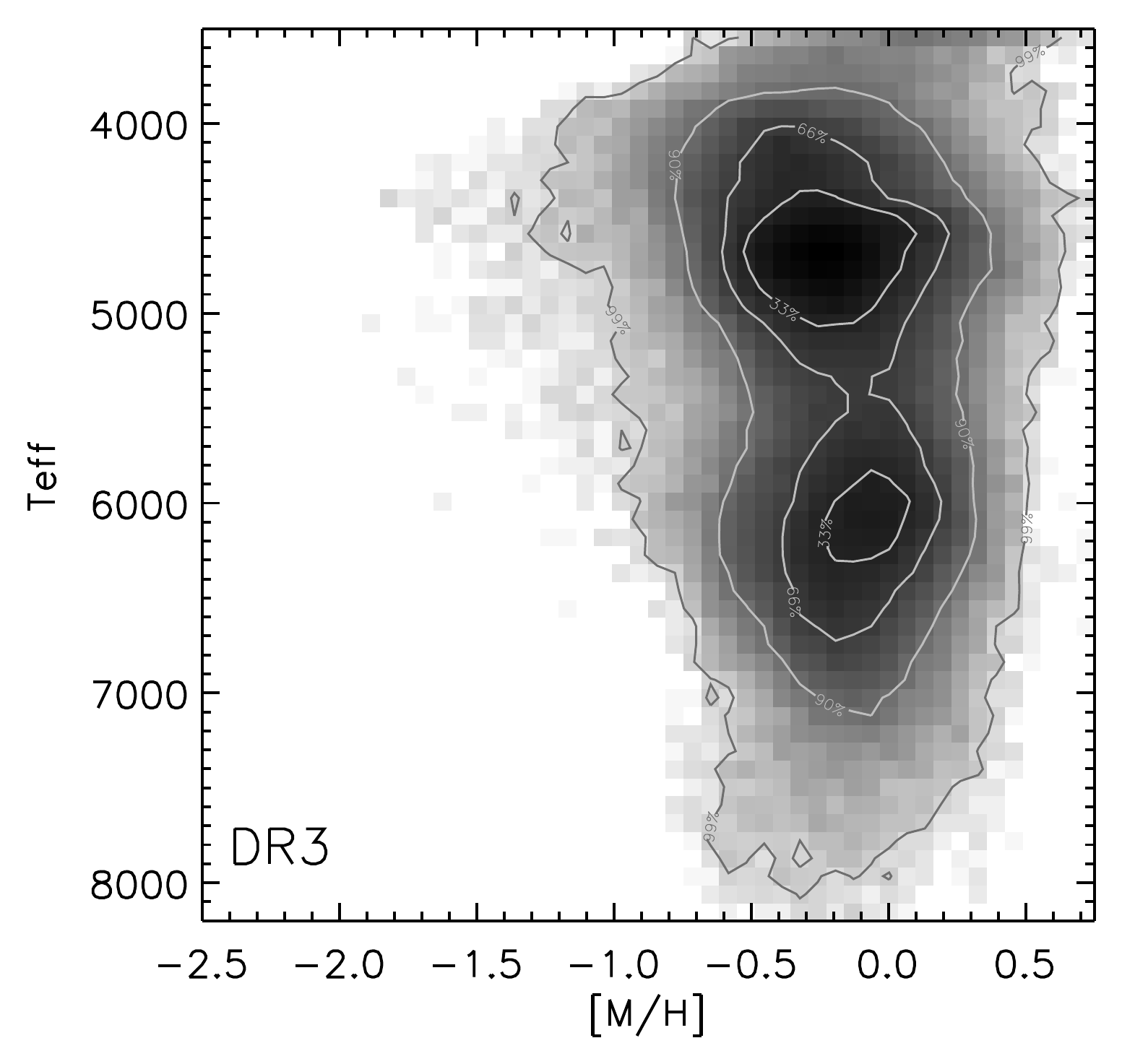} 
\end{array}$
\caption{Surface gravity (\logg, upper plots) and effective temperature (\teff, lower plots) versus metallicity for all the selected RAVE stars, defined in Sect.~\ref{sect:rejection_criteria}. The contour lines contain 33\%, 66\%, 90\% and 99\% of the total considered sample. The results for the  DR4 pipeline are on the left hand side, whereas the results from the DR3 pipeline are on the right hand side. For both pipelines the mean metallicity slightly decreases   for the lowest gravities, which is a signature of a different mixture in the probed Galactic populations. In addition, one can notice the \logg~ and the \teff~ trends which are present for the dwarfs analysed with the DR3 pipeline. }
\label{Fig:TeffLogg_vs_meta}
\end{center}
\end{figure*}

The DR4 and DR3\footnote{Calibrated according to Eq.~2 of \cite{Siebert11}, with $c_0 = 0.578, c_1 = 1.095, c_3 = 1.246, c_4 = -0.520$.} calibrated metallicity trends, as a function of the surface gravity and the effective temperatures can be seen in  Fig.~\ref{Fig:TeffLogg_vs_meta}. 
As far as the \teff\ dependencies are concerned, one can notice that the metallicity distribution functions (MDFs) of the DR4 pipeline get broader when the effective temperature lowers. In particular, the DR4 pipeline finds that the hottest stars have a narrow metallicity distribution with a mean value at slightly super-solar values, as expected for the young stars in the solar neighbourhood. 
This is not the case for the results of the DR3 pipeline, where metallicities as low as \meta$\sim -0.5$~dex are obtained. 
Furthermore, from the iso-contours of the \logg\ vs \meta, we can see that despite the mild pixelisation of the values, there are no trends of the metallicities as a function of the surface gravity for the dwarfs, as derived by the DR4 pipeline.
This is not the case for the DR3 pipeline results, where a shift is noticed.

\begin{figure*}[tbp]
\begin{center}
\includegraphics[width=0.9\textwidth,angle=0]{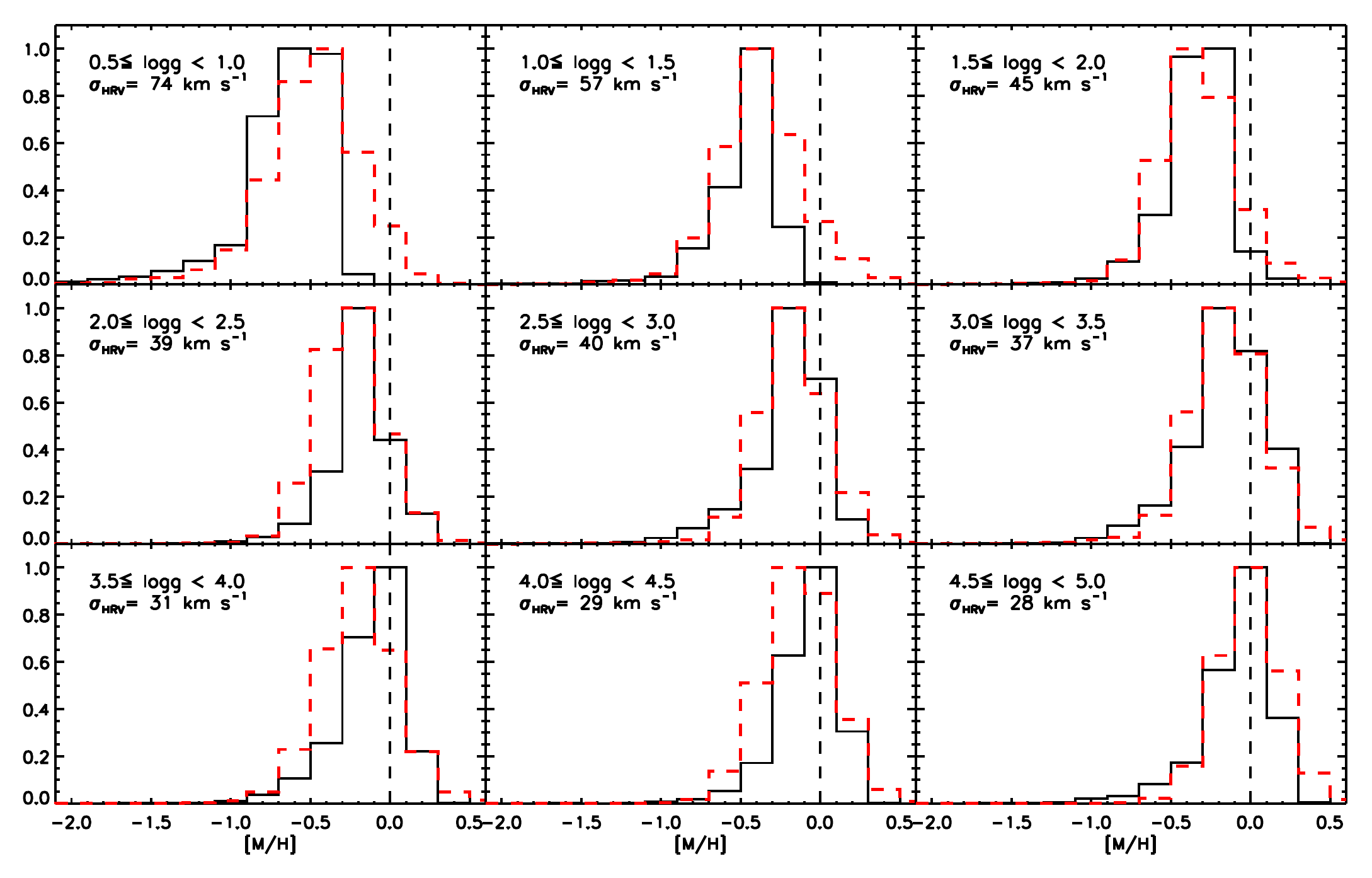}
\caption{Calibrated metallicity distribution functions for different surface gravity (\logg) bins.  The peak of the histograms have been normalised to unity. The mean radial velocity dispersion for the selected stars per surface gravity bin ($\sigma_{\rm HRV}$) are noted in the upper part of each box. RAVE DR4 results are plotted in black solid lines, RAVE DR3 ones in red dotted lines.}
\label{Fig:MDF_logg_bins}
\end{center}
\end{figure*}

In order to investigate whether  this shift is real and justified given our classical view of the Milky Way, we explored the stellar heliocentric radial velocities and the evolution of the MDFs  for different  surface gravity bins. Figure~\ref{Fig:MDF_logg_bins} shows the resulting histograms for the calibrated metallicities of the DR4  (in black solid lines) and the DR3 pipelines (red dashed lines).
The radial velocity dispersions of the selected stars have also  been reported inside each box. For the lower panels, corresponding to the dwarf stars (3.5$<$\logg$<$5~dex), the radial velocity  dispersion stays constant. Considering that each Galactic population (thin disc, thick disc and  halo) is characterised by a different velocity dispersion, the constant $\sigma_{V_{\rm HRV}}$ that is found indicates that the same proportions of  Galactic populations are probed for these gravity bins. As a result, the MDFs should not vary inside these bins. This is the case only for the DR4 MDFs, the DR3 ones shifting by 0.2~dex in this range of \logg. 
As far as the sub-giant and giant stars are concerned, a good agreement is found between the DR3 and DR4 MDFs, with a shift towards lower metallicities with decreasing \logg\ and at the same time an increase in the radial velocity dispersion. This  is in agreement with a change in the mixture of the probed Galactic populations as a function of the probed volume, passing from an old thin disc dominated population to the presence of more halo stars for the larger volume probed by the more luminous  giant stars.

\begin{figure*}[tbp]
\begin{center}
\includegraphics[width=0.9\textwidth,angle=0]{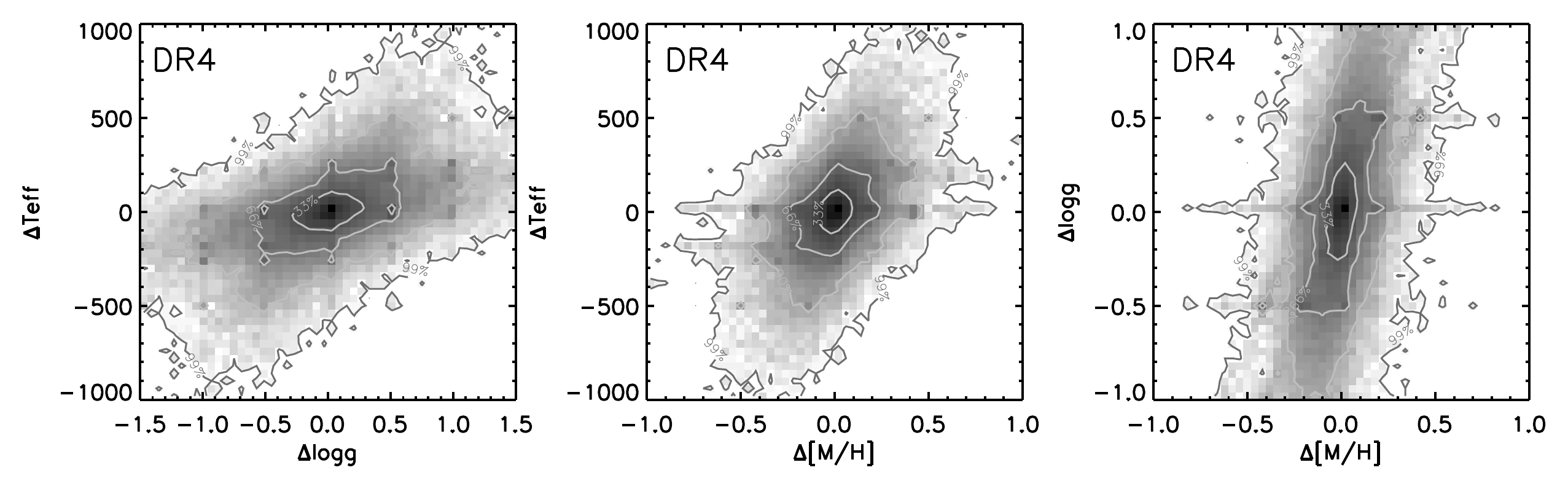}\\
\includegraphics[width=0.9\textwidth,angle=0]{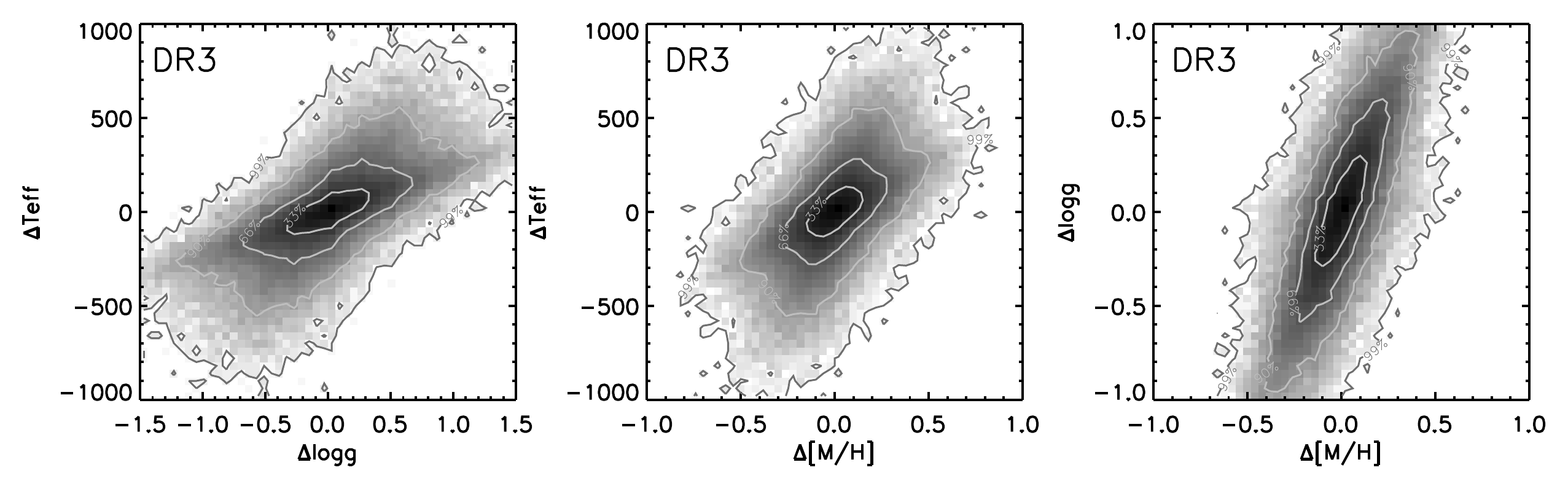}
\caption{Correlations between the derived atmospheric parameters (top: DR4 pipeline, bottom: DR3 pipeline) for the stars that have been observed several times by RAVE. The iso-contour levels  contain 33\%, 66\%, 90\% and 99\% of the total considered sample. See the text for an explanation.}
\label{Fig:Repeat_correl}
\end{center}
\end{figure*}

To show the correlations between the parameters, we select among the DR4 catalogue those stars that are observed multiple times, and for which several independent spectra and derived parameter sets are available. In the panels we plot the differences between the several determinations of the measured \teff, \logg~ and calibrated metallicities. 
Figure~\ref{Fig:Repeat_correl} shows the results for the stars with SNR$>20$~pixel$^{-1}$, for both  DR4 and  DR3 pipelines.
From that figure one can see 
that the new DR4 pipeline is more robust than the DR3 one, since the bulk of the stellar parameters show a very small discrepancy between the repeated observations, as well as   a negligible parameter correlation. This validates once more the robustness  of our calibration relation of Eq.~\ref{eq:bias_correction}.  
However, we note that correlations between the parameter estimations still exist for some stars. This is expected, due to the intrinsic spectral degeneracy: an underestimation of the \teff~leads to a similar underestimation of the \logg~ and the \meta.

\begin{figure*}[tbp]
\begin{center}
$\begin{array}{cc}
\includegraphics[width=0.35\textwidth,angle=0]{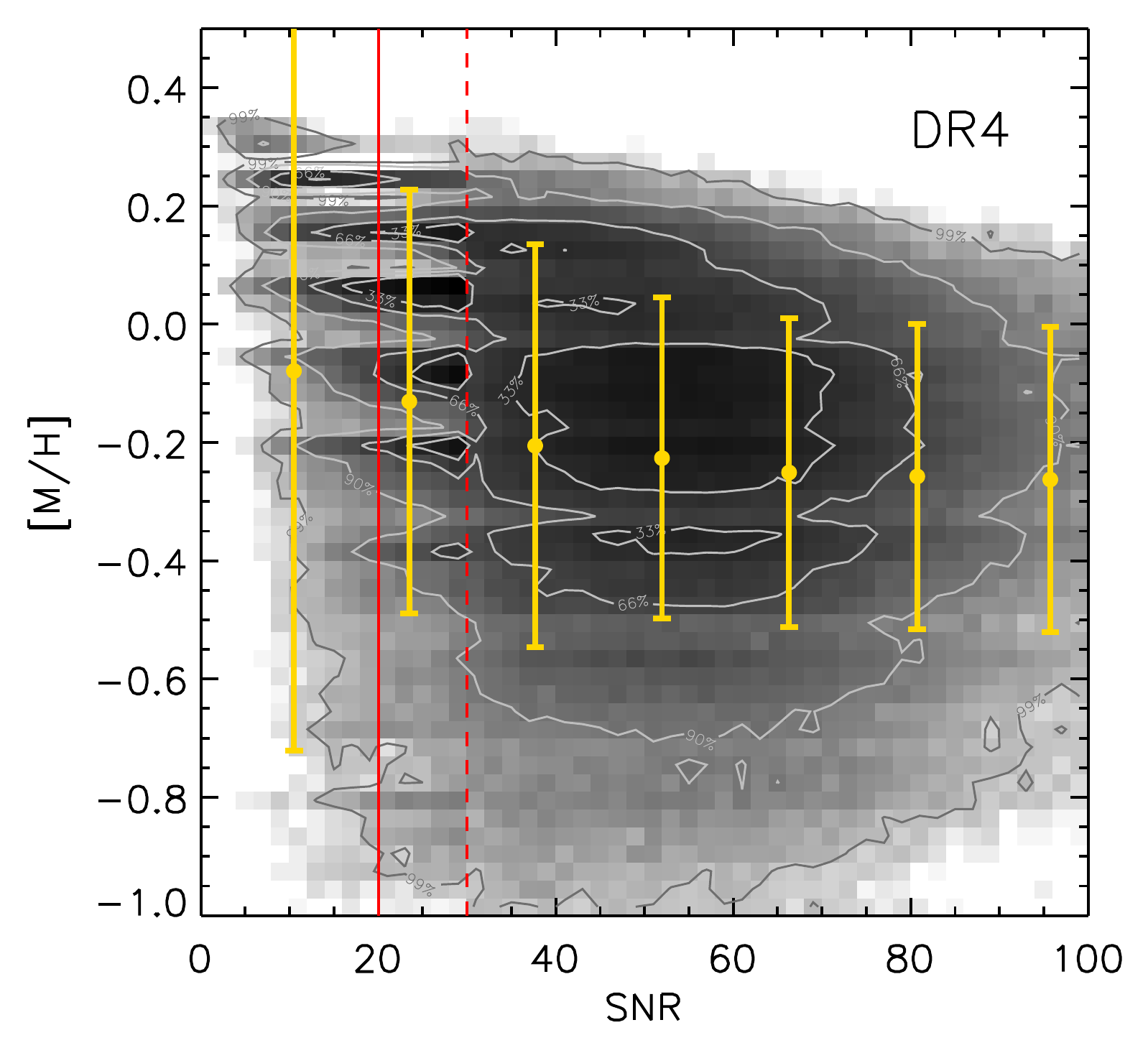} &
\includegraphics[width=0.35\textwidth,angle=0]{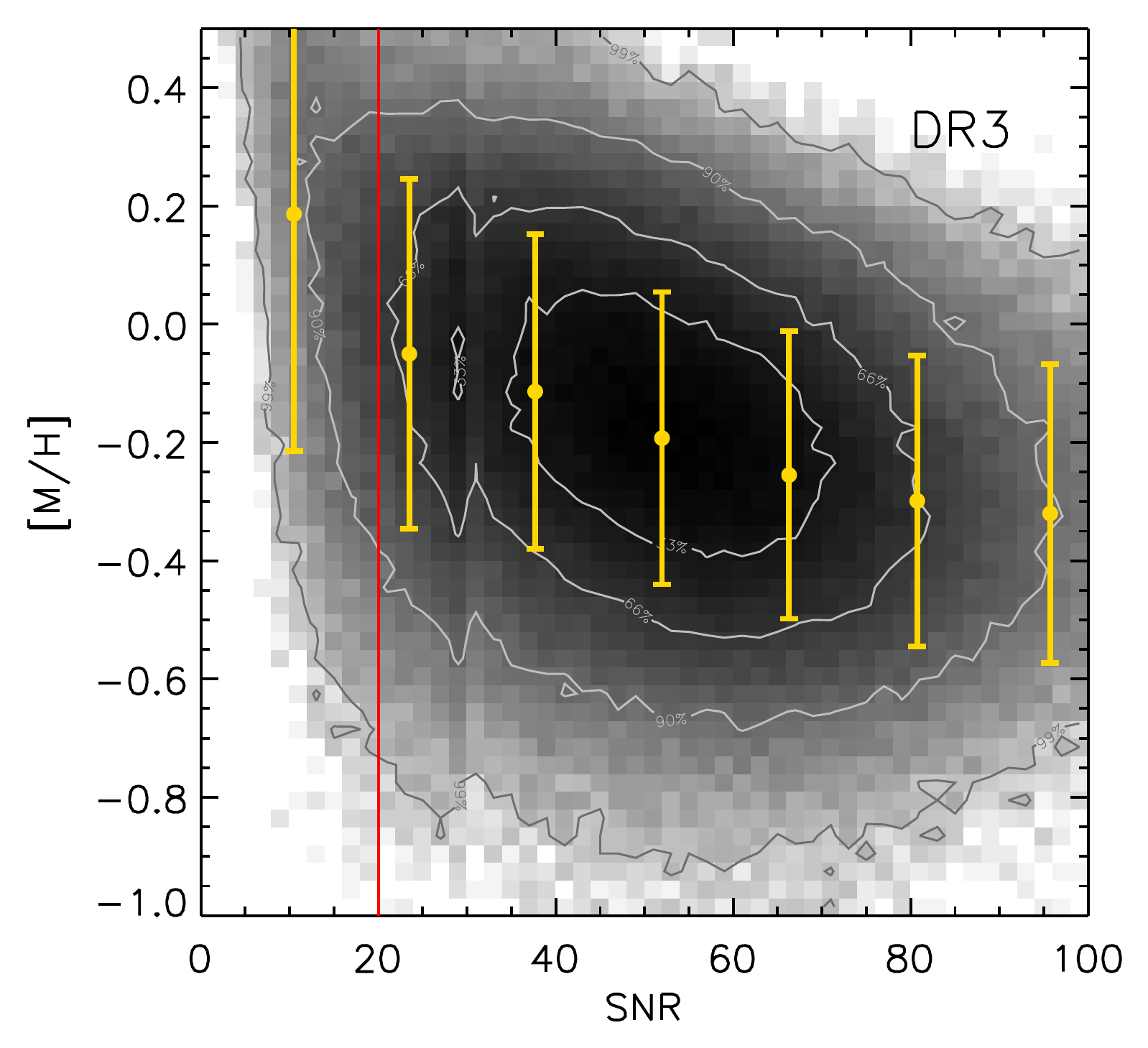}
\end{array}$
\caption{Calibrated \meta\ versus SNR for the selected RAVE stars with the DR4 pipeline (left hand side) and the DR3 one (right hand side). One can notice the pixelisation at low SNR (SNR$<30$~pixel$^{-1}$, red dashed line), due to the DEGAS algorithm.  The yellow points are calculated as the mean measured metallicity in different SNR bins, and the error bars represent the metallicity dispersion inside each bin. One can notice that the DR4 results (left hand side plot) are more stable to SNR changes, compared to the results  of the DR3 (right hand side plot). A red solid line is plotted at the SNR$=20$~pixel$^{-1}$ value, below which the parameters are not considered reliable.  }
\label{Fig:SNR_vs_meta}
\end{center}
\end{figure*}

The robustness of the DR4 pipeline in terms of better treatment of the spectral degeneracies can also be seen on Fig.~\ref{Fig:SNR_vs_meta}. The evolution of the mean metallicity as a function of the SNR (yellow points on Fig.~\ref{Fig:SNR_vs_meta})  shows no trends  down to  SNR$>15$~pixel$^{-1}$. Nevertheless, the cost for this better treatment is the pixelation of the results for the most metal-poor stars or the ones having low SNR. 
In particular, the pixelisation effect can be clearly seen in Fig.~\ref{Fig:SNR_vs_meta} below the threshold of SNR$=30$~pixel$^{-1}$. 
Indeed,  we recall that  for all the observed spectra DEGAS is first used to find the nominal template spectrum of the learning grid in order to re-normalise the observed one. Then, on one hand,  for the high SNR regime, the MATISSE algorithm is used on these optimally re-normalised spectra in order to obtain the final parameters. On the other hand, for the low SNR data ($<30$~pixel$^{-1}$) and at the boundaries of the learning grid, tests on synthetic spectra have shown that the projection method was giving less accurate results than the decision-tree. Given these facts,  \cite{Kordopatis11a} showed that DEGAS is preferred over the projection approach of MATISSE for SNR$<30$~pixel$^{-1}$, even if a pixelisation of the parameters is introduced.

\subsubsection{Chemical abundance reliability: element by element}
\label{sect:chemical_reliability}
\begin{figure}[t]
\begin{center}
\includegraphics[width=7cm]{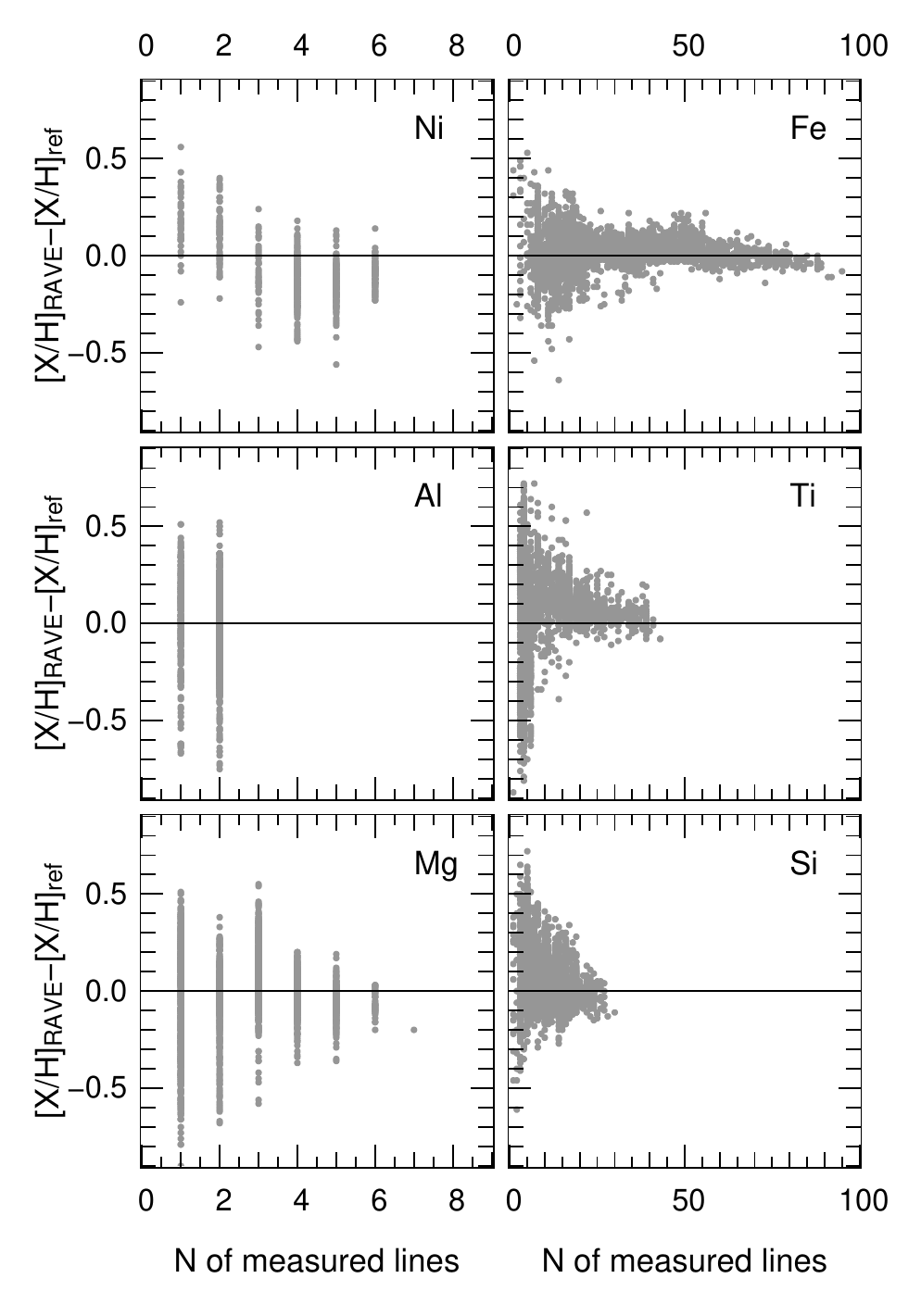}
\caption{Residuals between measured and expected abundances (y-axis) as
function of the number of measured lines (x-axis) for each element in the
test with synthetic spectra at SNR=100, 40, 20~pixel$^{-1}$.}
\label{Nlines_err}
\end{center}
\end{figure}

Besides the quality indicators described in Sect.~\ref{sec_catalog}, the number of measured absorption
lines for an element can be a good indicator of the reliability and
precision of the abundance estimation (as illustrated in
Fig.~\ref{Nlines_err}). We outline here a summary of the expected precision
of the abundance element by element.\\

{\em Magnesium} yields reliable results on synthetic and real
spectra with no significant correlation with stellar parameters. We expect
errors $\sigma_{\mbox{Mg}}\le0.15$~dex for SNR$>$40~pixel$^{-1}$ and
$\sigma_{\mbox{Mg}}\sim0.25$~dex for 20$<$SNR$<$40~pixel$^{-1}$.
{  Tests on synthetic spectra show that Magnesium suffers of a systematic
error when measured at low SNR. Such an error has not been confirmed with
real spectra}.\\

{\em Aluminium} gives a reliable abundance although obtained with only
two isolated lines. Abundance errors expected are
$\sigma_{\mbox{Al}}\le0.15$~dex for SNR$>$40~pixel$^{-1}$ and
$\sigma_{\mbox{Al}}\sim0.25$~dex for 20$<$SNR$<$40~pixel$^{-1}$.\\

{\em Silicon} is among the most reliably determined elements. 
Abundance errors expected are                           
$\sigma_{\mbox{Si}}\le0.15$~dex for SNR$>$40~pixel$^{-1}$ and                            
$\sigma_{\mbox{Si}}\sim0.25$~dex for 20$<$SNR$<$40~pixel$^{-1}$ with a small
overestimation of $\sim$0.1~dex.\\

{\em Titanium} gives reliable estimates at high SNR for cool giants
(\temp$<$5500K and \logg$<$3). We suggest  rejecting Ti abundances for
dwarf stars. Tests on synthetic and real spectra suggest an expected
error of $\sigma_{\mbox{Ti}}\le0.2$~dex for SNR$>$40~pixel$^{-1}$ and
$\sigma_{\mbox{Ti}}\sim0.3$~dex for 20$<$SNR$<$40~pixel$^{-1}$.\\

{\em Iron} gives robust and precise abundances thanks to its large number of
measurable lines at all stellar parameter values. Expected errors are 
$\sigma_{\mbox{Fe}}\le0.1$~dex for SNR$>$40~pixel$^{-1}$ and
$\sigma_{\mbox{Fe}}\sim0.2$~dex for 20$<$SNR$<$40~pixel$^{-1}$.\\

{\em Nickel} abundances have to be used with care because of the few lines that are measurable. From synthetic spectra we infer that Ni should be used for
cool stars only (\teff$<$5000K) and high SNR. In this regime, the abundances are
reliable (despite being underestimated by $\sim$0.1~dex) with an expected error of
$\sigma_{\mbox{Ni}}\sim0.25$~dex. The mean abundance correlates with the
number of measured lines ({\it i.e.} with SNR) as highlighted in
Fig.~\ref{Nlines_err}.\\

{\em $\alpha$-enhancement} is the average of [Mg/Fe] and [Si/Fe] and it
proved to be a robust estimation, particularly useful at low SNR, where the
measurements are more uncertain. The expected error is $\sim$0.15~dex
for SNR$>$40~pixel$^{-1}$ and $\sim$0.2~dex for
20$<$SNR$<$40~pixel$^{-1}$.\\

\begin{figure*}[t]
\begin{center}
\includegraphics[width=12cm]{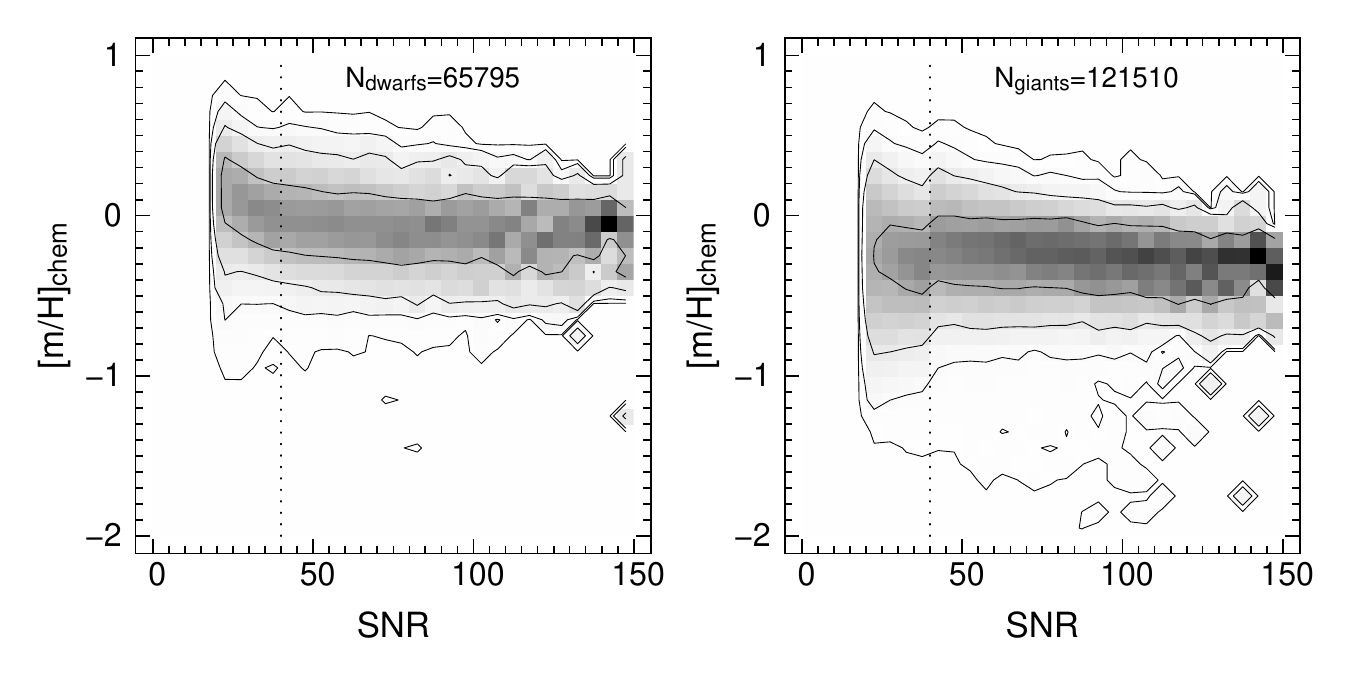}
\caption{\metc\ distribution for dwarfs (left panel) and giants (right panel)
as a function of signal-to-noise SNR. In order to highlight the shape of the
distribution at any SNR, the \metc\ distribution has been normalised for
every SNR bin. }
\label{mH_chem_STN}
\end{center}
\end{figure*}

Thanks to the improved line profile and the correction of the EW library
for the opacity of the neighbouring lines, the continuum re-normalisation has
improved. The new abundances are now
less affected by systematic errors than the previous ones and their
correlations with \teff\ are now negligible. On the other hand, 
the new continuum re-normalisation reveals the scarcity of information
for elements with weak and few visible lines like [Ca/H] (which has been 
dropped in this data release) and [Ti/H] on dwarf stars, which 
turns out to be not reliable.
A slight correlation between abundances and SNR is present,
as shown in Fig~\ref{mH_chem_STN}. Such correlation is negligible for giant
stars whereas for dwarfs stars \metc\  
\citep[computed with the formula given by][ see  Sect.~3.4 of \citetalias{Boeche11}]{ Salaris93}
increases by $\sim$0.1~dex from
SNR=80~pixel$^{-1}$ to SNR=40~pixel$^{-1}$. 
The different re-normalisation procedure for the two SNR regimes generates the step in average metallicity seen at SNR=40~pixel$^{-1}$.

\begin{figure}[tbh]
\begin{center}
\includegraphics[width=9cm]{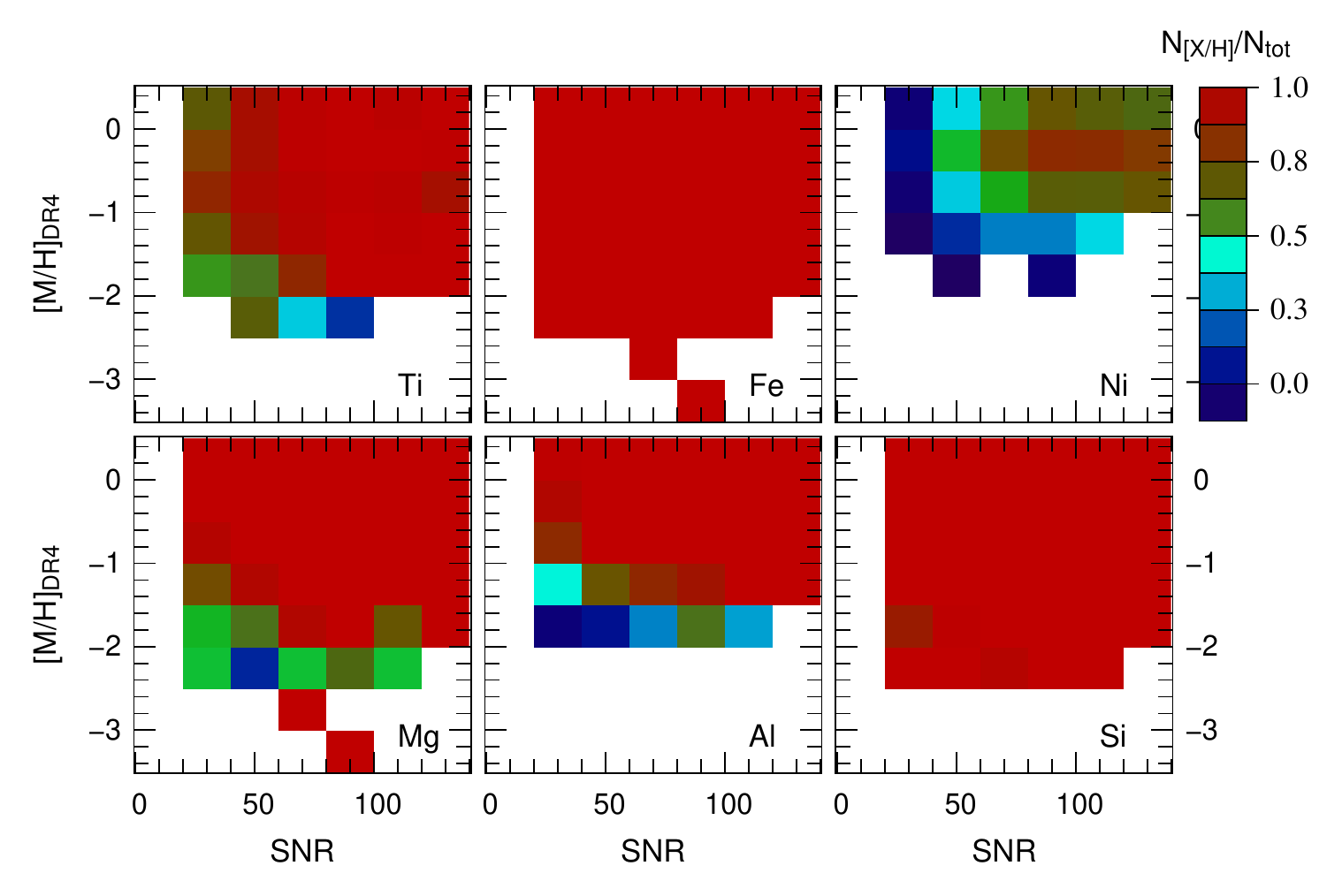}
\caption{The colours of the bins represent the number of spectra having
abundance estimation (N$_{X}$/N$_{tot}$) as a fraction of the total number of spectra as
a function of  metallicity and SNR. Each panel shows the distribution for the
identified element.}
\label{NX_Ntot_all}
\end{center}
\end{figure}

The accuracy of the RAVE abundances depends on many variables, often
inter-dependent in a non-linear way, which makes difficult the accuracy estimation of the individual abundances.
Indeed, on one hand the abundance accuracy depends on the number of measured absorption
lines. On the other hand, the number of measurable lines ({\it i.e.} strong enough to be identified in the noise) depends on SNR and on the stellar parameters.
In Fig.~\ref{Nlines_err} the dispersion of the residuals between measured
and expected abundances (for the sample of synthetic spectra) decreases
as the number of measured lines increases.
This number is a useful index of goodness of the abundance
accuracy, although it must considered together with SNR. 
In Fig.~\ref{NX_Ntot_all} we
illustrate the fraction of spectra having an abundance
estimation ({\it i.e.} at least one absorption line measured) for different elements, 
and how this fraction decreases when SNR
and/or metallicity decrease. This is a selection effect due
to the measurement process and it must be taken in
account during data analysis and interpretation.

\subsubsection{Comparison between \meta\ and \metc}

\begin{figure*}[t]
\begin{center}
\includegraphics[width=14cm,height=5cm]{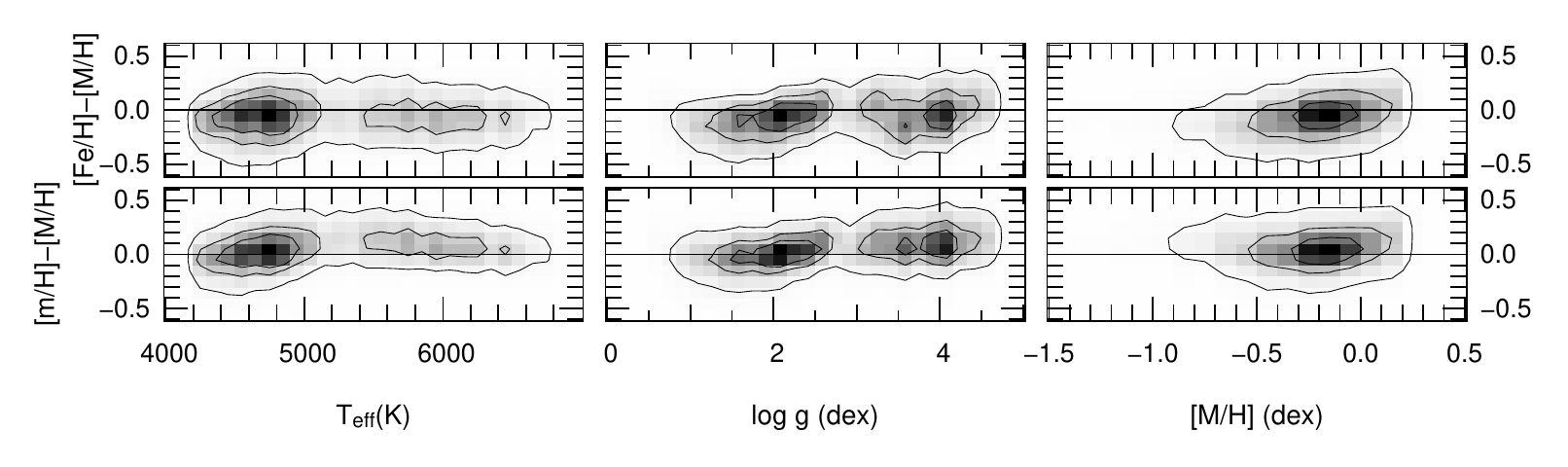}
\caption{Residuals between the DR4 metallicity (\meta), the chemical pipeline metallicity ([m/H]) and the measured iron abundance ([Fe/H]) as a function of the effective temperature, surface gravity and metallicity. The contour levels hold 34\%, 68\% and 95\% of the points, respectively.}
\label{MH-mchem_Teff_logg_MH}
\end{center}
\end{figure*}

In order to measure the chemical abundances, the RAVE chemical pipeline uses on one hand 
the estimation of \teff\ and \logg\ of the RAVE pipeline, and on the other hand \meta\ is
employed only as a first guess.  This means that the metallicity \metc\
(computed as explained in Sec~3.4 of \citetalias{Boeche11}) and
any elemental abundance provided by the chemical pipeline are independent of \meta.
It is, therefore, interesting to compare \meta\ with \metc\ as much as with
[Fe/H], because the latter is the element of reference used to calibrate \met.
In Fig.~\ref{MH-mchem_Teff_logg_MH} the residuals between \met, \metc\ and
[Fe/H] are shown as a function of \teff, \logg\ and \meta\ for spectra with SNR$>$40~pixel$^{-1}$.
In general, \meta\ appears to lie between [Fe/H] and \metc, in the order
[Fe/H]$\le$\met$\le$\metc. More specifically, there are some differences: for
dwarf stars this difference is slightly larger ($\sim$0.1~dex) than for
giants ($\le$0.05~dex). This can be due to the higher number of strong and narrow
absorption lines available in giants with respect to hot dwarfs, which allows
better measurements over the whole SNR range.
In addition, the [Fe/H] deviation can be due to the $\alpha$-enhanced stars which do not follow the enhancement relation of the learning grid (see Sec.~\ref{sect:synthetic_grid}) and for which [Fe/H]$\ne$\meta. Indeed, roughly 25\% of the stars with SNR$>$40~pixel$^{-1}$ deviate more than $1\sigma$ (0.15~dex) from the enhancements of the learning grid of Sec.~\ref{sect:synthetic_grid}. 

\section{Proper motions}
\label{sect:proper_motions}
In DR3, the proper motions were sourced from the PPMX \citep{Roeser08}, Tycho-2 \citep{Hog00}, SSS \citep{Hambly01}, and
UCAC2 \citep{Zacharias04} catalogs.
As already described in DR2,
the most precise available proper motion
was chosen for each object. In DR4, we  no longer follow this procedure
but publish a set of available proper motions for each object and
leave the selection to the user. The reason is the following:
the proper motion error bars published are of different origin,
they may be calculated either from the scatter
or from the weights of the individual positions. Hence, the
proper motion with the smallest formal error is not necessarily the most
accurate.
So, the DR4 catalogue lists proper motions from the following sources:
Tycho-2, UCAC2, UCAC3, UCAC4 \citep{Zacharias10, Zacharias13}, PPMX, PPMXL \citep{Roeser10} and SPM4 \citep{Girard11}.

\section{Radial velocities}
\label{sect:Vrad}
The radial  velocities for this  fourth data release  are based on  the RAVE
pipeline described in \citet{Siebert11},  therefore only a brief reminder of
the general features is given here.

The  radial  velocities  are  obtained using  a  standard  cross-correlation
algorithm in Fourier  space on the continuum subtracted  spectra.  First, an
estimate of  the radial velocity is  obtained using a subset  of 10 template
spectra.  This first estimate, with  an accuracy better than 5~\kms~ is used
to  put  the  spectrum in  the  zero  velocity  frame.   A new  template  is
constructed using  the full template  database using a  penalised chi-square
technique described in \citet{Zwitter08}.  The  new template is then used to
derive  the final,  more precise,  radial  velocity. The  internal error  is
obtained as the error on the determination of the maximum of the correlation
function. This part  is performed using the IRAF xcsao task.

\begin{figure}[btp]
\centering
\includegraphics[width=0.45\textwidth]{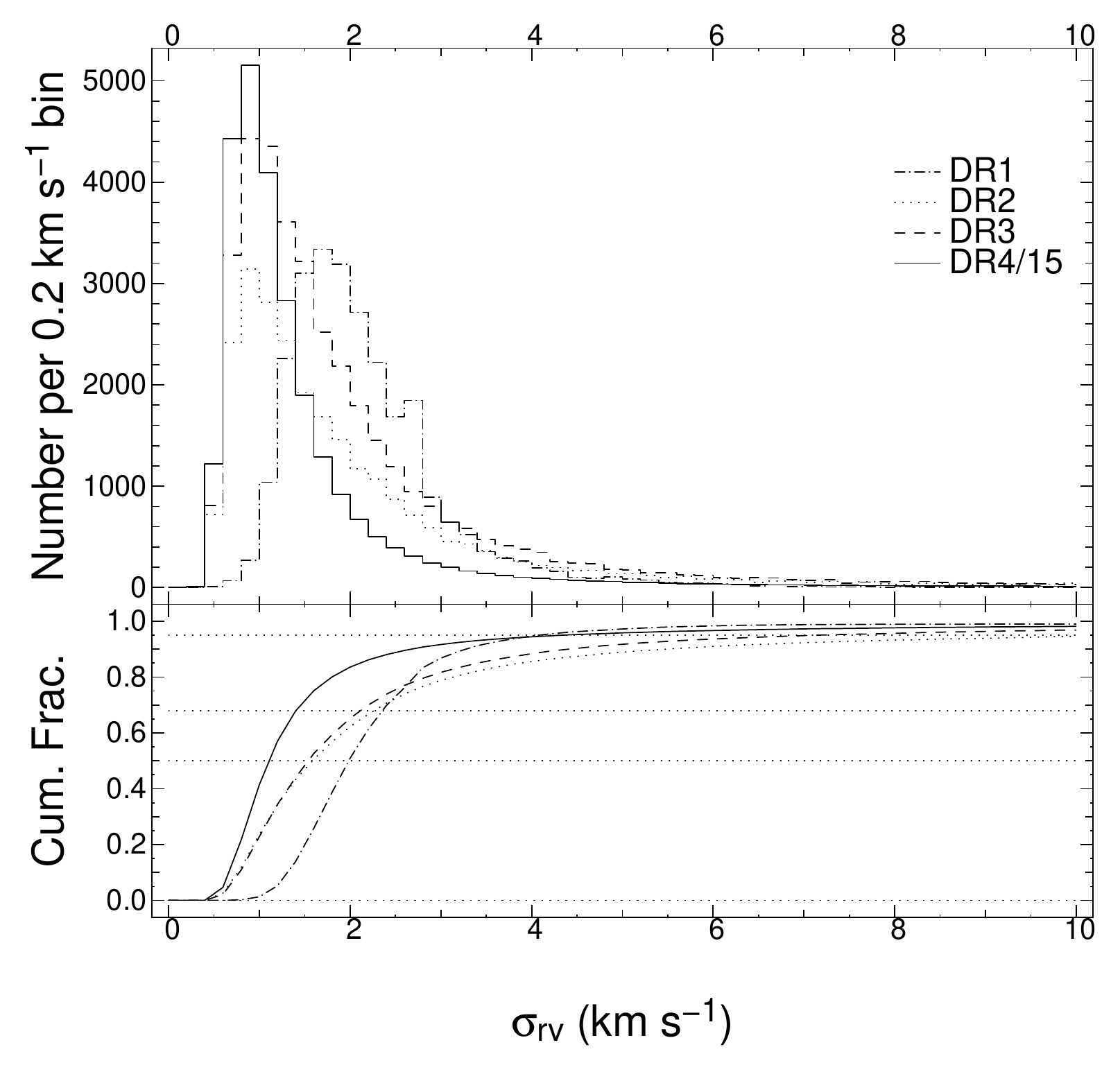}
\caption{{\bf Top:} histograms of the internal  radial velocity error for data new
  to each data release.  The bin size is 0.2~\kms.  For DR4, the number of
  stars per  bin is divided by 15  to compensate for the  increase in sample
  size. {\bf Bottom:} cumulative distributions.  The  dotted lines mark 50, 68 and
  95\% of the samples. }
\label{f:internal}
\end{figure}

The  histogram  distribution  of  the  internal  radial  velocity  error  is
presented  in  Fig.~\ref{f:internal}  top  panel. The  different  histograms
contain  data  new to  each  data  release as  indicated  in  the top  right
corner.  The
bottom   panel  of  Fig.~\ref{f:internal}   is  the   associated  cumulative
distribution of internal radial  velocity error. The  figure shows a  clear improvement  of the
quality of the radial velocity from  DR1 towards DR4, with a jump in quality
for data  new to DR4. From  DR2 to DR4\footnote{DR1 data  suffer from second
  order  contamination, which was corrected for  DR2 data,  hence  the radial
  velocity measurement  is not as  precise as for subsequent  releases.}, while
the mode of the distribution remains constant at $\sim 1$~\kms, the tail at
larger velocity errors  is consistently reduced with a  leap between DR3 and
DR4. Indeed,  while for DR3  68\% of the  data had internal errors  better than
2~\kms, for DR4  the 68\% limit goes down to  1.4~\kms.  The source of
this improvement is two-fold. First the  DR4 data are based upon a new input
catalog which uses DENIS $I$-band  magnitude instead of a pseudo-$I$ magnitude
constructed from Tycho-2  $B_T$ and $V_T$ photometry for  the bright part of
the catalog  and photographic $I$-band  from the SuperCosmos Sky  Survey for
the faint part.  This more  accurate photometry allows a better splitting of
the bright and faint sub-samples  which have different exposure times. Second,
at the telescope the signal-to-noise  ratio is now ``monitored'', and fields
with insufficient SNR benefit from longer exposure times to ensure a minimal
quality of  the data. The  combination of these two points allows  us to
considerably reduce the tail of the radial velocity error distribution.

\subsection{Repeat observations}

\begin{figure*}[btp]
\centering
\includegraphics[width=6cm]{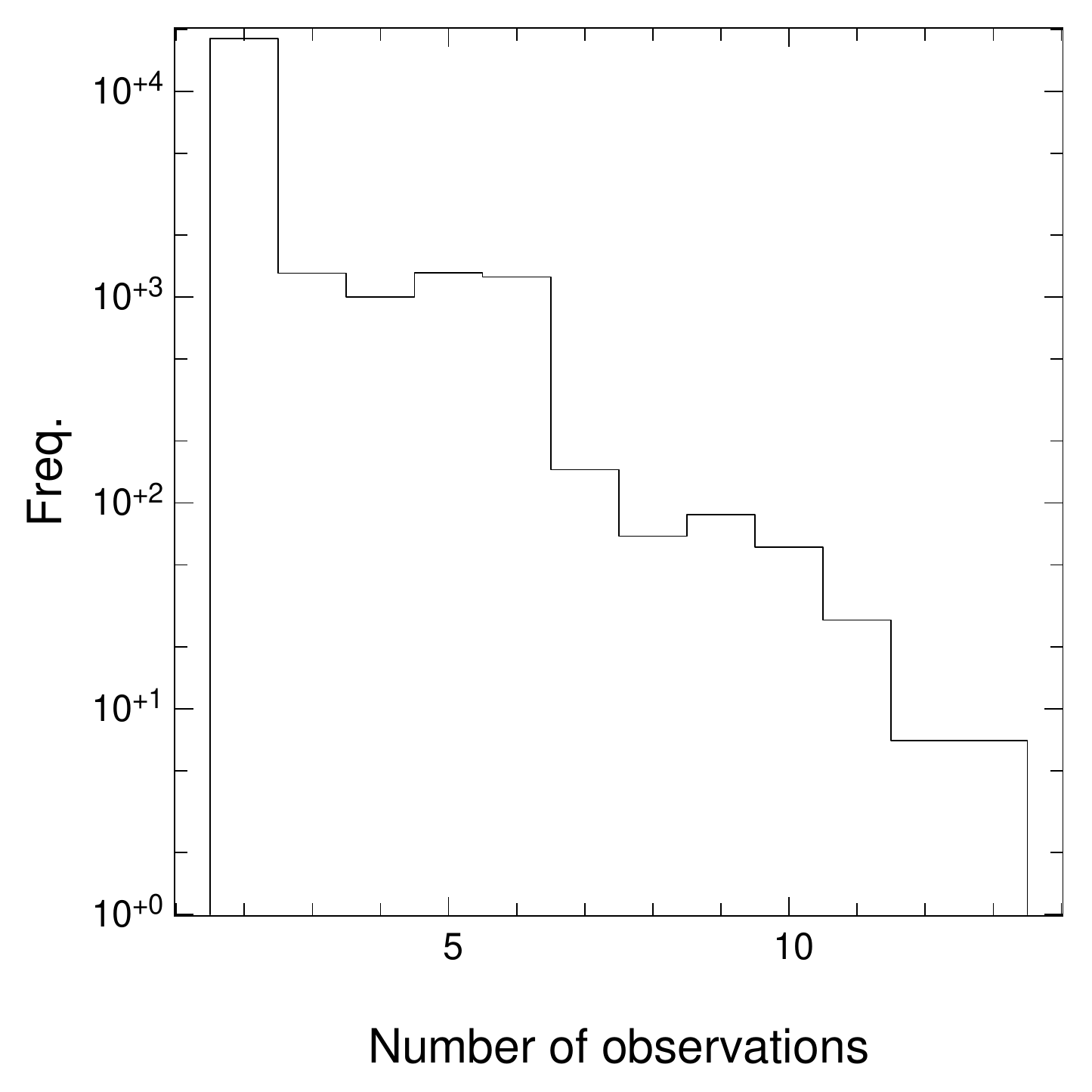}
\includegraphics[width=6cm]{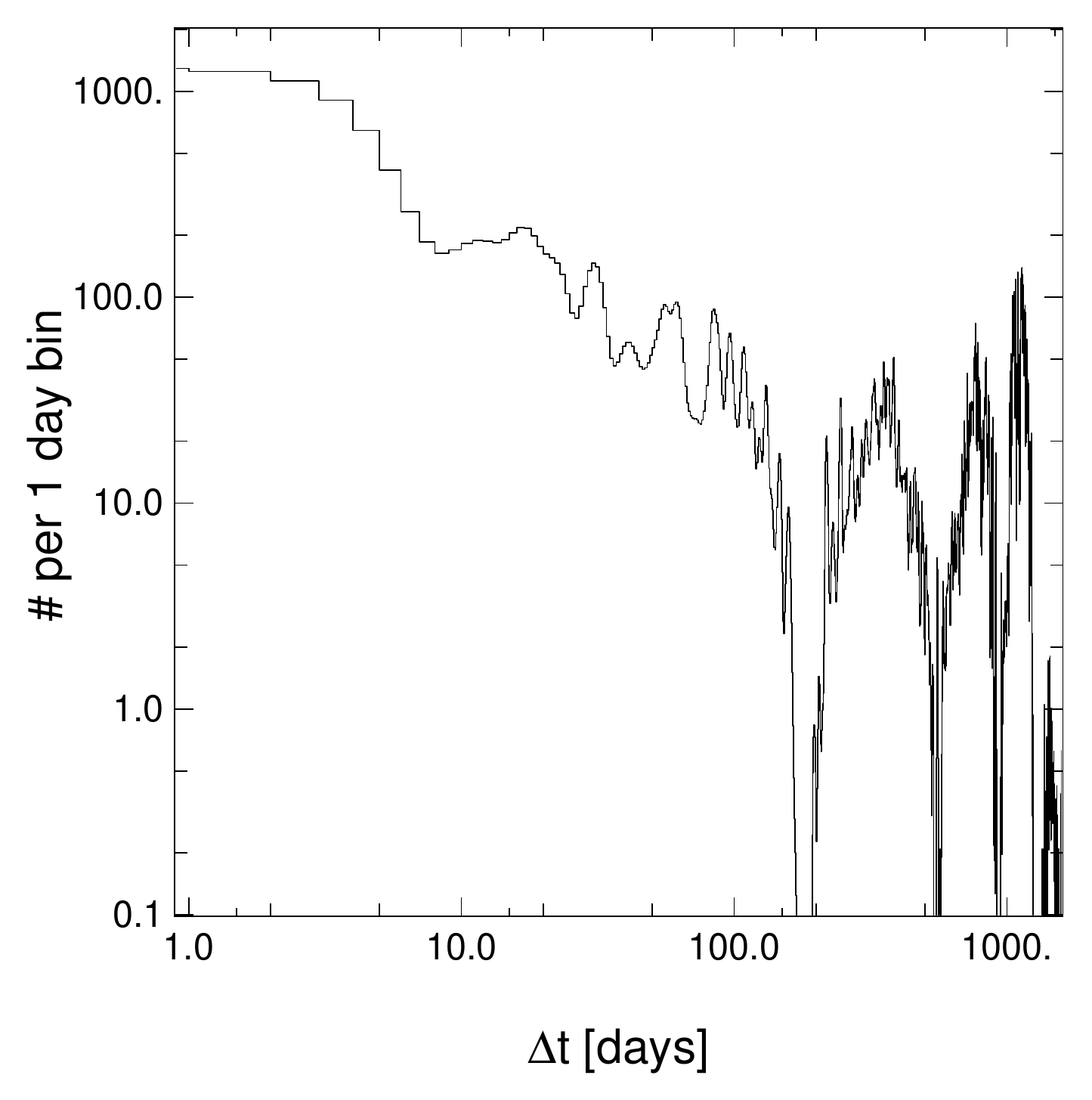}
\caption{Left: histogram  of the number  of observations for  stars observed
  more  than once.  Right: distribution  of the  time intervals  between the
  first observation of a star and its re-observation.}
\label{f:reobs_stat}
\end{figure*}

To verify  the quality of the  RAVE data, a  fraction of the survey  time is
devoted to multiple observations of RAVE targets with time intervals between
observations  ranging from  a few  hours  to 4  years. 23,288  stars in  the
present release belong  to this programme for a  total of 61,457 measurements,
some stars  having been observed  up to 13  times.  The distribution  of the
number  of  observations  per  star  is  presented  in  the  left  panel  of
Fig.~\ref{f:reobs_stat}. The  distribution of  the time interval  $\Delta t$
between  re-observations  is  presented  in  Fig.~\ref{f:reobs_stat},  right
panel, where the $\Delta t$ are  binned using intervals of one day.  As seen
from  this figure,  the  distribution is  not  random.  A  quasi-logarithmic
spacing  is  used to  sample  optimally the  possible  orbital  state of  the
spectroscopic binaries  with an enhancement of the  observations at specific
intervals of  one day, 2  weeks, 1, 2 and  3 months, 6  months and 1, 2, 3
and 4 years.

\begin{figure}[btp]
\centering
\includegraphics[width=0.5\textwidth]{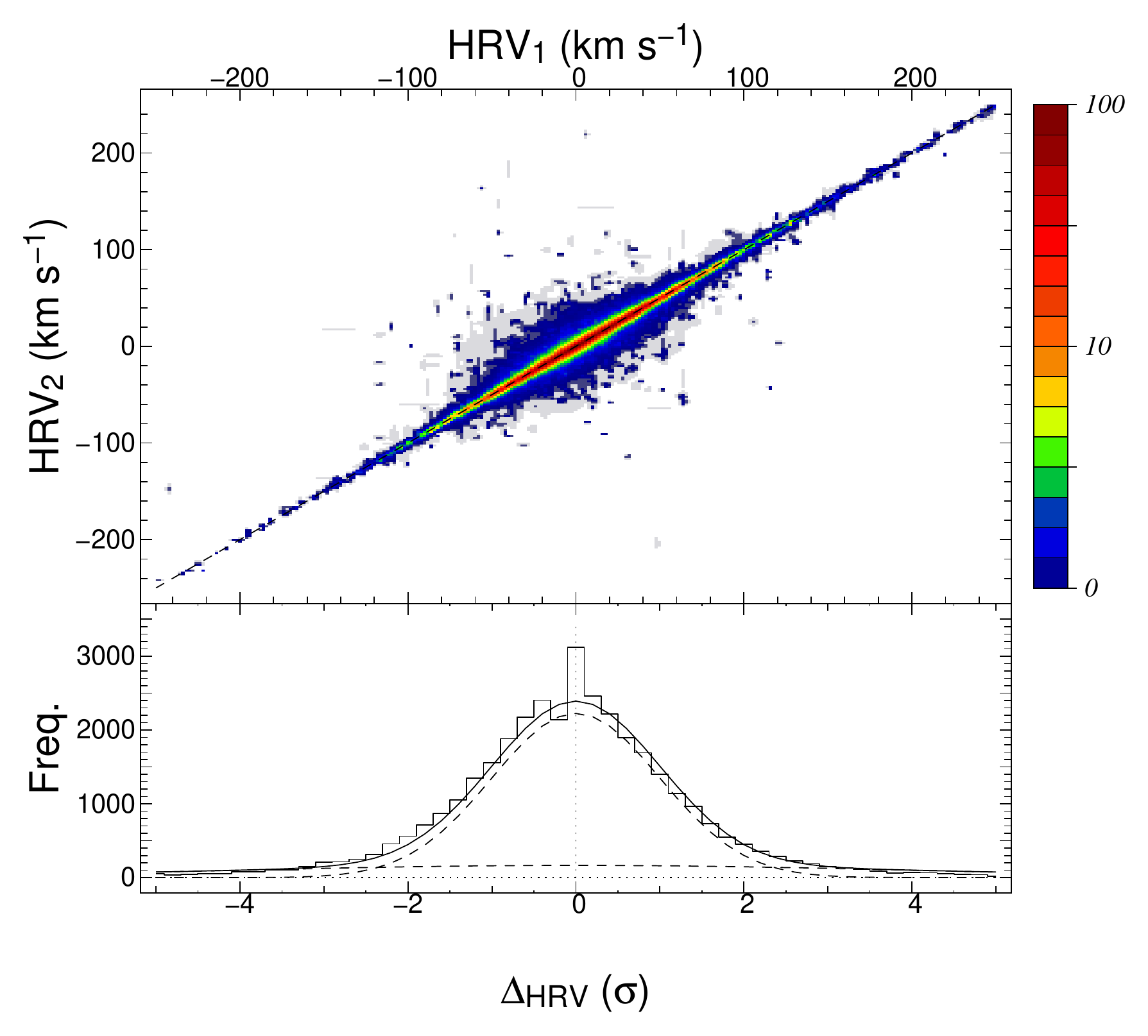}
\caption{Top: comparison  of the radial velocities  measured for re-observed
  RAVE targets. For  each star in this sample,  the measurement with highest
  SNR  is  used as  the reference  point on  the  x-axis  ($HRV_1$) while  other
  measurements  are  along  the  y-axis  ($HRV_2$).   Along  each  axis  the
  distribution  is obtained  by  convolving the  measurement  with a  Gaussian
  function  whose   dispersion  is  the  associated   internal  error.   The
  colour-coding follows  the resulting density  on a logarithmic  scale. The
  one-to-one relation is indicated by  the dashed line. Bottom: histogram of
  the radial  velocity differences normalised  to the error. The  plain black
  line is a  fit to the histogram assuming that it  is composed of two
  Gaussian populations  (see text).  From  this fit, the contribution  to the
  histogram of spectroscopic binaries and problematic spectra is 23\%.}
\label{f:repeats}
\end{figure}

The comparison  between the radial  velocity measurements for  the different
observations is given  in Fig.~\ref{f:repeats}. In the top  panel, the colour
coding follows the density per bin  of 2~\kms. For each star, the velocity
of the  observation with the highest  SNR is used as  the reference velocity
along  the x-axis  ($HRV_1$)  while subsequent  observations  are along  the
y-axis ($HRV_2$).   The measurements are convolved with  a Gaussian function
along each direction.  The dispersion  of the Gaussian is the internal error
associated with  the measurement along each  axis. The general trend follows closely
the one-to-one relation, showing no sign of any systematic effects.

Another  way to  look quantitatively  at the  radial velocity  difference is
presented  in the  bottom  panel of  Fig.~\ref{f:repeats},  which shows  the
histogram  of  the  radial  velocity  difference normalised  to  the  errors
$\Delta_{HRV}=(HRV_1-HRV_2) / (\sqrt{\sigma_1^2+\sigma_2^2})$.    If   our
measurements  were  perfect,  this  distribution should  follow  a  Gaussian
function  of  zero  mean  and  unit dispersion.   Binaries  and  problematic
measurements contribute to the tail and  can be assumed to follow a Gaussian
function with a  larger dispersion. In the  case of our RAVE data, a  sum of two
Gaussians is used  to fit the histogram, setting the  dispersion of the first
Gaussian function  to one.  Also,  we remove the  central bin from  the fit.
This central bin is mostly populated by repeat observations without delay in
time. The best fit is shown  as a continuous black line, the contribution of
each Gaussian function  being represented by a dashed  line.  Apart from the
central bin,  the fit  provides an adequate  representation of  the observed
histogram. The respective contributions of the two Gaussian functions is 77\%
for   the    standard   population   and   23\%    for   the   spectroscopic
binaries/problematic observations.  This fraction  is in agreement  with our
finding for DR3 data (26\%).

\subsection{Zero point offset}
\begin{figure}[tbp]
\centering
\includegraphics[width=7cm]{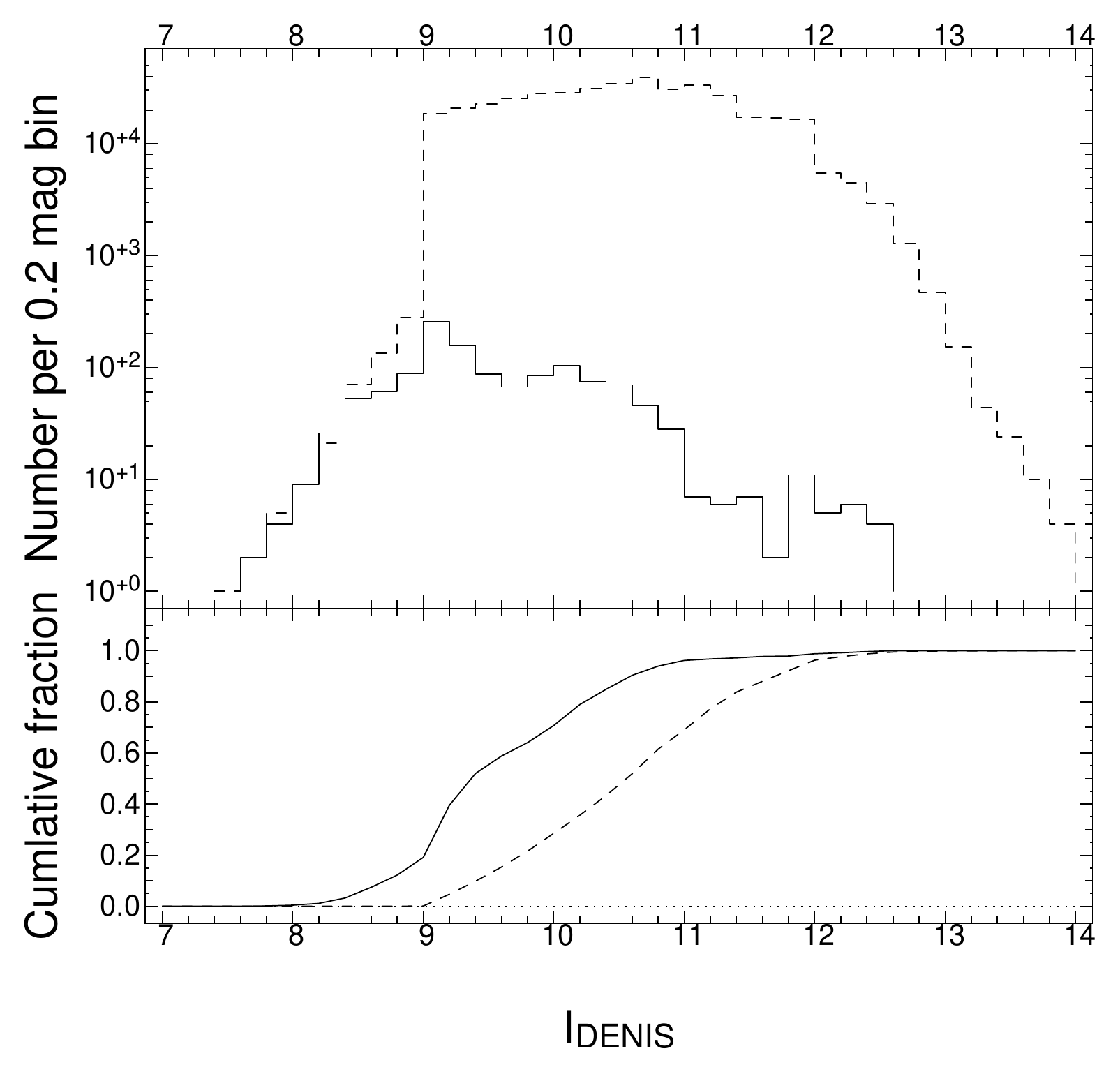}
\includegraphics[width=7cm]{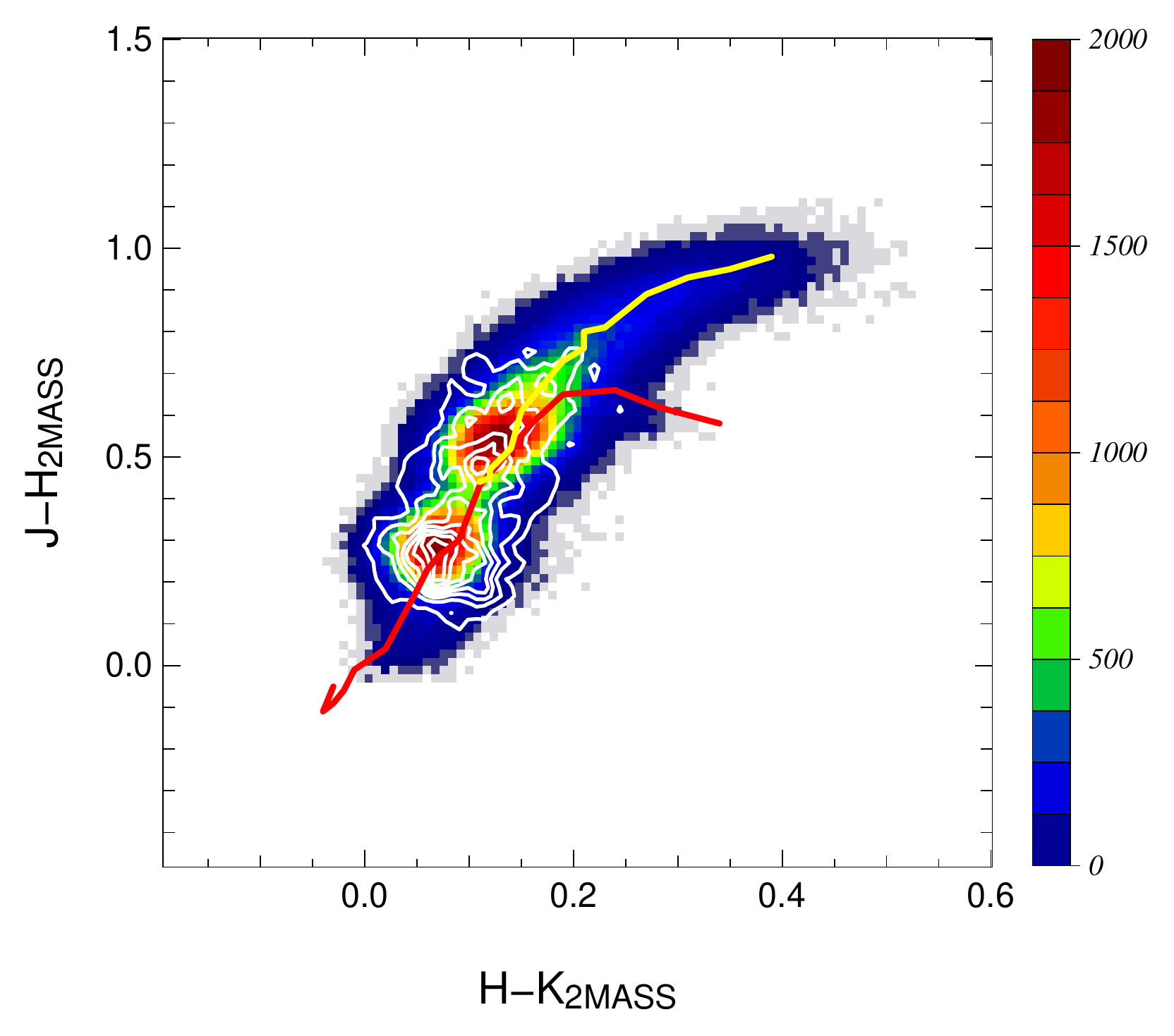}
\caption{Top: DENIS $I$ magnitude histogram  of the sub-sample of RAVE stars
  with external  radial velocity measurements  (plain line) compared  to the
  distribution  of DENIS  $I$ magnitude  for the  full RAVE  catalog (dashed
  line).  Bottom: 2MASS $J-H$ vs $H-K$ two dimensional histogram of the RAVE
  DR4 catalog with  a bin size of 0.02 magnitude on  each axis. The contours
  show the  location of the  sub-sample with external HRV  measurements. The
  thick  yellow  and   red  lines  are  fiducial  colours   from  Table~2  of
  \citet{Wainscoat92} for giant and dwarf stars respectively.}
\label{f:external_dist}
\end{figure}

As for the DR3 and previous releases, the radial velocity solutions
are corrected for potential zero-point offsets due to change in
temperature in the spectrograph room. The procedure uses the available
sky lines in the RAVE spectra to construct a smooth solution of the
zero point offset across the field plate. This procedure is fully described
in \citep{Siebert11} and the relevant measurements for each fibre are
given in the catalogue.
To verify the validity of our velocity zero point solution, comparison to independent
measurements  is made.   Our  comparison sample  comprises  data from  seven
different   sources:  the   Geneva-Copenhagen   Survey  \citep[GCS,][]{GCS},
\citet{Chubak12} data and  high-resolution echelle follow-up observations of
RAVE targets  at the  ANU 2.3m telescope,  Asiago observatory,  Apache Point
Observatory from \citet{Ruchti11} and Observatoire de Haute Provence using
the Elodie and  Sophie instruments. In total, the sample  of RAVE stars with
external   radial  velocity   measurements  contains   1265   stars.   Their
distributions in  DENIS $I$ magnitude  and 2MASS $J-H$ vs  $H-K$ color-color
diagram are  presented in Fig.~\ref{f:external_dist}.  Stars  with $I<9$ are
mostly custom RAVE observations of  bright GCS stars.  At fainter magnitudes
the  sample consists primarily  of high  resolution re-observations  of RAVE
targets.   In the  2MASS colour-colour  diagram, we  observe that  this sample
covers the  two main peaks (representing  dwarf and giant  stars) present in
the RAVE catalog.  However, dwarfs are over-represented compared to the RAVE
distribution due  to the large number  of GCS stars observed  in this sample
which are Hipparcos dwarfs.

The  summary  of  the  comparison   to  the  control  samples  is  given  in
Table~\ref{t:external}.  This  table reports the  number of objects  in each
sample,   the    total   number   of   observations,    the   mean   $\Delta
RV=RV_{DR4}-RV_{ext}$ and  the standard deviation.  $\Delta  RV$ is computed
using an  iterative $\sigma$-clipping technique to  remove the contamination
by spectroscopic binaries or  problematic measurements. No other quality cut
was applied on the samples.  The  clipping parameter and the number of stars
rejected for each sample is given in the last column of the table.

\begin{table}
\centering
\caption{Summary of the radial velocity comparison to external samples.}
\label{t:external}
\begin{tabular}{c c c c c }
Sample & $N_{\rm stars}$ & $N_{\rm obs}$ & $< \Delta RV>$ &  $\sigma_{\Delta RV}$ ($\sigma_{clip},n_{rej}$)\\
\hline
GCS        & 733 & 1024 &  0.28 & 1.72 (3,120) \\
Chubak     &  77 &   97 & -0.07 & 1.28 (3,2) \\
Ruchti     & 314 &  445 &  0.78 & 1.78 (3,34) \\
Asiago     &  25 &   47 & -0.22 & 2.95 (3,0) \\
ANU 2.3m   &  73 &  203 & -0.60 & 2.87 (3,18) \\
OHP Elodie &   9 &   13 &  0.29 & 0.40 (2.5,3) \\
OHP Sophie &  34 &   43 &  0.83 & 1.56 (3,4) \\
\hline
Full sample&1265 & 1872 &  0.20 & 1.52 (3,266)\\
\end{tabular}
\footnotetext{The mean difference in the radial velocities, $ \Delta RV$,  has been computed as following: $ \Delta RV = RV_{\rm DR4}-RV_{\rm ext}$}
\end{table}

\begin{figure}[tbp]
\centering
\includegraphics[width=0.45\textwidth]{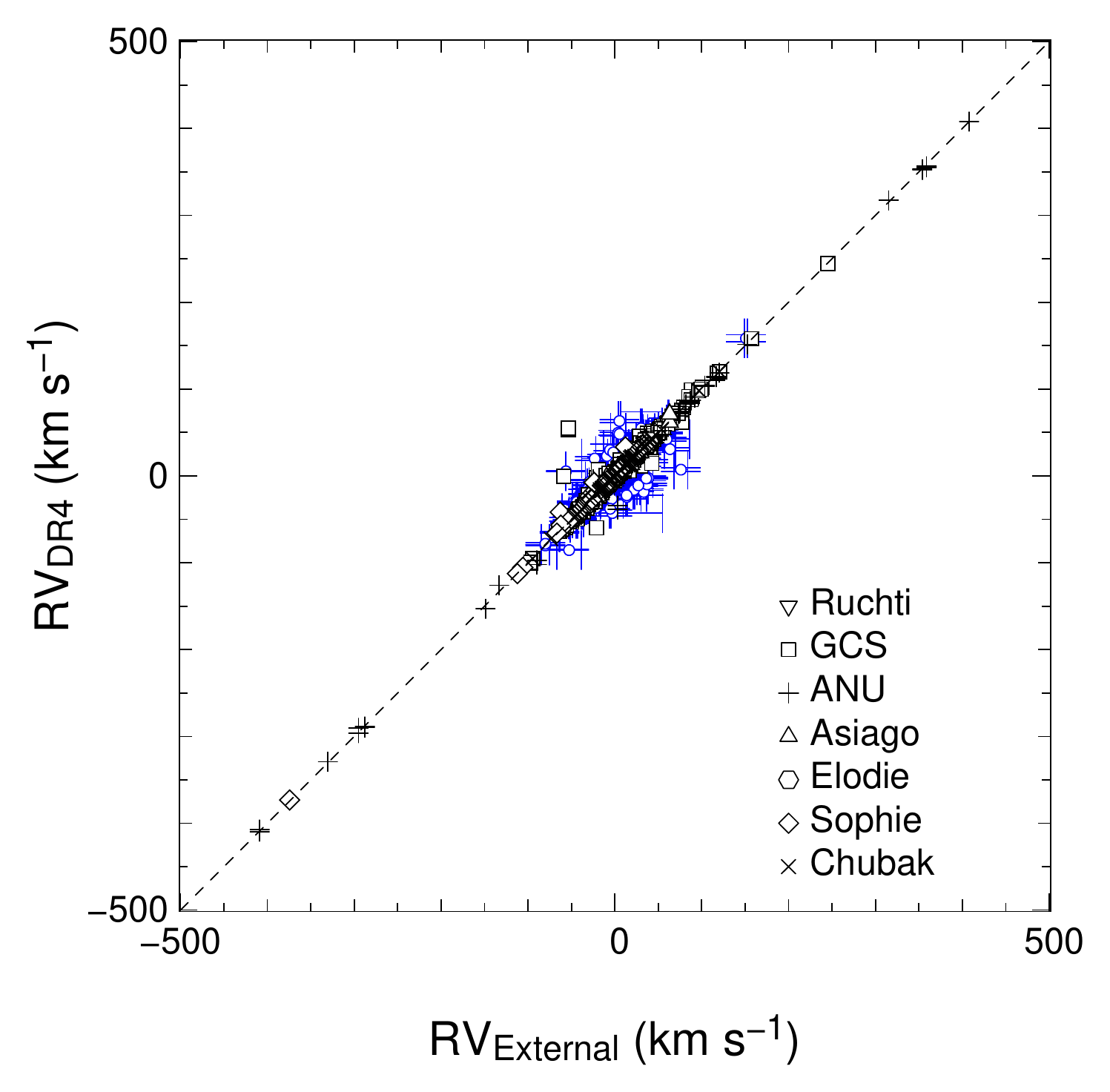}
\includegraphics[width=0.45\textwidth]{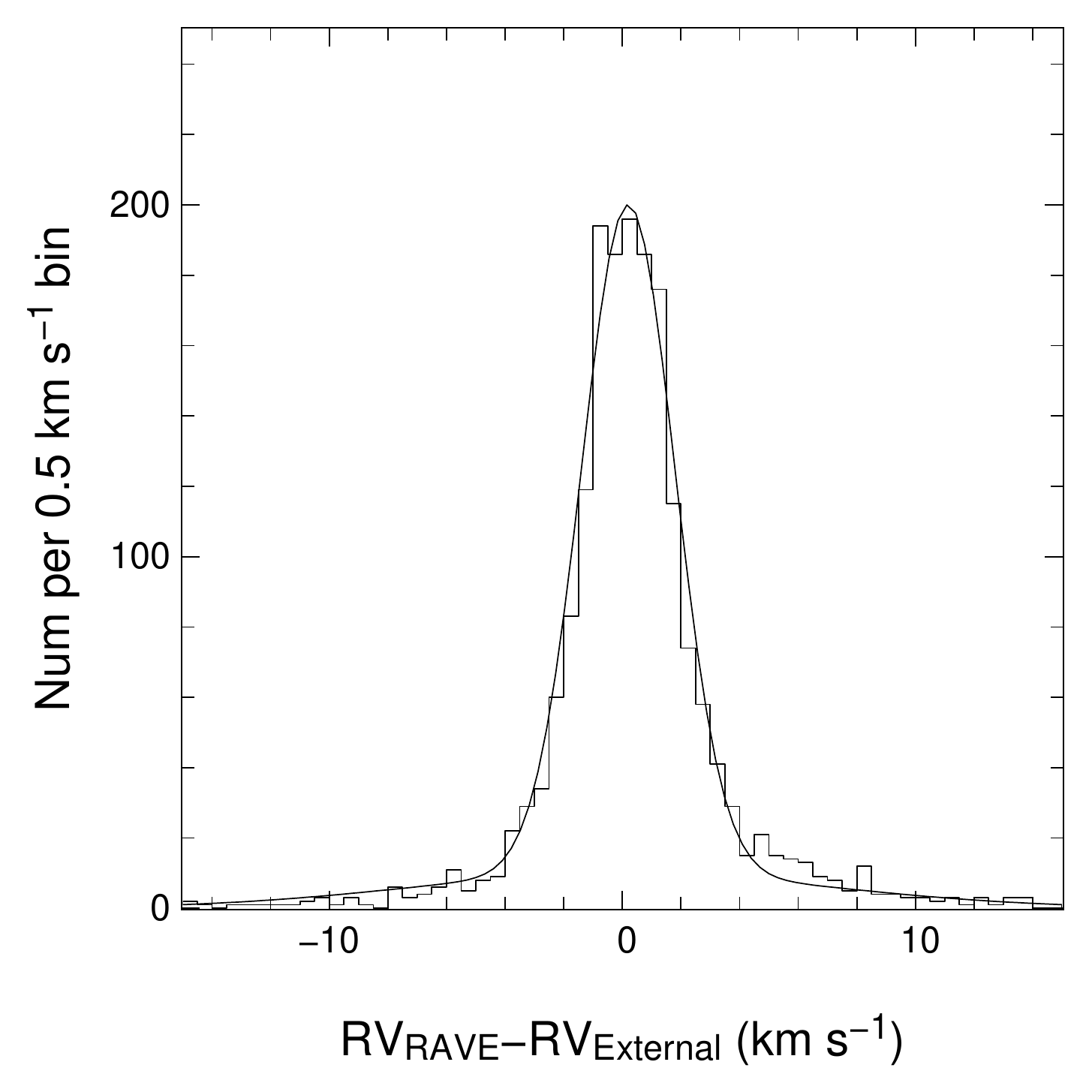}
\caption{Top:   Comparison   of   RAVE   radial  velocities   to   external
  measurements. The blue circles  are spectroscopic binaries detected in the
  high-resolution spectra.  The dashed  line marks the one to one relation.
  Bottom: Histogram of the radial  velocity difference.  The histogram can be
  modelled  using two  Gaussian functions  to  account for  normal stars  and
  binaries/problematic   spectra  (black   curve).  The   Gaussian  function
  recovering  the peak  is  at zero  mean  with a  standard  deviation of  1.5~\kms. 
  The  problematic spectra/binaries  contribution to this  sample is 5\%.}
\label{f:external}
\end{figure}

We note that  the agreement between RAVE and the  external sources is better
than 1~\kms~ in all the cases. Fig.~\ref{f:external} top panel presents the
direct  comparison of  the  DR4  radial velocities  to  the external  source
measurements.   The blue circles  are the  known spectroscopic  binaries and
show    a    broader    distribution    than   the    remainder    of    the
sample. Figure~\ref{f:external} bottom panel  shows the histogram of the radial
velocity difference.  This distribution can be adequately  reproduced by the
sum of two Gaussian functions,  the peak representing the single stars being
of zero mean with a dispersion  of 1.5~\kms, consistent with our expectation
for RAVE data.

\begin{figure}[tbp]
\centering
\includegraphics[width=0.45\textwidth]{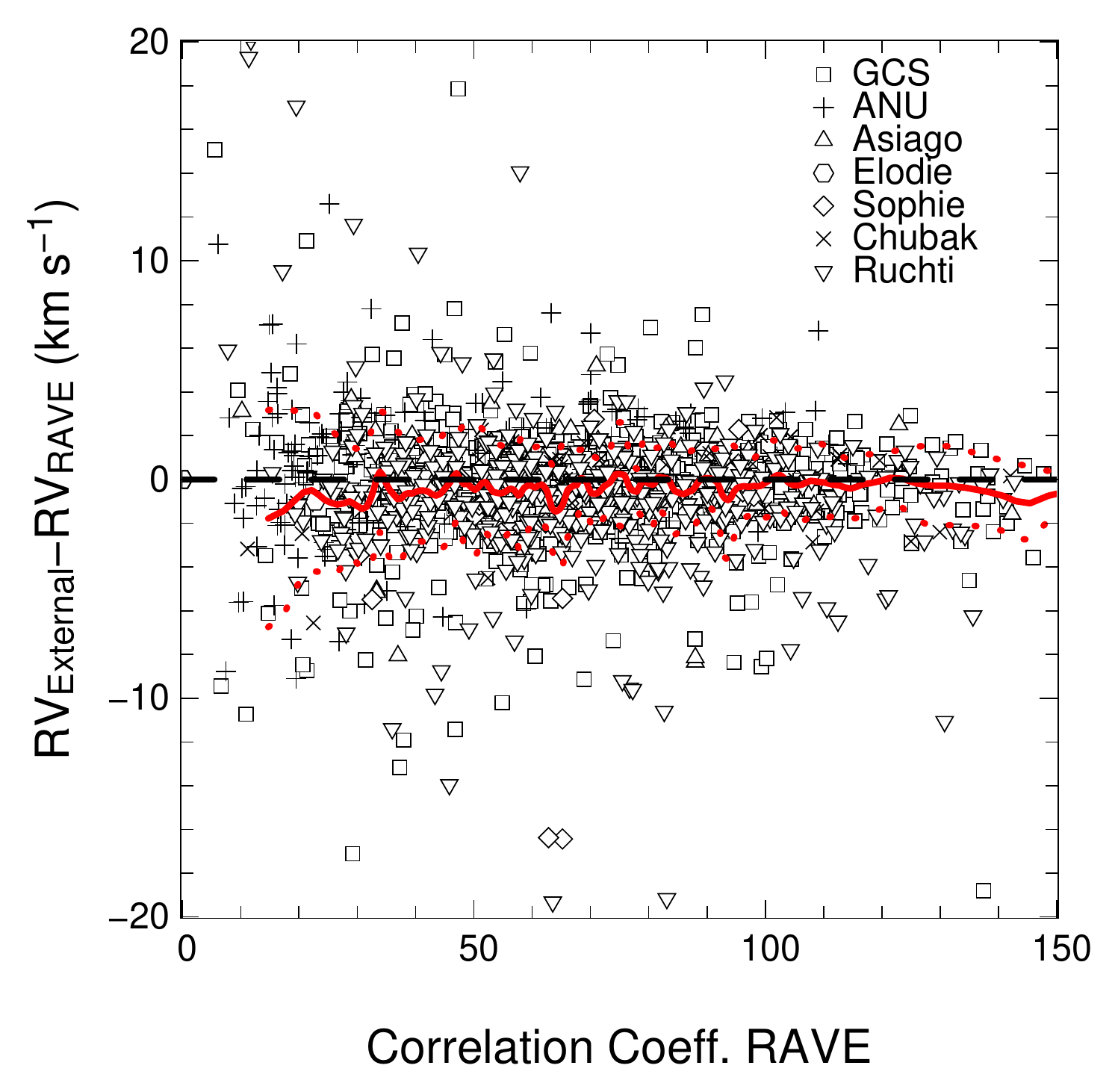}
\includegraphics[width=0.45\textwidth]{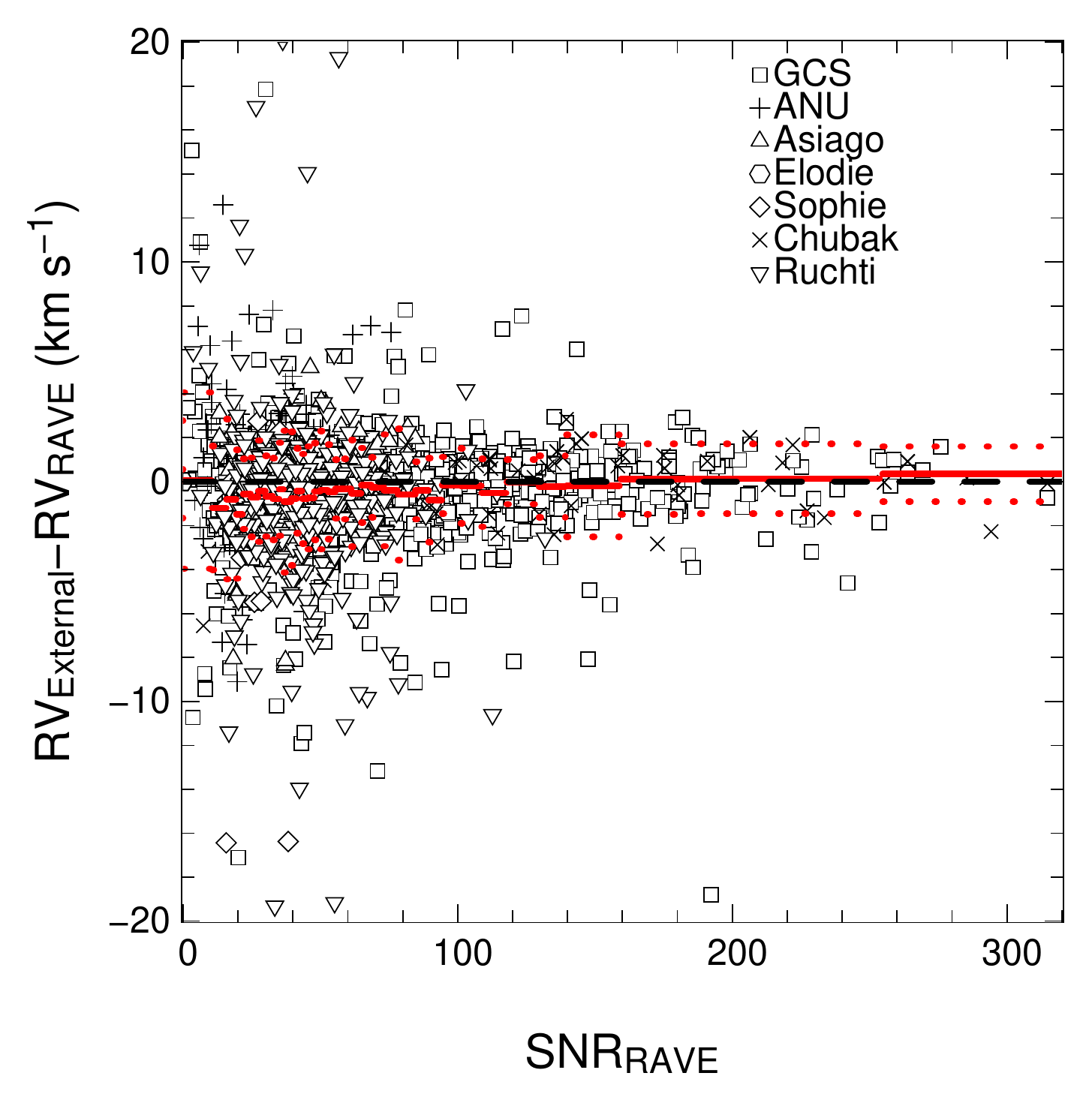}
\caption{Radial velocity difference as a function of Tonry-Davis correlation
coefficient (upper plot) and signal to noise (lower plot). The thick red line marks the
mean relation, the red dotted line the dispersion. The symbols indicate the
source of the external measurements.}
\label{f:external_test}
\end{figure}

As a final  test, we check the dependence of  the radial velocity difference
as  a   function  of  the   Tonry-Davis  correlation  coefficient   (R)  and
signal-to-noise of  the RAVE data  (Fig.~\ref{f:external_test}). Although an
increase in the  dispersion is observed for SNR$<$30~pixel$^{-1}$, or R
of 40,  the absence of an  apparent bias indicates that  our radial velocity
measurements are reliable.

\section{Stellar distances, ages and extinctions}
\label{sect:distances}
In the absence of parallaxes for the stars, the best way to obtain individual stellar distances is to project the atmospheric parameters on a set of theoretical isochrones and obtain the most likely value of the absolute magnitude of the stars. 
Up to now, the RAVE consortium has published a variety of studies using the distances inferred by red clump giants \citep[e.g.:][]{Siebert08, Veltz08, Williams13}, and 
developed three different methods in order to obtain the individual stellar line-of-sight distances \citep{Breddels10, Zwitter10, Burnet10}. Previous RAVE catalogues published the derived distances for the first two of these methods.

\begin{figure}[tbp]
\begin{center}
\includegraphics[width=0.45\textwidth]{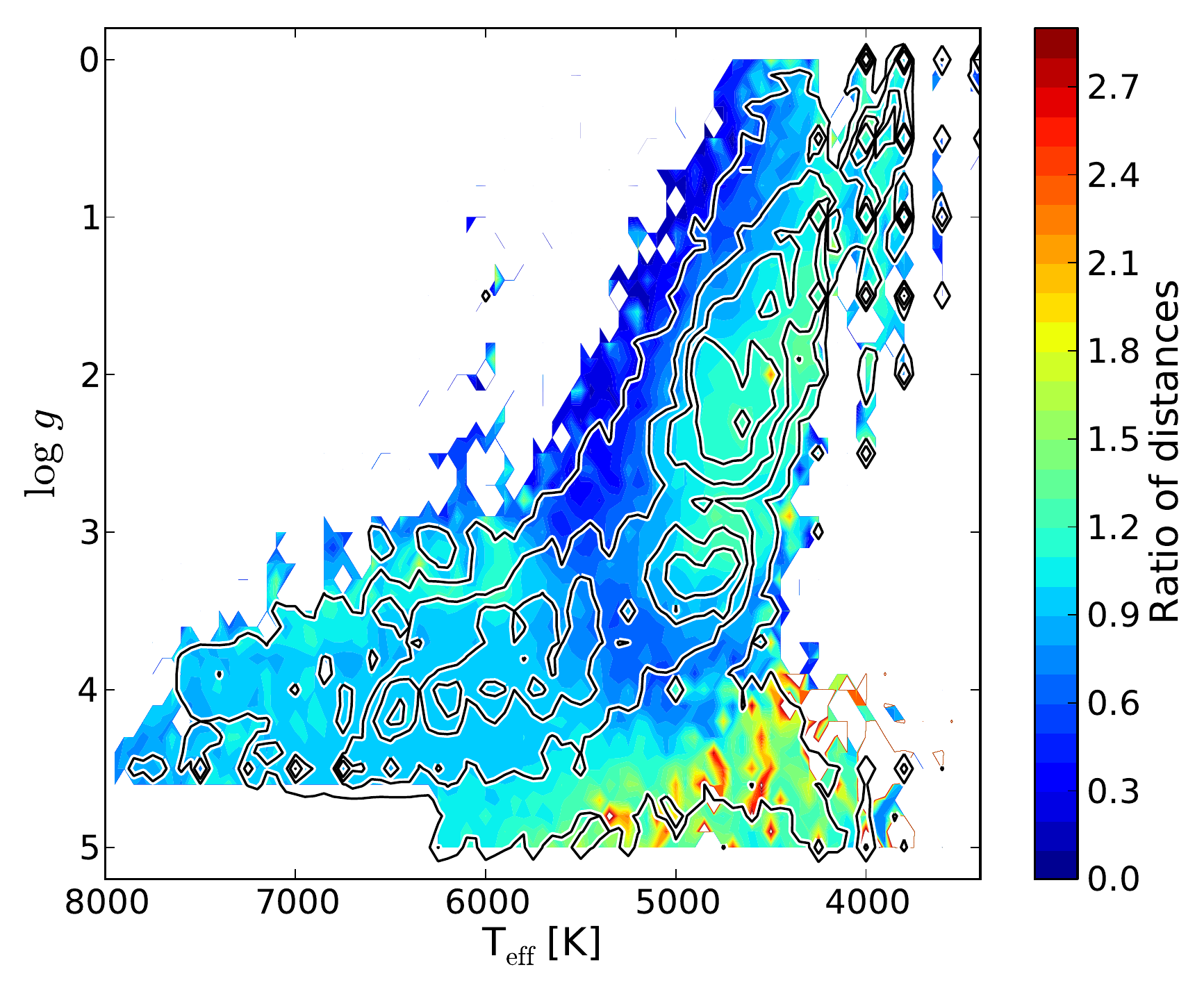} 
\caption{Comparison of  the average ratio of distances in the \teff--\logg\ plot for the stellar spectra with SNR$>$20~pixel$^{-1}$. The ratio is the \cite{Zwitter10} distance based on DR3 parameters divided by the \citet{Zwitter10} distance based on DR4 parameters.
The ratio is in filled contours, bin size is 50 K in temperature and 0.1 dex in gravity. Empty contours show occurrence of  \teff\ and \logg (expectation values) as calculated by DR4.}
\label{Fig:Zwitter_DR3_DR4}
\end{center}
\end{figure}

Figure~\ref{Fig:Zwitter_DR3_DR4} presents  for different regions of the \teff--\logg\ diagram, the ratio between the distances obtained using the \citet{Zwitter10} method with the DR3 parameters (Z10-DR3 hereafter) and with the DR4 ones (Z10-DR4 hereafter). It should be noted that Z10-DR3 distances have been obtained using only  internal errors for the atmospheric parameters, whereas Z10-DR4 consider the quadratic sum of the external uncertainties (Table~\ref{tab:external_errors}) and the internal ones (Table~\ref{tab:Internal_errors_noColorcuts} or Table~\ref{tab:Internal_errors_Colorcuts}).  One can see that  the largest deviations in distances occur in parts of the H--R diagram which are (and should be) scarcely populated. 
As an overall mean, the Z10-DR3 distances are 6\% larger than the Z10-DR4 ones, with the $1\sigma$ dispersion being at roughly 30\%. \\

For the present DR4 catalogue, we publish two sets of distances: one using the \citetalias{Zwitter10} method with the DR4 parameters, as well as another set obtained using a more robust algorithm, based on the Bayesian distance-finding method of \cite{Burnet10}. The improved algorithm is presented in \citet[B13 hereafter]{Binney13}, and now takes into account the interstellar extinction, as well as kinematic correction factors obtained by the method of \cite*{Schonrich12}.
The pipeline determines the probability distribution function of each star in the space of initial mass, age, metallicity, distance, extinction, etc., and from this distribution an age, a distance and an extinction are inferred from appropriate expectation values. The most reliable distance indicator turns out to be not the expectation value of the distance but the inverse of the expectation value of the parallax.
Here we discuss briefly the results for only the distances, but we refer the reader to B13 for a full  explanation of the method  as well as a detailed analysis of the reliability of the parameters. \\



The B13-DR4 spectrophotometric parallaxes have been computed for the stars which had spectra with  SNR$>10$~pixel$^{-1}$. For the targets that were observed several times, only the spectrum with the highest SNR has been used every time. The results have shown that the spectrophotometric parallaxes of Hipparcos stars are very satisfactory and are definitely improved by taking extinction into account. Nevertheless, an over-estimation of less than 10\% for the dwarfs and less than 20\% for the giants is evident. The method has also been tested on the open cluster spectra presented in Table~\ref{tab:testing_sets}, deriving very satisfactory distances, provided a cluster-specific age prior is used. Indeed, our data barely constrain the ages of stars, so a star's adopted age is heavily influenced by the age prior used, and the prior that is appropriate for field stars leads to excessive ages being assigned to most cluster stars, which tend to be young.

\begin{figure}[tbp]
\begin{center}
\includegraphics[width=0.4\textwidth,angle=0]{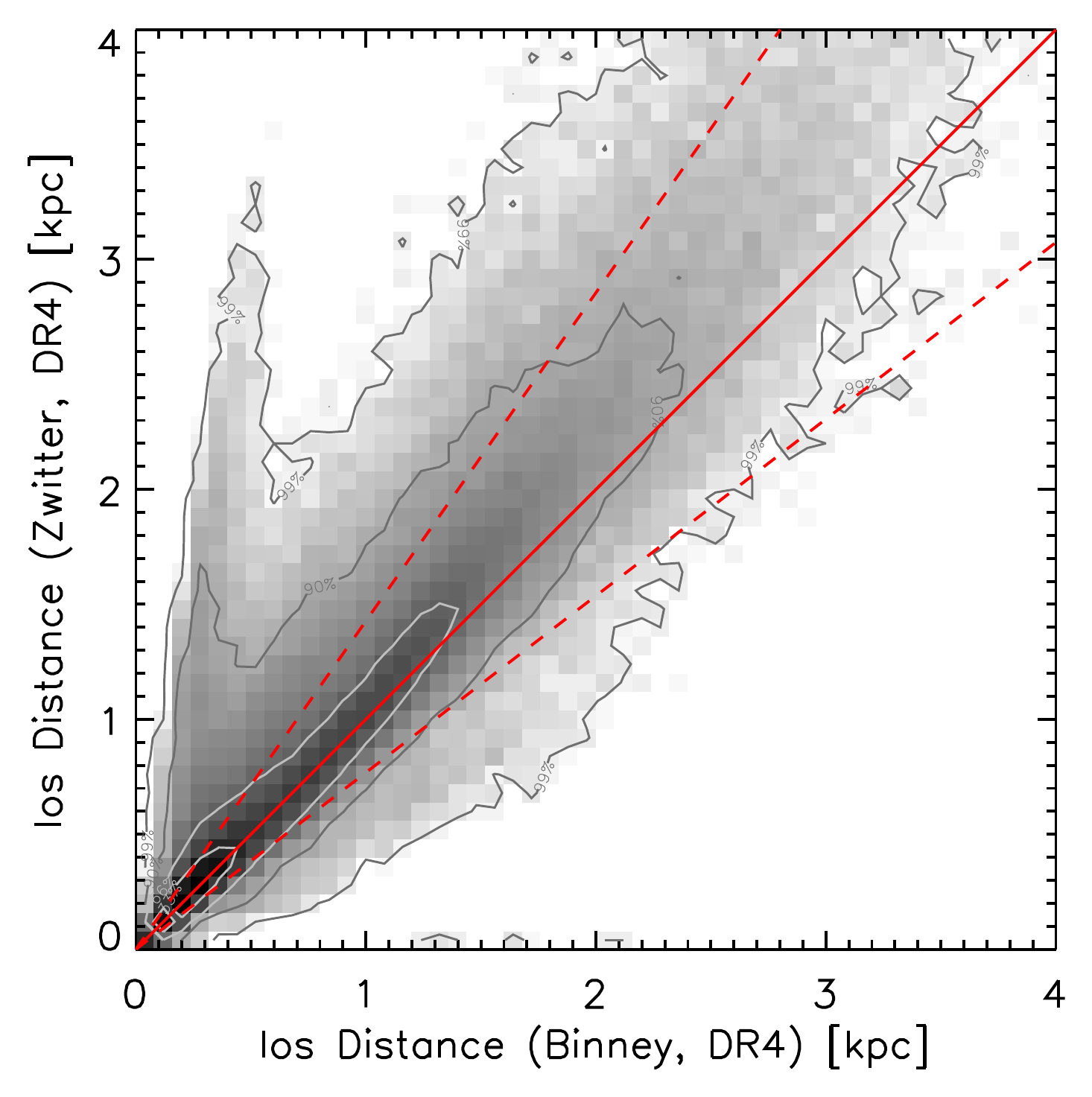} 
\caption{Comparison of the line-of-sight distance estimations obtained using the \cite{Binney13} method ($x-$axis) and the ones obtained using the  \citet{Zwitter10}  method combined with the Padova isochrones  ($y-$axis). Red line is the 1:1 relation and dashed lines show deviations of 30\% from unity. Contour lines hold 33\%, 66\%, 90\% and 99\% of the sample.}
\label{Fig:Zwitter_Binney}
\end{center}
\end{figure}

Figure~\ref{Fig:Zwitter_Binney} compares the distance estimations obtained using the Padova isochrones and the Z10-DR4 or the B13-DR4 methods.  
From this figure one can see that for the bulk of the targets, the distances obtained with the two methods are quite similar. We find that the median of the distribution defined by  Z10/B13 is equal to 1.02. However, for the most distant stars  ($D \gtrsim 2$~kpc) Z10-DR4  predicts larger distances than B13-DR4. Finally, the targets for which B13-DR4 predicts $D\sim 0.5$~kpc and Z10-DR4 derives $D>1.5$~kpc correspond to the most metal-poor giant stars of the sample, for which the atmospheric parameters have also the largest errors. 
 This disagreement is mainly due to the fact that Z10 projects the atmospheric parameters on the isochrones, without having any  {\it a priori} constraint on their expected output parameters, based on their position in the Galaxy, their radial velocity or proper motion. Hence, input atmospheric parameter values with large uncertainties, especially for metal-poor giant stars, naturally lead to erroneous distances and to the discrepancy shown in Fig.~\ref{Fig:Zwitter_Binney}. 

We conclude that the B13-DR4 distances are more robust and hence should give better results if the priors are satisfied.  



\section{Photometry from APASS}
\label{sect:APASS}
$B$$V$$g'$$r'$$i'$ photometric data of RAVE stars have been obtained as part
of the ongoing APASS survey (Henden et al.  2013, Munari et al.  2013a,b). The APASS photometric survey covers the whole sky, from Pole to Pole, with
ongoing observations from CTIO (Chile), for the southern hemisphere, and New
Mexico for the northern counterpart.  At both sites, a pair of twin remotely
controlled, small telescopes obtain simultaneous CCD observations during
dark- and grey-Moon time over five optical bands: $B$, $V$ \citep[tied to the
equatorial standards of][]{Landolt09} and $g'$,$r'$,$i'$ bands \citep[tied to the
158 primary standards given by][that define the Sloan
photometric system]{Smith02}.  The telescopes are 20cm f/3.6 astrographs feeding
Apogee U16m cameras (4096$\times$4096 array, 9 $\mu$m pixels), that cover a
field 2.9 deg wide with a 2.6 arcsec/pix plate factor.  The photometric
filters are of the dielectric multi-layer type and are produced by Astrodon. Transmission curves and photometric performances of Astrodon filters used in
the APASS survey are discussed and compared to more conventional types of
photometric filters in \citet{Munari12a} and \citet{Munari12b}. On average 80 fields are observed per night at each APASS location, 20 of
them being standard fields (Landolt, Sloan).

The APASS observations are obtained with fixed exposure times (different and
optimised for each photometric band), set to detect $V$=17 stars at SNR=5
on a single exposure.  Stars brighter than $V$=10~mag may saturate under optimal
seeing conditions.  At the time of writing, 90\% of the whole sky has been
covered, with 42 million stars measured on at least two distinct epochs.
Differential photometry within a given field is accurate to better than 0.01~mag, absolute photometry over the whole sky is currently accurate to better
than 0.025 mag (closely similar to 2MASS accuracy). APASS astrometric positions
are also highly accurate. Comparison with the positions given in the Carlsberg Meridian Catalog for the 118\,940 RAVE stars in common shows a distribution peaked at a separation of 0.105~arcsec, with the median value at 0.177~arcsec.
Although APASS DR7 is publicly available, its values are not published in RAVE DR4, because future APASS DR will provide better accuracy and coverage of RAVE DR4.  Clearly APASS photometry will significantly enhance analysis of RAVE data. We recommend users to adopt APASS photometry as it becomes available.


\section{Conclusions}
\label{sect:conclusion} 
The fourth public data release of the RAVE survey includes the stellar atmospheric parameters of \totalspectranumber~spectra obtained from April 2004 to December 2012. Compared to the previous catalogue of DR3, 
a new input catalogue, based on DENIS DR3 and 2MASS, is used to select the observed targets. The new input catalogue has the major new feature of extending to lower Galactic latitudes and providing more accurate astrometry, leading to fibre placement better matching stellar positions on the sky, which results in higher signal-to-noise spectra. 
In addition, 
the parameters have been revisited, thanks to a new pipeline, presented in \cite{Kordopatis11a}, and 809 reference spectra that allowed us to validate the effective temperatures and surface gravities, and calibrate the metallicities.  
The {  RAVE} stellar atmospheric parameters that are obtained with the new pipeline are free of any obvious systematics {  (no correlations between the derived parameters or as function of signal-to-noise)}, in particular for the overall metallicities of the stars.
The spectra with the lowest signal-to-noise ratio have a distribution function of atmospheric parameter values which shows a well-understood pattern of discretisation effects, but it has been shown that this discretisation does not alter the accuracy of the derived parameters.  
We show that the metallicity distribution functions of the observed stars shift towards lower metallicity values for the lower surface gravity bins, at the same time as the radial velocity dispersion  increases. This is in agreement with a change in the mixture of the probed Galactic populations as a function of the probed volume, passing from an old thin disc dominated population to the presence of more halo stars for the larger volume probed by the more luminous giant stars in low surface gravity bins. That is, at face value the distribution functions for derived stellar parameters are consistent with plausible astrophysical expectations.

In addition to the atmospheric parameters obtained with the new pipeline, those obtained with the DR3 pipeline are also published since they have been used in published analyses. However, they are of lower reliability than our DR4 dataset, so situations demanding their re-analysis should be rather rare.

The abundances of six individual elements, namely aluminium, silicon, titanium, iron, magnesium and nickel are published, using an improved version of the \cite{Boeche11} chemical pipeline. The reliability of these elemental abundances varies according to the effective temperature, surface gravity and metallicity of the star, the signal-to-noise ratio of the spectrum, and of course the element itself. 

The catalogue also includes the line-of-sight distances computed using the methods presented in \cite{Binney13} and \cite{Zwitter10} as well as the ages and the interstellar extinctions which are a sub-product of the \cite{Binney13} pipeline. Radial velocities,  photometric information, proper motions, and stellar binarity flags  complete  the DR4 catalogue entries.



\appendix

\section{Appendix material}
Table~\ref{tab:DR4} describes the contents of individual columns of
the Fourth Data Release catalog. The catalog is accessible online at
\url{http://www.rave-survey.org} and via the CDS VizieR service.

\begin{table}
\caption{Catalogue description} 

\label{tab:DR4}
\begin{center}
\begin{tabular}{llcclll}
\hline 
Col  &  Format   &   Units &   NULL & Label      &               Explanations \\           \hline
1       & char(32)     & -          & N     & RAVE\_OBS\_ID                 & Target designation                                                     \\
2       & char(18)     & -          & N     & RAVEID                      & RAVE target designation                                                \\
3       & double       & deg        & N     & RAdeg                       & Right ascension                                                        \\
4       & double       & deg        & N     & DEdeg                       & Declination                                                            \\
5       & double       & deg        & N     & Glon                        & Galactic longitude                                                     \\
6       & double       & deg        & N     & Glat                        & Galactic latitude                                                      \\
7       & float        & km/s       & N     & HRV                         & Heliocentric radial velocity                                           \\
8       & float        & km/s       & N     & eHRV                        & HRV error                                                              \\
9       & float        & (R+)       & N     & CorrelationCoeff            & Tonry \& Davis R correlation coefficient                                \\
10      & float        & -          & N     & PeakHeight                  & Height of correlation peak                                             \\
11      & float        & -          & N     & PeakWidth                   & Width of correlation peak                                              \\
12      & float        & km/s       & Y     & SkyRV                       & Measured HRV of Sky                                                    \\
13      & float        & km/s       & Y     & eSkyRV                      & Error Measured HRV of Sky                                              \\
14      & float        & (R+)       & Y     & SkyCorrelationCoeff         & Tonry \& Davis R Sky correlation coefficient                            \\
15      & float        & km/s       & Y     & CorrectionRV                & Zero point correction applied Radial Velocity                          \\
16      & char(5)      & -          & Y     & ZeroPointFLAG               & Quality flag for Zero point correction (Note 3)                        \\
17      & char(16)     & -          & Y     & ID\_TYCHO2                   & TYCHO2 Target designation                                              \\
18      & float        & arcsec     & Y     & Dist\_TYCHO2                 & Center distance to target catalog                                      \\
19      & char(2)      & -          & Y     & XidQualityFLAG\_TYCHO2       & Crossmatch quality flag (Note 4)                                       \\
20      & float        & mas/yr     & Y     & pmRA\_TYCHO2                 & Proper motion RA from TYCHO2                                           \\
21      & float        & mas/yr     & Y     & epmRA\_TYCHO2                & error Proper motion RA from TYCHO2                                     \\
22      & float        & mas/yr     & Y     & pmDE\_TYCHO2                 & Proper motion DE from TYCHO2                                           \\
23      & float        & mas/yr     & Y     & epmDE\_TYCHO2                & error Proper motion DE from TYCHO2                                     \\
24      & char(16)     & -          & Y     & ID\_UCAC2                    & UCAC2 Target designation                                               \\
25      & float        & arcsec     & Y     & Dist\_UCAC2                  & Center distance to target catalog                                      \\
26      & char(2)      & -          & Y     & XidQualityFLAG\_UCAC2        & Crossmatch quality flag (Note 4)                                       \\
27      & float        & mas/yr     & Y     & pmRA\_UCAC2                  & Proper motion RA from UCAC2                                            \\
28      & float        & mas/yr     & Y     & epmRA\_UCAC2                 & error Proper motion RA from UCAC2                                      \\
29      & float        & mas/yr     & Y     & pmDE\_UCAC2                  & Proper motion DE from UCAC2                                            \\
30      & float        & mas/yr     & Y     & epmDE\_UCAC2                 & error Proper motion DE from UCAC2                                      \\
31      & char(16)     & -          & Y     & ID\_UCAC3                    & UCAC3 Target designation                                               \\
32      & float        & arcsec     & Y     & Dist\_UCAC3                  & Center distance to target catalog                                      \\
33      & char(2)      & -          & Y     & XidQualityFLAG\_UCAC3        & Crossmatch quality flag (Note 4)                                       \\
34      & float        & mas/yr     & Y     & pmRA\_UCAC3                  & Proper motion RA from UCAC3                                            \\
35      & float        & mas/yr     & Y     & epmRA\_UCAC3                 & error Proper motion RA from UCAC3                                      \\
36      & float        & mas/yr     & Y     & pmDE\_UCAC3                  & Proper motion DE from UCAC3                                            \\
37      & float        & mas/yr     & Y     & epmDE\_UCAC3                 & error Proper motion DE from UCAC3                                      \\
38      & char(16)     & -          & Y     & ID\_UCAC4                    & UCAC4 Target designation                                               \\
39      & float        & arcsec     & Y     & Dist\_UCAC4                  & Center distance to target catalog                                      \\
40      & char(2)      & -          & Y     & XidQualityFLAG\_UCAC4        & Crossmatch quality flag (Note 4)                                       \\
41      & float        & mas/yr     & Y     & pmRA\_UCAC4                  & Proper motion RA from UCAC4                                            \\
42      & float        & mas/yr     & Y     & epmRA\_UCAC4                 & error Proper motion RA from UCAC4                                      \\
43      & float        & mas/yr     & Y     & pmDE\_UCAC4                  & Proper motion DE from UCAC4                                            \\
44      & float        & mas/yr     & Y     & epmDE\_UCAC4                 & error Proper motion DE from UCAC4                                      \\
45      & char(16)     & -          & Y     & ID\_PPMXL                    & PPMXL Target designation                                               \\
46      & float        & arcsec     & Y     & Dist\_PPMXL                  & Center distance to target catalog                                      \\
47      & char(2)      & -          & Y     & XidQualityFLAG\_PPMXL        & Crossmatch quality flag (Note 4)                                       \\
48      & float        & mas/yr     & Y     & pmRA\_PPMXL                  & Proper motion RA from PPMXL                                            \\
49      & float        & mas/yr     & Y     & epmRA\_PPMXL                 & error Proper motion RA from PPMXL                                      \\
50      & float        & mas/yr     & Y     & pmDE\_PPMXL                  & Proper motion DE from PPMXL                                            \\
51      & float        & mas/yr     & Y     & epmDE\_PPMXL                 & error Proper motion DE from PPMXL                                      \\
52      & char(10)     & -          & N     & Obsdate                     & Observation date yyyymmdd                                              \\
53      & char(14)     & -          & N     & FieldName                   & Name of RAVE field (RA/DE)                                             \\
54      & int          & -          & N     & PlateNumber                 & Number of field plate [1..3]                                           \\
55      & int          & -          & N     & FiberNumber                 & Number of optical fiber [1,150]                                        \\
56      & float        & K          & Y     & Teff\_K                      & Effective temperature (Note 1)                                         \\
57      & float        & K          & Y     & eTeff\_K                     & Error Effective temperature (Note 1)                                   \\
58      & float        & dex        & Y     & logg\_K                      & Log gravity (Note 1)                                                   \\
59      & float        & dex        & Y     & elogg\_K                     & Error Log gravity (Note 1)                                             \\
60      & float        & dex        & Y     & Met\_K                       & Metallicity [m/H](Note 1)                                              \\
61      & float        & dex        & Y     & Met\_N\_K                     & Metallicity [m/H](Note 1)                                              \\
62      & float        & dex        & Y     & eMet\_K                      & ErrorMetallicity [m/H] (Note 1)                                        \\
63      & float        & -          & Y     & SNR\_K                       & Signal to Noise value (Note 1)                                         \\
64      & float        & -          & Y     & Algo\_Conv\_K                 & Quality Flag for Stellar Parameter pipeline [0..4] (Note 1, Note 5)    \\
65      & float        & dex        & Y     & Al                          & Abundance of Al [Al/H]                                                 \\
66      & int          & -          & Y     & Al\_N                        & Number of used spectral lines for calculation of abundance             \\
67      & float        & dex        & Y     & Si                          & Abundance of Si [Si/H]                                                 \\
68      & int          & -          & Y     & Si\_N                        & Number of used spectral lines for calculation of abundance             \\
\end{tabular}
\end{center}
\end{table}

\setcounter{table}{6}
\begin{table}
\caption{Catalogue description (continued)} 
\begin{center}
\begin{tabular}{llcclll}
\hline 
Col  &  Format   &   Units &   NULL & Label      &               Explanations \\           \hline
69      & float        & dex        & Y     & Fe                          & Abundance of Fe [Fe/H]                                                 \\
70      & int          & -          & Y     & Fe\_N                        & Number of used spectral lines for calculation of abundance             \\
71      & float        & dex        & Y     & Ti                          & Abundance of Ti [Ti/H]                                                 \\
72      & int          & -          & Y     & Ti\_N                        & Number of used spectral lines for calculation of abundance             \\
73      & float        & dex        & Y     & Ni                          & Abundance of Ni [Ni/H]                                                 \\
74      & int          & -          & Y     & Ni\_N                        & Number of used spectral lines for calculation of abundance             \\
75      & float        & dex        & Y     & Mg                          & Abundance of Mg [Mg/H]                                                 \\
76      & int          & -          & Y     & Mg\_N                        & Number of used spectral lines for calculation of abundance             \\
77      & float        & -          & Y     & CHISQ\_c                     & ?2 of the Chemical Pipeline (Note 1)                                  \\
78      & float        & K          & Y     & Teff\_SPARV                  & Effective temperature (Note 1)                                         \\
79      & float        & dex        & Y     & logg\_SPARV                  & Log gravity (Note 1)                                                   \\
80      & float        & dex        & Y     & alpha\_SPARV                 & Metallicity (Note 1)                                                   \\
81      & char(16)     & -          & Y     & ID\_2MASS                    & 2MASS Target designation                                               \\
82      & float        & arcsec     & Y     & Dist\_2MASS                  & Center distance to target catalog                                      \\
83      & char(2)      & -          & Y     & XidQualityFLAG\_2MASS        & Crossmatch quality flag (Note 4)                                       \\
84      & double       & mag        & Y     & Jmag\_2MASS                  & J  magnitude                                                             \\
85      & double       & mag        & Y     & eJmag\_2MASS                 & Error J  magnitude                                                       \\
86      & double       & mag        & Y     & Hmag\_2MASS                  & H  magnitude                                                             \\
87      & double       & mag        & Y     & eHmag\_2MASS                 & Error H  magnitude                                                       \\
88      & double       & mag        & Y     & Kmag\_2MASS                  & K  magnitude                                                             \\
89      & double       & mag        & Y     & eKmag\_2MASS                 & Error K  magnitude                                                       \\
90      & char(16)     & -          & Y     & ID\_DENIS                    & DENIS Target designation                                               \\
91      & double       & arcsec     & Y     & Dist\_DENIS                  & Centre distance to target catalog                                      \\
92      & char(2)      & -          & Y     & XidQualityFLAG\_DENIS        & Crossmatch quality flag (Note 4)                                       \\
93      & double       & mag        & Y     & Imag\_DENIS                  & I  magnitude                                                             \\
94      & double       & mag        & Y     & eImag\_DENIS                 & Error I  magnitude                                                       \\
95      & double       & mag        & Y     & Jmag\_DENIS                  & J  magnitude                                                             \\
96      & double       & mag        & Y     & eJmag\_DENIS                 & Error J  magnitude                                                       \\
97      & double       & mag        & Y     & Kmag\_DENIS                  & K  magnitude                                                             \\
98      & double       & mag        & Y     & eKmag\_DENIS                 & Error K  magnitude                                                       \\
99      & char(16)     & -          & Y     & ID\_USNOB1                   & USNOB1 Target designation                                              \\
100     & double       & arcsec     & Y     & Dist\_USNOB1                 & Centre distance to target catalog                                      \\
101     & char(2)      & -          & Y     & XidQualityFLAG\_USNOB1       & Crossmatch quality flag (Note 4)                                       \\
102     & double       & mag        & Y     & B1mag\_USNOB1                & B1  magnitude                                                            \\
103     & double       & mag        & Y     & R1mag\_USNOB1                & R1  magnitude                                                            \\
104     & double       & mag        & Y     & B2mag\_USNOB1                & B2  magnitude                                                            \\
105     & double       & mag        & Y     & R2mag\_USNOB1                & R2  magnitude                                                            \\
106     & double       & mag        & Y     & Imag\_USNOB1                 & I  magnitude                                                             \\
107     & float        & mas        & Y     & parallax                    & Parallax (Note 4)                                                      \\
108     & float        & mas        & Y     & e\_parallax                  & Error Parallax (Note 4)                                                \\
109     & float        & kpc        & Y     & dist                        & Distance (Note 4)                                                      \\
110     & float        & kpc        & Y     & e\_dist                      & Error Distance (Note 4)                                                \\
111     & float        & -          & Y     & DistanceModulus\_Binney      & Distance modulus (Note 4)                                              \\
112     & float        & -          & Y     & eDistanceModulus\_Binney     & Distance modulus (Note 4)                                              \\
113     & float        & -          & Y     & Av                          & Extinction (Note 4)                                                    \\
114     & float        & -          & Y     & e\_Av                        & Error Extinction (Note 4)                                              \\
115     & float        & -          & Y     & age                         & Age (Note 4)                                                           \\
116     & float        & -          & Y     & e\_age                       & Error Age (Note 4)                                                     \\
117     & float        & MSun       & Y     & mass                        & Mass (Note 4)                                                          \\
118     & float        & MSun       & Y     & e\_mass                      & Error Mass (Note 4)                                                    \\

   \end{tabular}
   \end{center}
\end{table}

\setcounter{table}{6}
\begin{table}
\caption{Catalogue description (continued)} 
\begin{center}
\begin{tabular}{llcclll}
\hline 
Col  &  Format   &   Units &   NULL & Label      &               Explanations \\           \hline
119     & char(2)      & -          & Y     & c1                          & n.th minimum distance (Note 6)                                         \\
120     & char(2)      & -          & Y     & c2                          & n.th minimum distance (Note 6)                                         \\
121     & char(2)      & -          & Y     & c3                          & n.th minimum distance (Note 6)                                         \\
122     & char(2)      & -          & Y     & c4                          & n.th minimum distance (Note 6)                                         \\
123     & char(2)      & -          & Y     & c5                          & n.th minimum distance (Note 6)                                         \\
124     & char(2)      & -          & Y     & c6                          & n.th minimum distance (Note 6)                                         \\
125     & char(2)      & -          & Y     & c7                          & n.th minimum distance (Note 6)                                         \\
126     & char(2)      & -          & Y     & c8                          & n.th minimum distance (Note 6)                                         \\
127     & char(2)      & -          & Y     & c9                          & n.th minimum distance (Note 6)                                         \\
128     & char(2)      & -          & Y     & c10                         & n.th minimum distance (Note 6)                                         \\
129     & char(2)      & -          & Y     & c11                         & n.th minimum distance (Note 6)                                         \\
130     & char(2)      & -          & Y     & c12                         & n.th minimum distance (Note 6)                                         \\
131     & char(2)      & -          & Y     & c13                         & n.th minimum distance (Note 6)                                         \\
132     & char(2)      & -          & Y     & c14                         & n.th minimum distance (Note 6)                                         \\
133     & char(2)      & -          & Y     & c15                         & n.th minimum distance (Note 6)                                         \\
134     & char(2)      & -          & Y     & c16                         & n.th minimum distance (Note 6)                                         \\
135     & char(2)      & -          & Y     & c17                         & n.th minimum distance (Note 6)                                         \\
136     & char(2)      & -          & Y     & c18                         & n.th minimum distance (Note 6)                                         \\
137     & char(2)      & -          & Y     & c19                         & n.th minimum distance (Note 6)                                         \\
138     & char(2)      & -          & Y     & c20                         & n.th minimum distance (Note 6)                                         \\
\hline
\footnotetext{{\bf Notes.} (1): Originating from:
   \_K indicates values from Stellar Parameter Pipeline, 
   \_N\_K indicates a calibrated value,
   \_c indicates values from Chemical Pipeline,
   \_SPARV indicates values of Radial Velocity Pipeline (used in DR3 also).
(2): Cross-identification flag as follows:
   $A = 1$~association within 2 arcsec.
   $B = 2$~associations within 2 arcsec.
   $C =$~More than 2 associations within 2 arcsec.
   $D =$~Nearest neighbour more than 2 arcsec. away.
   $X =$~No association found (within 10 arcsec limit ).
(3): Flag value of the form FGSH, F being for the entire plate, G for the 50 
fibres group to which the fibre belongs. S flags the zero-point correction used: C 
for cubic and S for a constant shift. If H is set to * the fibre is close to a 15 
fibre gap. For F and G the values can be A, B, C, D, or E
    $A =$~dispersion around correction lower than 1km/s
    $B =$~dispersion between 1 and 2km/s
    $C =$~dispersion between 2 and 3km/s
    $D =$~dispersion larger than 3km/s
    $E =$~less than 15 fibres available for the fit.
(4): See \citet{Binney13}. 
(5): Flag of Stellar Parameter Pipeline     
   $0 =$~Pipeline converged. 
   $1 =$~no convergence.
   $2 =$~MATISSE oscillates between two values and the mean has been performed. 
   $3 =$~results of MATISSE at the boundaries or outside the grid and the DEGAS value has been adopted
   $4 =$~the metal-poor giants with SNR$<$20 have been re-run by degas with a scale factor (ie, internal parameter of DEGAS) of 0.40
(6): Morphological Flag 
   n.th minimum distance to base spectrum given by one of 
   the types {\it a,b,c,d,e,g,h,n,o,p,t,u,w} \citep[see][]{Matijevic12}.
     }
   \end{tabular}
   \end{center}
\end{table}

\bibliographystyle{apj} 
\bibliography{RAVE} 




\acknowledgments
We are most grateful to our referee, for his detailed and very relevant comments which
improved the quality of the presentation of the paper.
Funding for RAVE has been provided by: the Australian Astronomical Observatory;
the Leibniz-Institut f\"ur Astrophysik Potsdam (AIP); the Australian
National University; the Australian Research Council; the French National Research
Agency; the German Research Foundation (SPP 1177 and SFB 881); the
European Research Council (ERC-StG 240271 Galactica); the Instituto Nazionale
di Astrofisica at Padova; The Johns Hopkins University; the National Science
Foundation of the USA (AST-0908326); the W. M. Keck foundation; the Macquarie
University; the Netherlands Research School for Astronomy; the Natural
Sciences and Engineering Research Council of Canada; the Slovenian Research
Agency; the Swiss National Science Foundation; the Science \& Technology
Facilities Council of the UK; Opticon; Strasbourg Observatory; 
and the Universities of Groningen, Heidelberg and Sydney. 
This work was partly supported by the European Union FP7 programme through ERC grant number 320360.
The RAVE web site is at
\url{http://www.rave-survey.org}.
\end{document}